\documentclass{article}
\usepackage{ijcai23}
\usepackage{times}
\usepackage{soul}
\usepackage{url}
\usepackage[hidelinks]{hyperref}
\usepackage[utf8]{inputenc}
\usepackage[small]{caption}
\usepackage{graphicx}
\usepackage{amsmath}
\usepackage{amsthm}
\usepackage{booktabs}
\usepackage{algorithm}
\usepackage{algorithmic}
\usepackage[switch]{lineno}

\usepackage{import}
\usepackage{ijcai23_preamble}

\title{The Computational Complexity of Single-Player Imperfect-Recall Games\footnote{Published in \textit{Proceedings of the Thirty-Second International Joint Conference on Artificial Intelligence (IJCAI-23)}, Macao, 2023. (\href{http://www.ijcai.org}{http://www.ijcai.org})}}

\author{
Emanuel Tewolde$^1$
\and
Caspar Oesterheld$^1$
\and
Vincent Conitzer$^{1,2}$
\And
Paul W. Goldberg$^2$
\affiliations
$^1$ Foundations of Cooperative AI Lab (FOCAL), Carnegie Mellon University
\\
$^2$ University of Oxford
\emails
emanueltewolde@cmu.edu,
oesterheld@cmu.edu,
conitzer@cs.cmu.edu,
paul.goldberg@cs.ox.ac.uk
}

\begin{document}

\maketitle

\begin{abstract}
    We study single-player extensive-form games with imperfect recall, such as the Sleeping Beauty problem or the Absentminded Driver game. For such games, two natural equilibrium concepts have been proposed as alternative solution concepts to ex-ante optimality. One equilibrium concept uses \emph{generalized double halving} (GDH) as a belief system and \emph{evidential decision theory} (EDT), and another one uses \emph{generalized thirding} (GT) as a belief system and \emph{causal decision theory} (CDT). Our findings relate those three solution concepts of a game to solution concepts of a polynomial maximization problem:\ global optima, optimal points with respect to subsets of variables and Karush–Kuhn–Tucker (KKT) points. Based on these correspondences, we are able to settle various complexity-theoretic questions on the computation of such strategies. For ex-ante optimality and (EDT,GDH)-equilibria, we obtain NP-hardness and inapproximability, and for (CDT,GT)-equilibria we obtain CLS-completeness results.

\end{abstract}

\section{Introduction}
\label{sec:introduction}

Most formalisms that are used for reasoning under uncertainty and for decision making in AI -- for example, HMMs, (dynamic) Bayesian networks, influence diagrams, MDPs, POMDPs, multiagent versions of these -- assume what is known as {\em perfect recall}: the agent does not forget anything it knew before.  This may seem to be a very natural assumption: in the design of AI agents, generally we have plenty of reliable memory available.  Moreover, the property of perfect recall ensures various desirable properties in the context of extensive-form games, including polynomial-time solvability of two-player zero-sum games~\cite{KollerM92} (and hence, {\em a fortiori}, single-player games).  Finally, even when modeling humans -- as in, for example, behavioral game theory~\cite{Camerer03:Behavioral} --  in spite of our clearly imperfect memory, usually perfect-recall models are used.  So why use models with imperfect recall in AI?

It turns out there are a number of reasons why imperfect recall is relevant for AI agents; moreover, in cases where it is relevant, it is clear what the agent will and will not remember -- unlike in the case of human memory, which is harder to predict and consequently to model in standard representations of imperfect recall. 
Imperfect-recall games already appear in the AI literature in the context of solving very large games such as poker: one technique for solving such games is {\em abstraction} -- i.e., reducing the game to a smaller, simplified one to solve instead -- and this process can give rise to imperfect recall in the abstracted game~\cite{Waugh09:Practical,Lanctot12:No,Kroer16:Imperfect}.  But imperfect recall is
also of interest for other reasons.
First, we may deliberately choose to have our agents forget: for example, the agent may temporarily need access to data that is sensitive from a privacy perspective, and therefore best forgotten afterwards. \citet{conitzer23:focal} give the example of an AI driver assistant that can take over whenever the human car driver makes a major error. When that happens, the AI needs to reason about how good the human driver is in general, about whom it is not allowed to store information. An AI agent could also take the form of a highly distributed system operating across many nodes, where not all the nodes have access to the same information; hence, it may act at one node without having access to information that it did have available when acting at another node.  Relatedly, the same agent (in the sense of being based on the same source code) may be instantiated multiple times, for example by human users deploying it in multiple contexts. In such cases it can still be useful to consider this family of instantiations as a single agent, but again these instantiations will not all have access to the same information. Finally, again building on the previous case, an agent may be acting not only in the real world, but also in simulations; for example, it may be simulated by another agent that wants to ensure that another instantiation of the same agent will act in a trustworthy fashion in the real world later \cite{Kovarik23}.  In this case, the real-world instantiation of the agent will generally not have access to the information that the simulation had access to earlier.

Notably, we need to model this phenomenon as imperfect recall rather than merely as imperfect information. In single-agent \textit{perfect-recall} imperfect-information games (say, a POMDP), there is never a (strict) reason to randomize, whereas in imperfect-recall games the agent might have to randomize in order to perform well overall; cf. the Absentminded Driver example \cite{PiccioneR73} (Appendix \ref{app:two game ex}). For example, suppose we deploy a content recommendation system to many people’s phones, in an edge-computing sort of setup: We are not in constant communication with the phones, so the nodes of our system have to act independently each day before getting back in touch with us. Over the next day, we would like to experiment (in an optimal way) what kind of content to recommend. With a pure strategy, we would show all users the same content and learn very little from it. Instead, we would prefer to randomize the content shown on each phone, that is, use a mixed strategy. Therefore, this situation cannot be a perfect-recall game (even of imperfect information).

Being able to make decisions with imperfect recall also represents a technical frontier.  Many existing techniques inherently rely on perfect recall. Solving two-player zero-sum games becomes NP-hard as soon as one player has imperfect recall \cite{KollerM92}. Moreover, in these contexts, there remains controversy at the very foundations of how to do probabilistic reasoning and decision making. For example, the Sleeping Beauty problem~\cite{Elga00:Self} (Appendix \ref{app:two game ex}) asks one to give the probability of a state of the world in an imperfect-recall setting; some (Thirders) believe that the correct answer is $1/3$, and others (Halvers) believe it is $1/2$, see Section \ref{sec:belief systems}. Only recently has a clear picture started to emerge regarding how each of these positions can be combined with a corresponding form of decision theory to make good decisions~\cite{Hitchcock04:Beauty,Draper08:Diachronic,Briggs10:Putting,Conitzer15:Dutch,Oesterheld22:Can}; here we build on that recent conceptual work to define and study several foundational computational problems.

In this paper, we use extensive-form games to represent settings with imperfect recall.  Even though we are considering a single-agent setting, the extensive form is still especially natural to use to model imperfect-recall settings, specifically with the use of information sets.  Indeed, as we will discuss, game-theoretic phenomena such as notions of equilibrium naturally come up in the presence of imperfect recall even when there is just a single agent. Intuitively that is because it is more challenging for that agent to coordinate its actions with those it takes at other times. Moreover, randomization is in general necessary.  We consider behavior strategies, which map each information set to a probability distribution over actions.  Based on recent literature, we study three distinct solution concepts: (1) {\em ex ante} optimality, where the behavior strategy is one that maximizes expected utility at the outset; (2) equilibria based on causal decision theory and generalized thirding, in which an agent would not want to change its action at any information set, under the assumption that at all other game tree nodes (including ones in the same information set) the agent would follow the original strategy; and (3) equilibria based on evidential decision theory and generalized double halving, in which an agent would not want to switch to a different distribution over actions at a given information set, assuming that the agent would also use the {\em new} distribution at other nodes in that information set (but would use the original strategy at all other information sets).  

Section \ref{sec:prelims 1} and \ref{sec:prelims 2} define those solution concepts and cover previously known characterizations and hardness results. Section \ref{sec:main results} presents our novel results: First we show that the equilibria based on causal decision theory and generalized thirding are exactly the Karush–Kuhn–Tucker points of a corresponding utility maximization problem. This makes gradient descent methods applicable to the computation of such an equilibrium, and, relatedly, we derive that problems of finding such an equilibrium -- up to an inverse exponential precision -- are complete for the class CLS (Continuous Local Search). Finally, we derive various NP-hardness results for maximizing over the set of equilibria and for finding an equilibrium based on evidential decision theory and generalized double halving. Naturally, all these complexity results also have implications for learning or dynamics that converge to these solutions.

\section{Background for Imperfect-Recall Games}
\label{sec:prelims 1}

\subsection{Single-Player Extensive-Form Games with Imperfect Recall}
\label{sec:games with imperfect recall}

We first define single-player extensive-form games, allowing for imperfect recall. The concepts we use in doing so are standard; for more detail and background, see, e.g.,~
\citet{Fudenberg91:Game_theory},  \citet{NisanTV07} and \citet{PiccioneR73}.

\begin{defn}
A \emph{single-player extensive-form game with imperfect recall} (denoted $\Gamma$), sometimes also called an extensive decision problem with imperfect recall, consists of:
\begin{enumerate}[nolistsep]
    \item A rooted tree, with nodes $\nds$ and where the edges are labeled with \emph{actions}. The game starts at the root node $h_0$ and finishes at a leaf node, also called \emph{terminal node}. The terminal nodes in $\nds$ will be denoted as $\term$. The set $A_h$ refers to the set of actions available at a nonterminal node $h \in \nds \setminus \term$.
    \item A \emph{utility function} $u : \term \to \R$, where $u(z)$ represents the payoff that the player receives from finishing the game at terminal node $z$.
    \item A partition $\nds \setminus \term = \nds_{*} \sqcup \nds_{c}$ of nonterminal nodes into a set of the player's \emph{decision nodes} $\nds_{*}$ and a set of \emph{chance nodes} $\nds_{c}$. This partition indicates whether the single player or exogenous stochasticity determines the action at any given node.
    \item For each chance node $h \in \nds_{c}$, a fixed distribution $\Prob_c(\cdot \mid h)$ over $A_h$ according to which chance determines an action at $h$.
    \item A partition $\nds_{*} = \sqcup_{I \in \infs_{*}} I$ of the player's decision nodes into \emph{information sets} (``info sets'' for short). We require $A_h = A_{h'}$ for any nodes $h, h'$ in the same info set $I$.
\end{enumerate}
\end{defn}

Throughout this paper, we let $\ninfs := |\infs_{*}|$ denote the number of info sets. For computational purposes, we assume that a game $\Gamma$ is represented by its game tree structure of size $\Theta(|\nds|)$ (which includes the info set partition), and by a binary encoding of its chance node probabilities and its utility payoffs. The last two shall take on rational values only.

Any node $h \in \nds$ uniquely corresponds to a (node,action)-history $\hist(h)$ from root $h_0$ to $h$ in the game tree. Define functions $d$, $\nu$, $a$ such that the node history and action history from $h_0$ to $h$ consist of the sequences $\big( \nu(h, 0), \, \nu(h, 1), \, ... \, , \, \nu(h, d(h)) \big)$ and $\big( \act(h, 0), \, \act(h, 1), \, ... \, , \, \act(h, d(h) - 1) \big)$ respectively. In other words, function $d: \nds \to \N_0$ identifies the tree depth of a node, $\nu: \nds \times \N_0 \to \nds$ the node ancestor at a specified depth, and $\act: \nds \times \N_0 \to \sqcup_{h \in \nds} A_h$ the action ancestor at a specified depth. In particular, for all $h \in \nds$, we have $\nu(h, 0) = h_0$ and $\nu(h, d(h)) = h$. We restrict the domain of functions $\nu$ and $a$ to inputs $(h,k)$ with $k \leq d(h)$ and $k \leq d(h) - 1$ respectively, and note that $a$ maps $(h, k)$ into $A_{\nu(h,k)}$.

The depth of $\Gamma$ is defined to be the maximal depth of the leaf nodes. For notational convenience, we add a singleton info set to $\Gamma$ for each chance node in $\Gamma$. The collection of these info sets, each consisting of a single element in $\nds_c$, shall be denoted by $\infs_c$. For each nonterminal node $h \in \nds \setminus \term$, let $I_h \in \infs_* \sqcup \infs_c$ denote its info set. For each info set $I \in \infs_* \sqcup \infs_c$, let $A_I$ denote its action set.

\begin{figure}[t]
\centering
\resizebox{8.5cm}{6cm}{
\begin{tikzpicture}

\node at(4.5,5.5){$\bullet$};\node at(4.5,5.8){$h_0$};
\node at(2,4){$\bullet$};\node at(2,4.3){$h_1$};
\node at(3,2){$\bullet$};\node at(3.1,2.3){$h_2$};
\node at(5,4){$\bullet$};\node at(5.3,4.3){$h_3$};
\node at(7,4){$\bullet$};\node at(7,4.3){$h_4$};

\node at(0,2){$\bullet$};
\node at(1.5,2){$\bullet$};
\node at(4.5,2){$\bullet$};
\node at(5.5,2){$\bullet$};
\node at(6.5,2){$\bullet$};
\node at(7.5,2){$\bullet$};

\node at(2.5,0.5){$\bullet$};
\node at(3.5,0.5){$\bullet$};

\draw[thin,black](4.5,5.5)--(2,4);\draw[thin,black](4.5,5.5)--(5,4);\draw[thin,black](4.5,5.5)--(7,4);
\draw[thin,black](2,4)--(0,2);\draw[thin,black](2,4)--(1.5,2);\draw[thin,black](2,4)--(3,2);
\draw[thin,black](5,4)--(4.5,2);\draw[thin,black](5,4)--(5.5,2);
\draw[thin,black](7,4)--(6.5,2);\draw[thin,black](7,4)--(7.5,2);
\draw[thin,black](3,2)--(2.5,0.5);\draw[thin,black](3,2)--(3.5,0.5);

\node[blue] at(2.5,5.5){$I_1$};\node[blue] at(3.5,3.7){$I_2$};
\draw[blue,dashed,rounded corners=10pt](1.5,3.5)--(1.5,4.5)--(4,6)--(5,6)--(5,5.2)--(2.1,3.5)--cycle;
\draw[blue,dashed,rounded corners=10pt](2.5,1.5)--(2.5,2.3)--(4.7,4.5)--(7.5,4.5)--(7.5,3.5)--(5,3.5)--(3.2,1.5)--cycle;

\node at(0,1.7){$0$};\node at(1.5,1.7){$0$};
\node at(2.5,0.2){$5$};\node at(3.5,0.2){$0$};
\node at(4.5,1.7){$0$};\node at(5.5,1.7){$0$};
\node at(6.5,1.7){$0$};\node at(7.5,1.7){$1$};

\node at(3,5){$L$};\node at(5,4.8){$C$};\node at(6.1,4.8){$R$};
\node at(0.8,3.2){$L$};\node at(2,3.2){$C$};\node at(2.7,3.2){$R$};
\node at(4.5,2.8){$X$};\node at(5.5,2.8){$Y$};
\node at(6.5,2.8){$X$};\node at(7.5,2.8){$Y$};
\node at(2.5,1.2){$X$};\node at(3.5,1.2){$Y$};

\end{tikzpicture}
}
\caption{Running example of a single-player extensive-form game with imperfect recall. It has two info sets $I_1$ and $I_2$, and five nonterminal nodes.}
\label{fig:main game ex}
\end{figure}
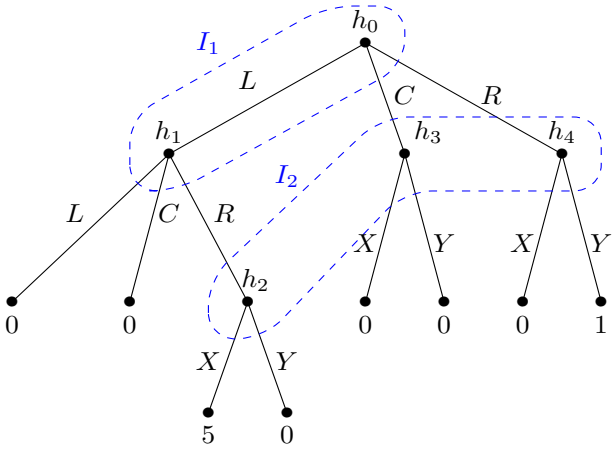

Nodes of the same info set are assumed to be indistinguishable to the player during the game (even though the player is always aware of the full game structure). There may be information about the history of play that the player holds at some node, and that the player forgets somewhere further down its subtree. For instance, consider the game in Figure~\ref{fig:main game ex}. Once the player arrives at node $h_3$, she cannot distinguish it from possibly being at node $h_2$. Thus she has already forgotten that she has only taken one action (action C) so far. In contrast to that, games with perfect recall have every info set reflect that the player remembers all her earlier actions. In particular, the player does not forget which info sets she entered in which order in the history of play.

With imperfect recall, it could furthermore be the case that multiple nodes of the same history (of some terminal node) belong to the same info set, as in info set $I_1$ in the game of Figure~\ref{fig:main game ex}. The inability of a player to distinguish between two nodes on the same history is a property that we will refer to as \textit{absentmindedness}; cf. the Absentminded Driver from \citet{PiccioneR73} (Appendix \ref{app:two game ex}).

Let $\Delta(A_I)$ denote the set of probability distributions over the actions in $A_I$. A (behavioural) \textit{strategy} $\mu : \infs_{*} \to \sqcup_{I \in \infs_{*}} \Delta(A_I)$ of the player assigns to each info set $I$ a probability distribution $\mu( \cdot \mid I) \in \Delta(A_I)$. At info set $I$, the player will then randomly draw an action according to $\mu( \cdot \mid I)$. By abuse of notation, we extend any strategy $\mu$ of the player to info sets $I_h \in \infs_c$ of chance nodes $h \in \nds_c$ by setting $\mu( \cdot \mid I_h) := \Prob_c(\cdot \mid h)$ there.

Given that the player is currently at node $\bar{h} \in \nds \setminus \term$ and that she plays according to strategy $\mu$, we can calculate the probability of reaching node $h \in \nds$ by multiplying the probabilities of the actions on the path from $\bar{h}$ to $h$:
\begin{align*}
    \Prob(h \mid \mu, \bar{h}) = \prod_{k=d(\bar{h})}^{d(h) - 1} \mu \big( \act(h, k) \mid I_{\nu(h, k)} \big) \quad \textnormal{if } \bar{h} \in \hist(h) 
\end{align*}
and $\Prob(h \mid \mu, \bar{h}) = 0$ otherwise. As a special case, we define the reach probability $\Prob(h \mid \mu) := \Prob(h \mid \mu, h_0)$ of a node $h \in \nds$ to be its reach probability from the root $h_0$ of $\Gamma$. Naturally, the reach probability of the root is $1$.

The expected utility payoff for being at node $h \in \nds \setminus \term$ and using strategy $\mu$ from then on can be determined by 
\\
$\U(\mu \mid h) := \sum_{z \in \term} \Prob(z \mid \mu, h) \cdot u(z)$.
Furthermore, let $\U : \mu \mapsto \U(\mu) := \U(\mu \mid h_0)$ be the function that takes a strategy $\mu$ of $\Gamma$ and returns the expected utility payoff of the player from following $\mu$ from the game start to termination. $\U(\mu)$ is also called the \textit{ex-ante} (expected) utility of $\mu$.

\subsection{Utility as a polynomial function}
\label{sec: Q is poly fct}

Fix an ordering $I_1, \ldots, I_\ninfs$ of the info sets in $\infs_{*}$ and denote $m_i := |A_{I_i}|$ for all $i \in [\ninfs]$. Moreover, fix an ordering $a_1, \ldots, a_{m_i}$ of the actions in $A_{I_i}$ for all $i \in [\ninfs]$.

We can uniquely describe a strategy $\mu$ of $\Gamma$ by the probability values that it assigns to each action $a_j$ at info set $I_i$, for $i\in [\ninfs]$ and $j\in[m_i]$. A strategy $\mu$ is a vector $\mu = (\mu_{ij})_{i,j} \in \bigtimes_{i = 1}^\ninfs \R^{m_i}$ such that each subvector $\mu_{i \cdot} = (\mu_{ij})_{j}$ lies in the simplex $\Delta(A_{I_i}) \equiv \Delta^{m_i - 1} := \{ y \in \R^{m_i} \, : \, y_j \geq 0 \, \forall j\, , \sum_{j = 1}^{m_i} y_j = 1\}$.
Therefore, the strategy space of $\Gamma$ is $\bigtimes_{i = 1}^\ninfs \Delta(A_{I_i}) \equiv \bigtimes_{i = 1}^\ninfs \Delta^{m_i-1}$.

The expected utility function $\U$ of a strategy $\mu$ can be fully written out as
\begin{align*}
\begin{aligned}
    \U(\mu) = \sum_{z \in \term} \Big( u(z) \cdot \prod_{k=0}^{d(z) - 1} \mu \big( \act(z, k) \mid I_{\nu(z, k)} \big) \Big) \, . 
\end{aligned}
\end{align*}
As noted by \citet{PiccioneR73}, this is a polynomial function in the variables $(\mu_{ij})_{i,j}$. Recall that at chance nodes, the probabilities are exogenously fixed constants.
Thus, the degree of the polynomial function $\U$ is upper-bounded by the maximum number of times the player of $\Gamma$ might have to take a decision in order to reach a terminal node. Note that polynomial $\U$ can be constructed in polynomial time in the encoding size of $\Gamma$.
\begin{ex}
    In the game of Figure~\ref{fig:main game ex}, we get $\ninfs = 2$, $m_1 = 3$, $m_2 = 2$. Let the actions be ordered as $(L,C,R)$ and $(X,Y)$. Then, for any point $\mu \in \R^{3} \times \R^{2}$, we have $\U(\mu) = 5 \mu_{11} \mu_{13} \mu_{21} + \mu_{13} \mu_{22}$.
\end{ex}

We show in Appendix \ref{app: poly fcts to games} that one can also reduce any multivariate polynomial $p : \bigtimes_{i = 1}^\ninfs \R^{m_i} \to \R$ to a single-player extensive-form game $\Gamma$ with imperfect recall such that its expected utility function satisfies $\U(\mu) = p(\mu)$ on $\bigtimes_{i = 1}^\ninfs \R^{m_i}$.

\subsection{(Computing) Ex-ante Optimal Strategies}
\label{sec: ex ante optimality}

Suppose we want to \emph{solve} a given game $\Gamma$. From a planning perspective, one would naturally search for a strategy that promises the highest payoff at a time before the player enters the game.

\begin{defn}
    We say a strategy $\mu^*$ is ex-ante optimal for $\Gamma$ if it solves
    \begin{align}
    \label{exantemaxprobl}
        \max_{\mu} \, \U(\mu) \quad \textnormal{s.t.} \quad \mu \in \bigtimes_{i = 1}^\ninfs \Delta(A_{I_i}) \, .
    \end{align}
\end{defn}

Due to \citet{KollerM92}, we can find an ex-ante optimal strategy for a single-player game with perfect recall in polynomial time. This will not be the case anymore in the presence of imperfect recall, as we will show next.

The class ZPP contains all those decision problems that can be solved in expected polynomial time by a randomized (Las Vegas) algorithm. Let OPT be an optimization problem with max and min values $\bar{q}$ and $\underline{q}$. Then, a fully polynomial-time approximation scheme (FPTAS) for OPT computes a solution to an instance of OPT with an objective value that is at most $\epsilon \cdot (\bar{q} - \underline{q})$ away from the optimal value. This computation must take polynomial time in $1 / \epsilon$ and the encoding size of the instance. For a more precise definition, see \citet{Klerk08}.

\begin{prop}
\label{hardness and inapprox of ex ante opt}
Consider the problem that takes a game $\Gamma$ and target value $t \in \Q$ (encoded in binary) as inputs and asks whether there is a strategy $\mu$ for $\Gamma$ with ex-ante expected utility $\U(\mu) \geq t$. This problem is NP-hard. Moreover:
\begin{enumerate}[nolistsep]
    \item[(1.)] Unless NP = ZPP, there is no FPTAS for this problem. NP-hardness and conditional inapproximability hold even if the game instance $\Gamma$ has a tree depth of $3$ and only one info set.
    \item[(2.)] NP-hardness holds even if $\Gamma$ has no absentmindedness, a tree depth of $4$ and the player has $2$ actions per info set.
    \item[(3.)] NP-hardness holds even if $\Gamma$ has no absentmindedness, a tree depth of $3$ and the player has $3$ actions per info set.
\end{enumerate}
\end{prop}

\citet{Gimbert20} shows that this problem of deciding whether a given target value can be achieved in~$\Gamma$ is in fact $\exists \mathbb{R}$-complete (a complexity class described in the last paragraph of this section).

An early NP-hardness result of that kind was given by \citet{KollerM92}. Note that finding the ex-ante optimal strategy of $\Gamma$ is at least as hard as this NP-hard decision problem of whether a target value can be achieved in~$\Gamma$. On the other hand, with an efficient solver of the decision version, one can recover the optimal ex-ante utility $\U^* := \max_{\mu} \U(\mu)$ through binary search.

A proof of Proposition \ref{hardness and inapprox of ex ante opt} can be found in Appendix \ref{app: ex ante opt}. Result (1.)\ is based on our reduction from polynomials to games and with the known hardness of maximizing a polynomial function over a simplex \cite{Klerk08}. Result (3.)\ reduces from the hard problem of finding an optimal joint strategy in multiplayer common payoff games \cite{ChuH01}. The proofs reveal that NP-hardness remains even if the encoding size of the chance node probabilities and utility values are in $O(|\nds|)$.

As for complexity upper bounds, consider the complexity class $\exists \mathbb{R}$ called
\emph{the existential theory of the reals} \cite{RENEGAR1992255,SchaeferS17}. It consists of all those problems that reduce to deciding whether a sentence of the following form is true: $\exists x_1 \ldots \exists x_n F(x_1, \ldots x_n)$, where the $x_i$ are real-valued variables and where $F$ is a quantifier-free formula that may contain equalities and inequalities of real polynomials. $\exists \mathbb{R}$ lies in between NP and PSPACE \cite{Shor90,Canny88}. By Section \ref{sec: Q is poly fct}, it is straight-forward to see that the decision problem of Proposition \ref{hardness and inapprox of ex ante opt} is contained in $\exists \mathbb{R}$. As of now, it is unclear whether NP membership can be shown; in part because easy-to-encode games may only admit ex-ante optimal strategies that take on irrational numbers  (see Appendix \ref{app:irrational solutions}). But we can decide an approximate version of the problem in Proposition \ref{hardness and inapprox of ex ante opt} in NP time; namely, when it is allowed to incorrectly return ``yes'' to the problem instance $(\Gamma, t, \epsilon)$ if there exists a strategy profile $\mu$ with $\U(\mu) \geq t - \epsilon$. Here, $\epsilon > 0$ represents an inverse-exponential precision parameter.

\section{Equilibria in Imperfect-Recall Games}
\label{sec:prelims 2}

Proposition \ref{hardness and inapprox of ex ante opt} shows a strong obstacle to finding or approximating ex-ante optimal strategies for single-player extensive-form games with imperfect recall. In light of these limitations, we will relax the space of solutions to \textit{equilibrium} strategies. This solution concept argues that, whenever the player finds herself in an info set, she has no influence over which actions she chooses at other info sets. Therefore, at an equilibrium strategy $\mu$, the player will play the best action at each info set, assuming that she has been playing according to $\mu$ up to the current decision point and that she will continue to do so at future decision points. Prior work has given a detailed description of viable equilibrium concepts in single-player games with imperfect recall \cite{PiccioneR73,Briggs10:Putting,Oesterheld22:Can}. We will consider two well-motivated equilibrium concepts that have been proposed and where an ex-ante optimal strategy also constitutes an equilibrium. In games without absentmindedness, these two equilibrium concepts coincide. In games with absentmindedness, the concepts differ in how expected utilities are evaluated for an action $a$ at a current info set $I$, given that the player plays according to strategy $\mu$ anywhere ``else''. Computing such expected utilities requires
\begin{enumerate}[nolistsep]
\item A Belief System: A method to form beliefs (i.e., a probability distribution) over being at a specific node/history of $\Gamma$ given that the player is at info set $I$; and
\item A Decision Theory: An understanding of how an action choice at the current node affects the freedom to choose an action at other nodes of \textit{the same} info set.
\end{enumerate}

In the sequel, let $I$ be the player's current info set at which she finds herself during play while playing $\mu$ in game $\Gamma$.

\subsection{Decision Theories}

Causal Decision Theory (CDT) postulates that the player can take an action $\alpha \in \Delta(A_I)$ at the current node without violating that the player has been playing according to $\mu$ at past arrivals at $I$, or that she will be playing according to $\mu$ at future arrivals at $I$. The intuition behind CDT is that the player's choice to deviate away from $\mu$ at the current node does not \textit{cause} any change in behaviour at any other node of the same info set $I$.

In contrast to that, Evidential Decision Theory (EDT) postulates that if the player takes an action $\alpha \in \Delta(A_I)$ at the current node, then she will have also deviated to $\alpha$ whenever she arrived in $I$ in past play, and she will be deviating to $\alpha$ whenever she arrives in $I$ again in future play. Indeed, EDT argues that the choice to deviate to $\alpha$ now is evidence for the player taking the same deviation choice in the past and future.

Denote with $\mu_{I \, \mapsto \alpha}$ an EDT deviation, i.e., the strategy of $\Gamma$ that plays according to $\mu$ at every info set except at the info set $I \in \infs_{*}$ where it plays according to $\alpha \in A_I$. By contrast, a CDT deviation may result in different actions taken at the same info set. This might not constitute a valid strategy that the player could have picked before the game started.

\begin{ex}
    Consider the game in Figure~\ref{fig:main game ex} and suppose the player enters the game with the strategy $\mu = (R, X)$. Say, upon visiting info set $I_1$, the player plans to deviate from the $\mu$-prescribed action $R$ to the action $L$ this one time only. Then, CDT argues that the player will stick to her $\mu$-prescribed action R at the other node of $I_1$, leading to one of the two action histories $(L,R,X)$ or $(R,X)$. EDT, on the other hand, argues that such a deviation will then happen at both nodes of $I$, leading to the action histories (L,L).
\end{ex}

\subsection{Self-locating Belief Systems}
\label{sec:belief systems}

Let $I^{\first} \subseteq I$ refer to those nodes $h \in I$ that are the first node of their history to enter info set $I$. Define the reach probability and (expected) visit frequency of $I$ under $\mu$ as $\Prob(I \mid \mu) := \sum_{h \in I^{\first}} \Prob(h \mid \mu)$ and $\Fr(I \mid \mu) := \sum_{h \in I} \Prob(h \mid \mu)$. Note that the reach probability and the visit frequency can only differ in games with absentmindedness, and that the visit frequency can be greater than $1$. However, we have in general that $\Prob(I \mid \mu) > 0$ if and only if $\Fr(I \mid \mu) > 0$. Finally, denote with $\chi: P \rightarrow \{0,1\}$ the function that takes a Boolean property $P$ as input and evaluate $1$ if and only if $P$ is true.

The first belief system argues that one should focus on the visit frequencies:

\begin{defn}
    Let $I$ be an info set with $\Fr(I \mid \mu) > 0$ under $\mu$, and let $h \in \nds_*$ be a player node. Then, Generalized Thirding (GT) determines the probability of the player to be at $h$, given that she uses $\mu$ and is currently in $I$, through
        \[ \Prob_{\GT}(h \mid \mu, I) := \chi(h \in I) \cdot \frac{\Prob(h \mid \mu)}{\Fr(I \mid \mu)} \, . \]
\end{defn}

The second belief system argues that one should rather focus on the reach probabilities. Note that the statement $I \cap \hist(z) \neq \emptyset$ evaluates as true if and only if $I$ occurs in history $\hist(z)$ of terminal node $z \in \term$ at least once. 

\begin{defn}
    Let $I$ be an info set with $\Prob(I \mid \mu) > 0$ under $\mu$, and let $z \in \term$ be a terminal node. Then, Generalized Double Halving (GDH) determines the probability of the player being on the path $\hist(z)$ to terminal node $z$, given that she uses $\mu$ and is currently in $I$, through
    \[\Prob_{\GDH}( \, \hist(z) \mid \mu, I \,) := \chi(I \cap \hist(z) \neq \emptyset) \cdot \frac{\Prob(z \mid \mu)}{\Prob(I \mid \mu)} \, . \]
\end{defn}

GT and GDH were introduced as ``consistency'' and ``z-consistency'' by \cite{PiccioneR73}.

With the current definitions, GT and GDH assign probabilities to different type of events (to be at player node versus to be in the history of a terminal node). In Appendix \ref{app:comparing GT with GDH}, we phrase GT and GDH in each other's language. In the language of GDH, GT assigns the event of being in history $\hist(z)$ of a terminal node $z$ a higher probability if the reach probability of $z$ under $\mu$ is higher (same as GDH) and if $\hist(z)$ visits info set $I$ very often (whereas GDH only cares about $I$ being visited at least once by $\hist(z)$).

\begin{ex}
\label{ex thirders halvers}
    Consider the game in Figure~\ref{fig:main game ex} again and suppose the player enters the game with the strategy $\mu = (\frac{1}{2}L + \frac{1}{2}R, X)$. Say, the player observes to be in info set $I_1$. Then a GT player believes to be at the node $h_0$ in the history $(R,X)$ with probability $\frac{1}{3}$ whereas a GDH player believes to be at $h_0$ in $(R,X)$ with probability $\frac{1}{2}$. The names ``Halving'' and ``Thirding'' originate from this contrast but for a different example called Sleeping Beauty \cite{Elga00:Self} (Appendix \ref{app:two game ex}).
\end{ex}

\subsection{Two equilibrium concepts}
\label{sec:eq def}

Any claims made in this section are proven in Appendix 
\ref{app: cdt eqs vs edt eqs}.

We start with the equilibrium concept that uses Causal Decision Theory and Generalized Thirding. Denote with $h \circ a$ the child node reached in $\Gamma$ by following action $a \in A_h$ from player node $h \in \nds_*$. Then $\U(\mu \mid h \circ a)$ is the expected utility the player receives from being at $h$, playing $a$ now, and playing according to $\mu$ afterwards.

\begin{defn}
\label{defn cdtgt exp util}
    Let the player currently be at an info set $I$ with $\Fr(I \mid \mu) > 0$, and let $\alpha \in \Delta(A_I)$ be a mixed action. Then, the \emph{(CDT,GT)-expected utility} of playing $\alpha$ now and according to $\mu$ otherwise is 
    \begin{align*}
        \EU_{\CDT,\GT}(\alpha \mid \mu, I) := \sum_{h \in I} \Prob_{\GT}&(h \mid \mu, I)
        \\
        &\cdot \Big( \sum_{a \in A_I} \alpha(a) \cdot \U(\mu \mid h \circ a) \Big) \, .
    \end{align*}
\end{defn}
Note that the inner sum collapses to $\U(\mu \mid h \circ a)$ if the considered mixed action $\alpha$ is a pure action $a$.
\begin{defn}
\label{CDT GT eq def}
    We say a strategy $\mu^*$ of $\Gamma$ is a \emph{(CDT,GT)-equilibrium} if for all info sets $I \in \infs_*$ with $\Fr(I \mid \mu^*) > 0$ under $\mu^*$, we have 
    \begin{align*}
        \mu^*(\cdot \mid I) \in \argmax_{\alpha \in \Delta(A_I)} \EU_{\CDT,\GT}(\alpha \mid \mu^*, I) \, .
    \end{align*}
\end{defn}

Alternatively, we can use the easier-to-check condition that for all info sets $I \in \infs_*$ with $\Fr(I \mid \mu^*) > 0$ and all pure actions $a \in A_I$ with $\mu^*(a \mid I) > 0$, we have
\begin{align}
\label{pure action CDT GT eq def}
    a \in \argmax_{a' \in A_I} \EU_{\CDT,\GT}(a' \mid \mu^*, I) \, .
\end{align}

Next, we introduce the equilibrium concept that uses Evidential Decision Theory and Generalized Double Halving. 

\begin{defn}
\label{GDH exp util}
    Let the player currently be at an info set $I$ with $\Prob(I \mid \mu) > 0$, and let $\alpha \in \Delta(A_I)$ be a mixed action. Then, the \emph{(EDT,GDH)-expected utility} of playing $\alpha$ now and according to $\mu$ otherwise is
    \begin{align*}
        \EU_{\EDT,\GDH}(\alpha \mid \mu, I) := \sum_{z \in \term} \Prob_{\GDH}( \, \hist(z) \mid \mu_{I \, \mapsto \alpha}, I \,) \cdot u(z) \, .
    \end{align*}
\end{defn}

The GDH belief probabilities in Definition \ref{GDH exp util} are well-defined due to $\Prob(I \mid \mu) = \Prob(I \mid \mu_{I \, \mapsto \alpha})$.

\begin{defn}
\label{EDT eq}
    We say a strategy $\mu^*$ of $\Gamma$ is a \emph{(EDT,GDH)-equilibrium} if for all info sets $I \in \infs_*$ with $\Prob(I \mid \mu^*) > 0$ under $\mu^*$, we have
    \begin{align}
    \label{EDT GDH eq def}
        \mu^*(\cdot \mid I) \in \argmax_{\alpha \in \Delta(A_I)} \EU_{\EDT,\GDH}(\alpha \mid \mu^*, I) \, .
    \end{align}
\end{defn}

For (EDT,GDH), it is not sufficient to only check for optimality of pure actions that are in the support of $\mu^*$. For instance, take the game in Figure~\ref{fig:main game ex} and suppose the player enters the game with the strategy $\mu = (C, X)$. Say, the player observes to be in info set $I_1$. Then action C is optimal among pure actions \{L, C, R\}. But the player would strictly benefit from deviating to mixed action $\frac{1}{2} \text{L} + \frac{1}{2} \text{R}$.

Finally, observe that in games without absentmindedness, the following notions coincide: CDT and EDT, GT and GDH, and Definitions \ref{defn cdtgt exp util} and \ref{GDH exp util}. In particular, both equilibrium concepts coincide:

\begin{lemma}\label{lem:no-absent}
In games without absentmindedness, a strategy $\mu$ is a (CDT,GT)-equilibrium if and only if it is an (EDT,GDH)-equilibrium.
\end{lemma}

\subsection{Equilibria from the Ex-Ante Perspective}
\label{sec:eq ex ante charact}

Recall from Section \ref{sec: Q is poly fct} that the (ex-ante) strategy utility function $\U$ of $\Gamma$ is a polynomial function from $\bigtimes_{i = 1}^\ninfs \R^{m_i}$ to $\R$. In this section, we give characterizations for (CDT,GT)- and (EDT,GDH)-equilibria in terms of $\U$, as presented by \citet{Oesterheld22:Can} and \citet{PiccioneR73}. We reprove these results in the appendix since our setup and end goal differs slightly. 

Polynomial $\U$ is continuously differentiable in $\mu \in \bigtimes_{i = 1}^\ninfs \R^{m_i}$. For $i \in [\ninfs]$ and $j \in [m_{i}]$, let $\nabla_{i j} \, \U$ stand for the partial derivative in direction $(i, j)$, that is, the linear change of $\U$ at a point $\mu$ if you infinitesimally increase its $\mu( a_{j} \mid I_i)$ value. 

\begin{lemma}
\label{derivative equals CDT deviation}
    Let $I_i$ be an info set, $a_j \in A_{I_i}$ an action, and $\mu \in \bigtimes_{i = 1}^\ninfs \Delta(A_{I_i})$ a strategy. Then:
    \begin{enumerate}[nolistsep] 
        \item $\nabla_{ij} \, \U(\mu) = 0 \quad $ if \, $\Fr(I_i \mid \mu) = 0$, and 
        \item $\nabla_{ij} \, \U(\mu) = \Fr(I_i \mid \mu) \cdot \EU_{\CDT,\GT}(a_j \mid \mu, I_i)$ otherwise.
    \end{enumerate}
\end{lemma}

Note that an infinitesimal increase of $\mu( a_{j} \mid I_i)$ means in a game-theoretic sense that the decision for action $a_j$ is made slightly more probable at \emph{every} node of info set $I_i$. This resembles an EDT type of deviation power but restricted to small deviations that stay close to the current action profile $\mu$. Then, Lemma \ref{derivative equals CDT deviation} says that a CDT deviation -- rescaled by $\Fr(I_i \mid \mu)$ -- accurately captures the linear (=dominant) effect of such a ``local EDT deviation''. 

\begin{lemma}
\label{EDT eq to NE of Poly}
    Strategy $\mu \in \bigtimes_{i = 1}^\ninfs \Delta(A_{I_i})$ of $\Gamma$ is an (EDT,GDH)-equilibrium if and only if for all $i \in [\ninfs]$:
    \begin{align*}
        \mu_{i \cdot} \in \argmax_{y \in \Delta(A_{I_i})} \, \, \U(\mu_{1 \cdot}, \ldots, \mu_{i-1 \cdot}, y, \mu_{i + 1 \cdot}, \ldots, \mu_{\ninfs \cdot}) \, .
    \end{align*}
\end{lemma}

One possible interpretation of Lemma \ref{EDT eq to NE of Poly} is that (EDT,GDH)-equilibria of $\Gamma$ are exactly the Nash equilibria of an $\ninfs$-player simultaneous and identical-interest game $G$: Each player $i$ shall have the continuous action space $\Delta(A_{I_i})$ and the (single) utility payoff, a function of the chosen action profile $(\mu_{i \cdot})_{i=1}^\ninfs \in \bigtimes_{i = 1}^\ninfs \Delta(A_{I_i})$, shall be the polynomial $\U$.

\subsection{Computational Considerations}
\label{sec:comp aspects}

One of our main results addresses the complexity of finding a (CDT,GT)-equilibrium. There are problem instances where all (CDT,GT)-equilibria take on irrational numbers even though the game is easy to encode (Appendix \ref{app:irrational solutions}). Therefore, we relax our search to $\epsilon$-approximate (CDT,GT)-equilibria where $\epsilon$ is an inverse-exponential numerical precision parameter.

\begin{defn}
\label{approx cdt eq def}
    An instance of the problem {\sc (CDT,GT)-equilibrium} consists of a single-player extensive-form game $\Gamma$ with imperfect recall and a precision parameter $\epsilon > 0$ encoded in binary. A solution consists of a strategy $\mu$ for $\Gamma$ that satisfies for all $I \in \infs_{*}$ with $\Fr(I \mid \mu) > 0$:
    \[ 
        \EU_{\CDT,\GT}\big( \mu( \cdot \mid I) \mid \mu, I \big) \geq \max_{a' \in A_I} \EU_{\CDT,\GT}(a' \mid \mu, I) - \epsilon \, .\]
\end{defn}
There is also an alternative notion of being close to an equilibrium, called $\epsilon$-well-supported (CDT,GT)-equilibrium. It instead requires condition (\ref{pure action CDT GT eq def}) to be satisfied up to $\epsilon$ precision. An analysis of when both approximation concepts are polynomial-time related can be found in Appendix~\ref{app: proofs of main results}.

We will give hardness results and restricted membership results for {\sc (CDT,GT)-equilibrium} for the class CLS (Continuous Local Search). CLS was introduced by \citet{DaskalakisP11} who noted that it contains various important problems of continuous local optimization that belong both to PPAD and PLS. PPAD \cite{Pap94} is well-known as the class that captures the complexity of many problems of Nash equilibrium computation (\cite{DGP09,CDT09} and much subsequent work), while PLS (Polynomial Local Search \cite{JPY88}) represents the complexity of many problems of discrete local optimization. Recently, \citet{FGHS23} showed that CLS is \emph{equal} to the intersection of PPAD and PLS, indicating that CLS-hardness is quite a reliable notion of computational difficulty. In addition, the hardness of CLS can also be based on the cryptographic assumption of indistinguishability obfuscation \cite{HubacekY17}. \citet{FGHS23} also showed that a version of the KKT point search problem is CLS-complete. Subsequent work has established CLS-completeness of mixed Nash equilibria of congestion games \cite{BabichenkoR21} and solutions to a certain class of contests \cite{EGG22}. We can characterize CLS through any of its complete problems.

We will mainly be interested in KKT points. Consider a general non-linear maximization problem 
\begin{align}
\label{general max problem}
    \max_{x \in \R^n} \, f(x) \quad \textnormal{s.t.} \quad B x + b \leq 0 \, , \, C x + c = 0
\end{align}
where $f : \R^n \to \R$ is continuously differentiable, $B \in \R^{m \times n}$, $b \in \R^m$, $C \in \R^{\ninfs \times n}$, and $c \in \R^\ninfs$, and the domain is bounded. A point $x \in \R^n$ is then said to be a KKT point for~(\ref{general max problem}) if there exist KKT multipliers $\tau_1, \ldots, \tau_m, \kappa_1, \ldots, \kappa_\ninfs \in \R$ such that $Bx + b \leq 0$ and $Cx + c = 0$, and $\forall j \in [m] \, : \, \tau_j \geq 0$, $ \forall j \in [m] \, : \, \tau_j = 0$ or $B_{j \cdot}x + b_j = 0$, and 
\[
    \nabla \, f(x) = \sum_{j = 1}^m \tau_j \cdot (B_{j \cdot})^T + \sum_{i = 1}^\ninfs \kappa_i \cdot (C_{i \cdot})^T = 0 \, .
\]
The KKT conditions are necessary first-order conditions for a point to be a local optimum of (\ref{general max problem}). Furthermore, feasible stationary points satisfy the KKT conditions.

\section{Main Results}
\label{sec:main results}

To our knowledge, the results of this section are all novel unless explicitly stated otherwise. All proofs can be found in Appendix \ref{app: proofs of main results}.

\subsection{Complexity of the Search Problems}

First, we use Lemma \ref{derivative equals CDT deviation} to give a characterization of (CDT,GT)-equilibria in terms of ex-ante utility. For that, recall the ex-ante maximization problem (\ref{exantemaxprobl}).

\begin{thm}
\label{CDT GT eq iff KKT point}
    Strategy $\mu \in \bigtimes_{i = 1}^\ninfs \Delta(A_{I_i})$ of $\Gamma$ is a (CDT,GT)-equilibrium if and only if $\mu$ is a KKT point of (\ref{exantemaxprobl}).
\end{thm}

\begin{figure}[t]
    \centering
    \includegraphics[width=6.5cm]{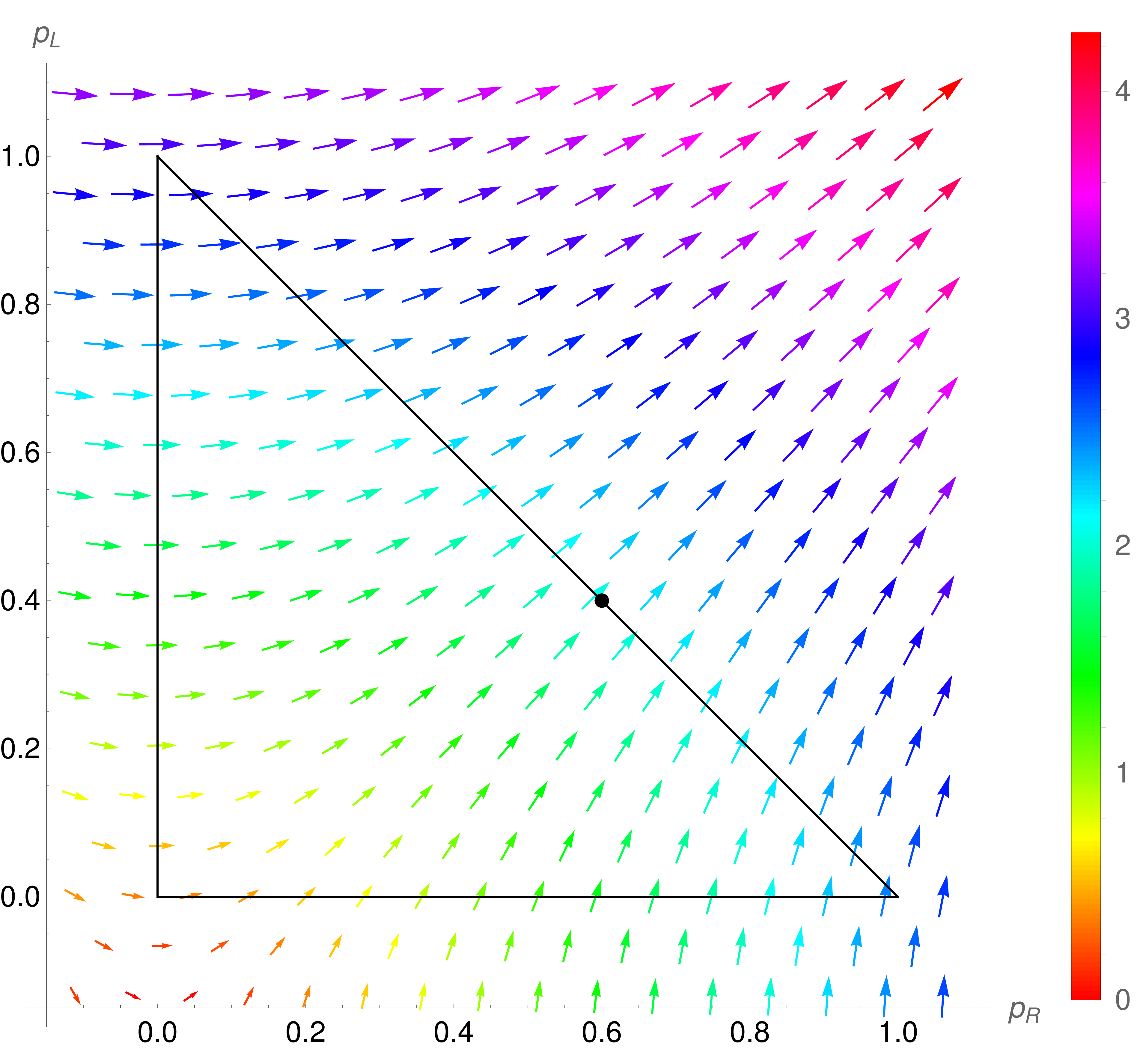}
    \caption{A plot of the gradient vector field $\nabla \, \U$ of the strategy utility function $\U$. The underlying game is the one from Figure~\ref{fig:main game ex} except where the player nodes of info set $I_2$ are replaced by chance nodes that choose actions $X$ and $Y$ with chance probabilities $\frac{1}{2}$ each. Then the player only has to choose a mixed action for info set $I_1$. The plot is in 2D for visualization convenience: The x-axis and y-axis represent the probabilities put on action $R$ and $L$ respectively. The action simplex $\Delta(\{L,C,R\})$ becomes a right triangle with the point $(0,0)$ corresponding to pure action $C$. The gradient coloring represents the vector length.
    \\
    There is no KKT point in the interior of the (projected) simplex because the gradient does not vanish there. The KKT points on the boundary of the simplex are those where the gradient is directed perpendicularly outwards of the boundary constraint (except corner points, whose gradient only needs to lie in the positive cone of the boundary constraint directions). Thus $( 0.6 \cdot L + 0.4 \cdot R )$ is the only KKT point. A game-theoretic analysis of the underlying game $\Gamma$ also yields $( 0.6 \cdot L + 0.4 \cdot R )$ as the only (CDT,GT)-equilibrium.}
    \label{fig:kkt visual}
\end{figure}

We visualize Theorem \ref{CDT GT eq iff KKT point} in Figure \ref{fig:kkt visual}. 
This result also reveals a method to find (CDT,GT)-equilibria, namely by applying Gradient Descent on $\U$. Note that in continuous optimization, there can be KKT points that are not locally optimal. An analogous effect can also happen in games with imperfect recall: In the game of Figure~\ref{fig:main game ex}, strategy $\mu = (C, X)$ is a (CDT,GT)-equilibrium. But it is not a local optimum because for any $\epsilon>0$ a shift from $\mu( \cdot \mid I_1) = C$ to $\epsilon \cdot (\frac{1}{2}\text{L}+\frac{1}{2}\text{R}) + (1 - \epsilon) \cdot \text{C}$ would yield the player ex-ante utility $5 \cdot \frac{\epsilon}{2} \cdot \frac{\epsilon}{2} > 0$. However, from a (CDT,GT) standpoint, the player should be satisfied with her choice at $I_1$. In terms of the original definition of CDT, this is because deviating exactly \textit{once} in $I_1$ never suffices to attain a utility of $5$. In terms of Lemma~\ref{derivative equals CDT deviation} and Theorem~\ref{CDT GT eq iff KKT point}, the issue is that the first order effect of increasing the probabilities of $R$ and $L$ is $0$.

The three solution concepts considered in this paper form an inclusion hierarchy, a result shown by \citet{Oesterheld22:Can} [cf.~\citeauthor{PiccioneR73}, \citeyear{PiccioneR73}]:

\begin{lemma}
\label{equilibrium hierarchy}
    An ex-ante optimal strategy of a game $\Gamma$ is also an (EDT,GDH)-equilibrium. An (EDT,GDH)-equilibrium is also a (CDT,GT)-equilibrium.
    
    In particular, any single-player extensive-form game $\Gamma$ with imperfect recall admits an (EDT,GDH)-equilibrium and a (CDT,GT)-equilibrium.
\end{lemma}

The implication chain of Lemma \ref{equilibrium hierarchy} does not hold in the reverse direction: Consider the game in Figure \ref{fig:main game ex}. Then strategy $\mu = (C,X)$ is a (CDT,GT)-equilibrium, but not an (EDT,GDH)-equilibrium. Moreover, strategy $\mu' = (R,Y)$ is an (EDT,GDH)-equilibrium with ex-ante utility $1$. This is not ex-ante optimal because strategy $\mu'' = (\frac{1}{2} L + \frac{1}{2} R,X)$ achieves the (optimal) ex-ante utility $5/4$.

The second part of Lemma \ref{equilibrium hierarchy} holds because ex-ante optimal strategies always exist. This is in contrast to the multi-player setting where Nash equilibria may not exist in the presence of imperfect recall. Moreover, Lemma \ref{equilibrium hierarchy} implies that finding an ex-ante optimal strategy must be at least as hard as finding an (EDT,GDH)-equilibrium which must be at least as hard as finding a (CDT,GT)-equilibria. For the latter, we get the following classification:
\begin{thm}
\label{CLS hardness}
    {\sc (CDT,GT)-equilibrium} is CLS-hard. CLS-hardness holds even for games restricted to:
    \begin{enumerate}[nolistsep]
        \item[(1.)]\label{two-actions} a tree depth of $6$ and the player has $2$ actions per info set,
        \item[(2.)] no absentmindedness and a tree depth of $6$, and
        \item[(3.)] no chance nodes, a tree depth of $5$, and only one info set.
    \end{enumerate}
    \smallskip
    The problem is in CLS for the subclass of problem instances of {\sc (CDT,GT)-equilibrium} where a lower bound on positive visit frequencies in $\Gamma$ is easily obtainable.
\end{thm}
All CLS results of Theorem \ref{CLS hardness} also hold analogously for the search problem of an approximate well-supported (CDT,GT)-equilibrium. We prove (1.)\ by a reduction from finding a KKT point of a polynomial function over the hypercube. For (2.)\ and (3.), we reduce from finding a Nash equilibrium of a polytensor identical interest game. Both search problems we reduce from were shown to be CLS-complete by \citet{BabichenkoR21}.

The CLS membership in Theorem \ref{CLS hardness} implies that, unless NP = co-NP, the considered problem cannot be hard for the class NP \cite{MegiddoP91}. We only prove CLS membership for those games $\Gamma$ where we can construct a lower bound value $\lambda > 0$ that satisfies $\Fr(I \mid \mu) = 0$ or $\Fr(I \mid \mu) \geq \lambda$ for all strategies $\mu$ and info sets $I$ in $\Gamma$. That is because if $\Fr(I \mid \mu) > 0$ is too small, the approximation error may explode when transitioning from the ex-ante perspective $\nabla_{ij} \, \U(\mu)$ to the \emph{de se} perspective $\EU_{\CDT,\GT}(a_j \mid \mu, I_i)$. Fortunately, such a lower bound exists and is polynomial-time computable for many well-known imperfect-recall games, such as the Absentminded Driver, game variants of the Sleepy Beauty problem, and all the games used in the CLS-hardness results of Theorem \ref{CLS hardness}. Thus, the computation of an approximate (CDT,GT)-equilibrium is CLS-complete in those games that admit such a lower bound on positive visit frequencies. Statement (2.)\ shows in particular that absentmindedness is not the reason for CLS hardness. With Lemma \ref{lem:no-absent}, this implies
\begin{cor}
In games without absentmindedness where a lower bound on positive visit frequencies is easily obtainable, it is CLS-complete to find an $\epsilon$-(EDT,GDH)-equilibrium.
\end{cor}

The authors are not aware of any complexity classification for the problem of \emph{finding} an approximate (EDT,GDH)-equilibrium in games that may have absentmindedness -- even though Lemma \ref{EDT eq to NE of Poly} gives a nice optimization characterization of (EDT,GDH)-equilibria. Nonetheless, we are able to give conditional inapproximability results for (EDT,GDH)-equilibria with the next theorem.

\subsection{Complexity of the Decision Problems}

Next, we show that maximizing expected utility in an info set or maximizing over the space of equilibria is NP-hard. In the following problem formulations, any target value $t \in \Q$ shall be encoded in binary.

\begin{thm}
\label{dec probs of eqs are np hard}
    The following problems are all NP-hard. Unless NP = ZPP, there is also no FPTAS for these problems.
    \begin{enumerate}[nolistsep]
    \item[(1a.)] Given $\Gamma$ and $t \in \Q$, is there a (CDT,GT)-equilibrium of $\Gamma$ with ex-ante utility $\geq t$?
    \item[(1b.)] Given $\Gamma$, an info set $I$ of $\Gamma$ and $t \in \Q$, is there a (CDT,GT)-equilibrium $\mu$ with $\Fr( I \mid \mu) > 0$, and such that the player has a (CDT,GT)-expected utility $\geq t$ upon reaching $I$?
    \item[(1c.)] Given $\Gamma$, an info set $I$ of $\Gamma$ and $t \in \Q$, is there a strategy $\mu$ of $\Gamma$ with $\Fr( I \mid \mu) > 0$, and such that the player has a (CDT,GT)-expected utility $\geq t$ upon reaching $I$?
    \item[(2a.)] Given $\Gamma$ and $t \in \Q$, is there an (EDT,GDH)-equilibrium of $\Gamma$ with ex-ante utility $\geq t$?
    \item[(2b.)] Given $\Gamma$, an info set $I$ of $\Gamma$ and $t \in \Q$, is there an (EDT,GDH)-equilibrium $\mu$ with $\Prob( I \mid \mu) > 0$, and such that the player has an (EDT,GDH)-expected utility $\geq t$ upon reaching $I$?
    \item[(2c.)] Given $\Gamma$, an info set $I$ of $\Gamma$ and $t \in \Q$, is there a strategy $\mu$ of $\Gamma$ with $\Prob( I \mid \mu) > 0$, and such that the player has an (EDT,GDH)-expected utility $\geq t$ upon reaching $I$?
    \item[(3a.)] Given $\Gamma$ and $t \in \Q$, do all (EDT,GDH)-equilibria of $\Gamma$ have ex-ante utility $\geq t$?
    \item[(3b.)] Given $\Gamma$, an info set $I$ of $\Gamma$ and $t \in \Q$, do all (EDT,GDH)-equilibria $\mu$ with $\Prob( I \mid \mu) > 0$ yield the player an (EDT,GDH)-expected utility $\geq t$ upon reaching~$I$?
    \end{enumerate}
\end{thm}

All the results of Theorem \ref{dec probs of eqs are np hard} follow from Proposition \ref{hardness and inapprox of ex ante opt}. Therefore, NP-hardness and conditional inapproximability remain for problems of the form (-a.)\ even if we restrict the game instances as described in Proposition \ref{hardness and inapprox of ex ante opt}. The same holds for problems of the form (-b.)\ and (-c.)\ except that we have to add one information set and one tree depth level to the game instances. Hardness of decision problem (3a.)\ also relies on the observation that in games with one info set only, any (EDT,GDH)-equilibrium is also ex-ante optimal (cf. Lemma \ref{EDT eq to NE of Poly}). 

From (3a.)\ we obtain in particular that, unless NP = ZPP, there is no FPTAS for the search problem of an (EDT,GDH)-equilibrium in games with imperfect recall\footnote{Note that FPTAS even allow for an approximation up to an inverse-\emph{polynomial} precision $\epsilon$.}. To compare this to Theorem \ref{CLS hardness}, we remark that this conditional inapproximability result for (EDT,GDH)-equilibria (and ex-ante optimal strategies) is obtained even for games where a lower bound on positive visit frequencies is easily obtainable.

The decision problems of the form (1-.), (2a.), and (2c.)\ are all members of the complexity class $\exists \mathbb{R}$, and therefore, in particular, in PSPACE. On one hand, this is because (CDT,GT)-expected utilities and (EDT,GDH)-expected utilities can be described as rational functions (fractions of polynomial functions). Furthermore, this is because the alternative definition (\ref{pure action CDT GT eq def}) of a (CDT,GT)-equilibrium gives rise to polynomially many comparisons of polynomial functions, and, for (2a.), because ex-ante optimal strategies are (EDT,GDH)-equilibria.

\section{Conclusion}
\label{sec:conclusion}

Games of imperfect recall have traditionally often been considered a theoretical curiosity; it is hard to model settings with human actors as imperfect-recall games, because, while most of us frequently forget things, we do not {\em reliably} forget things according to well-specified rules.  For AI agents, however, this is no longer true; moreover, because they can be instantiated many times, sometimes in simulation, one instantiation will generally not know what another knew earlier.  All this motivates the {\em computational} study of games of imperfect recall, which we initiated here for the single-player case.  We are aided in this endeavor by recent conceptual work that specifies and motivates several natural solution concepts, and we based our work on these.  Standard polynomial-time algorithms such as ones based on the sequence form are known to no longer work in the presence of imperfect recall. In this paper we found various complexity-theoretic evidence that indeed, single-player imperfect-recall games are hard to solve. Some of this evidence is, intriguingly, based on the complexity class CLS whose careful study is only very recent.  On the positive side, we also provided insights into solving such games by drawing close connections to several problems about maximizing polynomial functions.

There remain many avenues for future work.  What can be said about these computational problems for representation schemes other than the extensive form? Are there special cases of imperfect-recall games that can be solved more efficiently, whether they are single-player or multi-player?
One may also ask whether our results give insight into the more conceptual questions.  For example, to the extent that (CDT,GT)-equilibria are (under reasonable complexity assumptions) easier to compute than (EDT,GDH)-equilibria, does that provide support for using the former solution concept, at least for certain purposes? We hope that the work we have done in this paper can serve as a springboard for further research into this fascinating and important topic.

\section*{Acknowledgements}
We are grateful to Manolis Zampetakis, Vojtěch Kovařík and the anonymous reviewers for their valuable feedback on this project.
Emanuel Tewolde, Caspar Oesterheld and Vincent Conitzer thank the Cooperative AI Foundation, Polaris Ventures (formerly the Center for Emerging Risk Research) and Jaan Tallinn's donor-advised fund at Founders Pledge for financial support. Paul Goldberg is currently supported by a JP Morgan faculty award.

\bibliographystyle{named}
\bibliography{ijcai23}

\,
\clearpage
\appendix

\section{On Section \ref{sec: Q is poly fct}: Reductions from Polynomial Maximization to Games}
\label{app: poly fcts to games}

We consider problems of optimizing polynomials $p : \bigtimes_{i = 1}^\ninfs \R^{m_i} \to \R$, where each subset of $m_i$ variables are constrained to lie in the standard $(m_i-1)$-simplex.

At the end of Section \ref{sec: Q is poly fct}, we mention that we can reduce a polynomial $p : \bigtimes_{i = 1}^\ninfs \R^{m_i} \to \R$ to a single-player extensive-form game $\Gamma$ with imperfect recall such that $p = \U^{\Gamma}$ on \, $\bigtimes_{i = 1}^\ninfs \R^{m_i}$. Let us show this by giving two variants of the same reduction idea.

A general polynomial function $p : \bigtimes_{i = 1}^\ninfs \R^{m_i} \to \R$ of degree $d$ can be uniquely represented in terms of the standard monomial basis $\big\{ \prod_{i,j = 1}^{\ninfs,m_i} x_{ij}^{D_{ij}} \big\}_{D \in \MB( d, \bm{m} )}$. Here, we summarized $(m_i)_{i=1}^\ninfs$ to the vector $\bm{m}$, and we denote with $\MB( d, \bm{m} )$ the set
\[ 
    \{ D = (D_{ij})_{ij} \in \bigtimes_{i=1}^\ninfs \N_0^{m_i} \, : \, \sum_{i,j = 1}^{\ninfs,m_i} D_{ij} \leq d \}
\]
of all variations to draw up to $d \in \N$ elements out of the set $\{ (i,j) \}_{i,j = 1}^{\ninfs,m_i}$, with replacement and without regard to draw order. Each element $D \in \MB( d, \bm{m} )$ captures the variable degrees of the monomial basis element $\prod_{i,j = 1}^{\ninfs,m_i} x_{ij}^{D_{ij}}$ that it represents.

Throughout this paper (specifically, this appendix), if the instance to a problem contains a polynomial function $p : \bigtimes_{i = 1}^\ninfs \R^{m_i} \to \R$, then we assume it to be represented as a binary encoding of $\ninfs, (m_i)_{i=1}^\ninfs$, and the polynomials coefficients $(\lambda_D)_{D \in \MB( d,\bm{m} )}$, which need to be rational. This is also called the Turing (bit) model. Denote $\supp(p) := \{ D \in \MB( d,\bm{m} ) \, : \, \lambda_D \neq 0 \})$  and, for each $D \in \MB( d,\bm{m} )$, $\supp(D) := \{ (i,j) \text{ with } i \in [\ninfs], j \in [m_i] \, : \, D_{ij} > 0 \})$ as well as $|D| := \sum_{i,j = 1}^{\ninfs,m_i} D_{ij}$. Let $\supp(D)^{\textnormal{ms}}$ be the multiset that contains $D_{ij}$ many copies of the element $(i,j)$ in it for each $(i,j) \in \supp(D)$ (in multisets, duplicate elements are allowed). Then $|\supp(D)^{\textnormal{ms}}| = |D|$.

Given such a polynomial function, let us construct a corresponding single-player extensive-form game $\Gamma$ with imperfect recall. It shall have info sets $I_i$ for $i \in [\ninfs]$ with action sets $A_{I_i} := \{a_1, \ldots, a_{m_i} \}$, and a tree depth of up to $d+1$. Let the root $h_0$ be a chance node that has one outgoing edge to depth $1$-node $h_D$ for each monomial index $D \in \supp(p)$. An outgoing edge is drawn uniformly at random. First, handle the special case of $D$ being the zero vector. There, $h_D$ will be a terminal node with a utility payoff of $|\supp(p)|$. So consider $D \neq \0$ from now on, where $h_D$ will be a nonterminal node. Denote the subtree rooted at $h_D$ with $T_D$. Keep in mind that it is associated with monomial $\prod_{i,j = 1}^{\ninfs,m_i} x_{ij}^{D_{ij}}$. We will now build $T_D$ depth layer by depth layer, until a tree of depth $|D|$, by lexicographically going through the set $\supp(D)^{\textnormal{ms}}$. Depth $k-1$ of $T_D$ corresponds to the $k$-th element of $\supp(D)^{\textnormal{ms}}$, referred to as $\big( i(k),j(k) \big)$. We present two variants with how to continue building $\Gamma$:

{\bf Variant 1:} There will be at most one nonterminal node $h$ of depth $k-1$ of $T_D$ (it would be $h_D$ on depth $0$). Assign $h$ to info set $I_{i(k)}$. Create $m_{i(k)}$ outgoing edges out of $h$ into a new node of depth $k$, and the edges shall be labeled with $\{a_1, \ldots, a_{m_{i(k)}} \}$ of $A_{I_{i(k)}}$. The created node $h \circ a_{j(k)}$ shall be non-terminal and the created nodes $h \circ a_j$ for $j \neq j(k)$ shall be terminal with utility payoff $0$. With this procedure, subtree $T_D$ will have depth $\sum_{(i,j) \in \supp(D)} D_{ij}$. The nonterminal node of the last depth layer has action history $\big( a_{j(k)} \in A_{I_{i(k)}} \big)_{k \in [|D|]}$ in $T_D$. Now reverse the fact that it is nonterminal, and make it terminal instead. Denote it as $z_D$ and assign it a utility of $\lambda_D \cdot |\supp(p)|$.

{\bf Variant 2:} This variant does not have terminal nodes in depth layers $0, \ldots, |D| - 1$. Each node $h$ of depth $k-1$ of $T_D$ shall belong to info set $I_{i(k)}$ (Depth $0$ only has one node, namely $h_D$). Create $m_{i(k)}$ outgoing edges out of each $h$ into a new node of depth $k$. The edges shall be labeled with $\{a_1, \ldots, a_{m_{i(k)}} \}$ of $ \in A_{I_{i(k)}}$. The nodes of the last depth layer (depth $|D|$) shall be terminal nodes. One of those nodes has the action history $\big( a_{j(k)} \in A_{I_{i(k)}} \big)_{k \in [|D|]}$ in $T_D$, which we will refer to as $z_D$. Assign it a utility of $\lambda_D \cdot |\supp(p)|$. Assign all other terminal nodes of $T_D$ a utility of $0$.
\\

In both variants, we have that any point $x \in \bigtimes_{i = 1}^\ninfs \R^{m_i}$ of $p$ that is also in $\bigtimes_{i = 1}^\ninfs \Delta^{m_i-1}$ naturally comprises a strategy $\mu$ in $\Gamma$ with probabilities $\mu( a_j \mid I_i) = x_{ij}$. Moreover, the strategy utility function $\U$ of both variants of $\Gamma$ satisfies
\begin{align*}
    \U(\mu) &= \sum_{z \in \term} \Prob(z \mid \mu) \cdot u(z) = 
    \\
    &= \sum_{D \in \supp(p)} \Prob(z_D \mid \mu) \cdot \lambda_D \cdot |\supp(p)| 
    \\
    &= \sum_{D \in \supp(p)} \bigg[ \Big( \frac{1}{|\supp(p)|} \cdot \prod_{D_{ij} > 0} x_{ij}^{D_{ij}} \Big) \cdot \lambda_D \cdot |\supp(p)| \bigg] 
    \\
    &= \sum_{D \in \MB( d,\bm{m} ) } \lambda_D \cdot \prod_{i,j} x_{ij}^{D_{ij}}
    \\
    &= p(x)
\end{align*}
for corresponding $x$ and $\mu$. This extends to $\U = p$ on all $\bigtimes_{i = 1}^\ninfs \R^{m_i}$. 

The construction of the first variant of $\Gamma$ takes polynomial time in the encoding size of $p : \bigtimes_{i = 1}^\ninfs \R^{m_i} \to \R$. That is, because the game tree has size $\leq |\supp(p)| \cdot \text{deg}(p) \cdot (\max_i m_i)$. The second variant of $\Gamma$ can have exponential game tree size in general. But, if we know that polynomial instances $p$ have fixed degree $d$ for example, then the tree size is $\leq |\supp(p)| \cdot (\max_i m_i)^d$ which makes the whole construction of $\Gamma$ polynomial time in the size of the input instances again. An advantage of the second variant is that in that game, the reach probability $\Prob(I \mid \mu)$ and visit frequency $\Fr(I \mid \mu)$ of an info set $I \in \infs$ are independent of the used strategy $\mu$. That is because a subtree $T_D$ is chosen by nature at the beginning, and within $T_D$, any outcome path visits the exact same info sets with the same multiplicity. The reach probability of info set $I_i$ becomes 
\[
    \Prob(I_i) = \Prob(I_i \mid \mu) = \sum_{D \in \supp(p)} \frac{1}{|\supp(p)|} \cdot \chi ( \sum_{j \in [m_i]} D_{ij} \geq 1 )
\]
and its visit frequency becomes 
\begin{align}
\label{poly fct to game v2 freqs}
    \Fr(I_i) = \Fr(I_i \mid \mu) = \sum_{D \in \supp(p)} \frac{1}{|\supp(p)|} \cdot \sum_{j \in [m_i]} D_{ij} \, .
\end{align}
We will make use of this observation later when it comes to games that admit an easy-to-compute lower bound on positive visit frequencies.

\section{On Section \ref{sec: ex ante optimality}: Proof of Proposition \ref{hardness and inapprox of ex ante opt}}
\label{app: ex ante opt}

We recall and prove Proposition \ref{hardness and inapprox of ex ante opt}:

\begin{prop*}
Consider the problem that takes a game $\Gamma$ and target value $t \in \Q$ (encoded in binary) as inputs and asks whether there is a strategy $\mu$ for $\Gamma$ with ex-ante expected utility $\U(\mu) \geq t$. This problem is NP-hard. Moreover:
\begin{enumerate}[nolistsep]
    \item[(1.)] Unless NP = ZPP, there is no FPTAS for this problem. NP-hardness and conditional inapproximability hold even if the game instance $\Gamma$ has a tree depth of $3$ and only one info set.
    \item[(2.)] NP-hardness holds even if $\Gamma$ has no absentmindedness, a tree depth of $4$ and the player has $2$ actions per info set.
    \item[(3.)] NP-hardness holds even if $\Gamma$ has no absentmindedness, a tree depth of $3$ and the player has $3$ actions per info set.
\end{enumerate}
\end{prop*}

\begin{proof}
(1.)
\\
Appendix \ref{app: poly fcts to games} gives a reduction from maximizing polynomial $p : \bigtimes_{i = 1}^\ninfs \R^{m_i} \to \R$ over the product of simplices to maximizing ex-ante utility in a single-player extensive-form game $\Gamma$ with imperfect recall (take the first variant of the reduction). \citet{Klerk08} gives a survey on maximizing polynomials over popular compact domains. Consider the decision problem that takes as an instance a polynomial function $p : \Delta^{m-1} \to \R$ and a target value $t \in \Q$, and that has to decide whether there exists a point $x \in \Delta^{m-1}$ with $p(x) \geq t$. \citet{Klerk08} note that this problem is NP-hard and has no FPTAS unless NP=ZPP, even if $p$ is known to be quadratic functions only. They derive the conditional inapproximability from \citet{Hastad96}. In terms of our reduction, such polynomials reduce to a game of depth $3$ and one info set. Moreover, a instance of the polynomial problem is a yes instance if and only if the reduced instance of ex-ante utility game problem of this proposition is a yes instance.

(2.)
\\
Reduce from 3SAT. Let $x_1,\ldots,x_\ninfs$ be the variables of a 3CNF formula $\phi$ having $n$ clauses. We construct the corresponding game instance $\Gamma$ as follows. Each variable $x_i$ has corresponding info set $I_i$ whose nodes have 2 outgoing edges with action labels $T,F$. The root of the tree is a chance node that selects amongst $n$ subtrees, corresponding to the clauses, with equal probability $1/n$. Each subtree is a binary tree of depth 3. If the clause $C$ associated with the subtree contains variables $x_i,x_j,x_k$, then the root node of the subtree shall belong to info set $I_i$, its two children belong to info set $I_j$, and their four children belong to info set $I_k$. Each leaf has a path that is associated by truth assignments of $x_i,x_j,x_j$. A leaf shall yield a utility of $1$ if the associated truth assignment satisfies the clause $C$, else value 0. The utility target value for the game instance shall be $1$.

Then, $\phi$ is satisfiable if and only if there is a strategy in the corresponding game with ex-ante utility $1$. Note that every clause subtree is reached with positive probability and that an ex-ante utility of $1$ means that any realization of the candidate strategy must only reach terminal node with utility payoff $1$. For the backward implication, say $\mu$ is a behavioural (=randomizing) strategy with an ex-ante utility of $1$. Take any pure-action strategy $\mu'$ that only uses actions that are in the support of $\mu$. Then $\mu'$ makes a satisfying truth assignment for $\phi$.

(3.)
\\
Consider the following type of $2$-player games $G$: A game $G$ consists of a family $(G_s)_{s \in S}$ of $2$-player simultaneous games where $|S| \in \N$ is finite. There is a probability distribution over $S$ determining which game the two players will play. There is a partition $S = \sqcup_{I \in \infs_*} I$ for player $1$ and $S = \sqcup_{J \in \mathcal{J}_*} J$ for player $2$ representing the info sets of each player within which the respective player cannot differentiate. For any $I \in \infs_*$, games $G_s$ with $s \in I$ have the same action set $A_I$ for player $1$. Analogously, for any $J \in \mathcal{J}_*$, games $G_s$ with $s \in J$ have the same action set $A_J$ for player $2$. First, nature draws $s \in S$ according to the probability distribution, then each player simultaneously takes an action in $G_s$. Such an outcome $(s, a \in A_{I_s}, a' \in A_{J_s})$ yields each player the same payoff $u(s, a, a')$. A strategy for player $1$ (resp. player $2$) returns a mixed action $\alpha \in \Delta(A_I)$ to each set $I \in \infs_*$ (resp. an $\alpha \in \Delta(A_J)$ to each set $J \in \mathcal{J}_*$.

\citet{ChuH01} show that given a game $G$ as described above, it is NP-hard to decide whether there is a strategy profile $(\mu_1, \mu_2)$ that yields a payoff $\geq 1$ in $G$. After a closer inspection of their reduction from 3SAT, we can see that their NP-hardness result holds even for game instances where each player only has up to three actions in each $G_s$ and where all the payoffs are as simple as $0$ or $3$. They also show that such a game $G$ can be transformed into an extensive-form game in polynomial time. We will use almost the same transformation in our upcoming reduction from their problem to our problem of interest.

Take an instance $G$ as described above. Assume further that $G$ has up to three actions in each $G_s$.  Create a single-player extensive-form game $\Gamma$ with imperfect recall in the following way: $\Gamma$ has depth $3$ and information sets $\infs_* \cup \mathcal{J}_*$. The root of $\Gamma$ is a chance node that chooses a subtree $T_s$ associated with $s \in S$ according to the given probability distribution over $S$. The root of a subtree $T_s$ shall be assigned to info set $I_s \in \infs_*$ and it shall have $A_{I_s}$ outgoing edges. Each node of subtree $T_s$ of subtree depth level $1$ shall be assigned to info set $J_s \in \mathcal{J}_*$ and they shall have $A_{J_s}$ outgoing edges. Each node of the whole game $\Gamma$ of depth level $3$ shall be a terminal node, and if its associated action history is $(s, a \in A_{I_s}, a' \in A_{J_s})$, then it shall yield the single player a payoff of $u(s, a, a')$. Also, let the target value $t$ for $\Gamma$ be $1$. A strategy $(\mu_1, \mu_2)$ of $G$ shall correspond to the strategy $\mu$ of $\Gamma$ that returns $\mu_1(I) \in \Delta(A_I)$ for info sets $I \in \infs_* \subset \infs_* \cup \mathcal{J}_*$ and returns $\mu_1(J) \in \Delta(A_J)$ for info sets $J \in \mathcal{J}_* \subset \infs_* \cup \mathcal{J}_*$

Then, there is a strategy profile $(\mu_1, \mu_2)$ of $G$ that yields a payoff $\geq 1$ if and only if its corresponding strategy $\mu$ in $\Gamma$ yields an ex-ante utility $\geq t$.

Note that in this reduction, $\Gamma$ has no absentmindedness, a tree of depth $3$, and at most three actions for each player info set. 

\end{proof}

\section{Popular Game Examples of Imperfect Recall}
\label{app:two game ex}

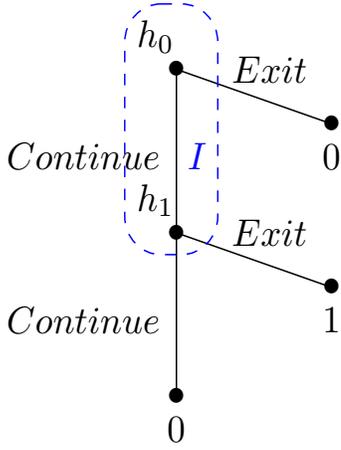
\begin{figure}[t]
\centering
\resizebox{5cm}{6cm}{
\begin{tikzpicture}

\node at(3,3.5){$\bullet$};\node at(2.8,3.8){$h_0$};
\node at(3,2){$\bullet$};\node at(2.8,2.3){$h_1$};

\node at(4.5,3){$\bullet$};
\node at(4.5,1.5){$\bullet$};
\node at(3,0.5){$\bullet$};

\draw[thin,black](3,3.5)--(4.5,3);\draw[thin,black](3,3.5)--(3,2);
\draw[thin,black](3,2)--(4.5,1.5);\draw[thin,black](3,2)--(3,0.5);

\node[blue] at(3.2,2.7){$I$};
\draw[blue,dashed,rounded corners=10pt](2.5,4.1)--(3.4,4.1)--(3.4,1.8)--(2.5,1.8)--cycle;

\node at(4.5,2.7){$0$};\node at(4.5,1.2){$1$};\node at(3,0.2){$0$};

\node at(3.9,3.5){$Exit$};\node at(2.1,2.7){$Continue$};
\node at(3.9,2){$Exit$};\node at(2.1,1.2){$Continue$};

\end{tikzpicture}
}
\caption{Absentminded driver game. The driver reaches his destination by taking the second of two exits from
the road he is driving on. Unfortunately, when approaching an exit, he forgets whether he has passed another exit before. Any pure strategy yields payoff 0. His best strategy is to exit with probability $\frac{1}{2}$ which gives him a payoff of $\frac{1}{4}$.}
\label{fig:adgame}
\end{figure}

\begin{figure}[t]
\centering
\resizebox{6cm}{5cm}{
\begin{tikzpicture}

\node at(3,3.5){$\bullet$};\node at(3,3.8){$h_0$};
\node at(1,2.5){$\bullet$};\node at(0.9,2.8){$h_1$};
\node at(5,2.5){$\bullet$};\node at(5.2,2.8){$h_2$};
\node at(5,1.5){$\bullet$};\node at(5.3,1.6){$h_3$};

\node at(1,0.5){$\bullet$};
\node at(5,0.5){$\bullet$};

\draw[thin,black](3,3.5)--(1,2.5);\draw[thin,black](1,2.5)--(1,0.5);
\draw[thin,black](3,3.5)--(5,2.5);\draw[thin,black](5,2.5)--(5,1.5);\draw[thin,black](5,1.5)--(5,0.5);

\node[blue] at(3,2.1){$I$};
\draw[blue,dashed,rounded corners=10pt](0.5,2.2)--(0.5,3)--(5.5,3)--(5.5,1.2)--(4.8,1.4)--cycle;

\node at(1,0.2){$0$};\node at(5,0.2){$0$};

\node at(1.9,3.3){$\frac{1}{2}$};\node at(4.1,3.3){$\frac{1}{2}$};

\end{tikzpicture}
}
\caption{Sleeping Beauty. In the beginning, a coin is flipped in secret from the game participant, the sleeping beauty. Next, she is put to sleep, and woken up. If the coin flip turned out heads, the game ends. If the coin flip turned out tails, she will be put to sleep again, and woken up again, after which the game ends. At any point she wakes up, she forgot whether she has been awake before. When woken up (= arriving in $I$), what should her belief probabilities be on the coin flip having turned out heads (= her being in $h_1$)? Generalized Thirding argues it should be $\frac{1}{3}$ whereas Generalized Double Halving argues it should be $\frac{1}{2}$.}
\label{fig:sbgame}
\end{figure}
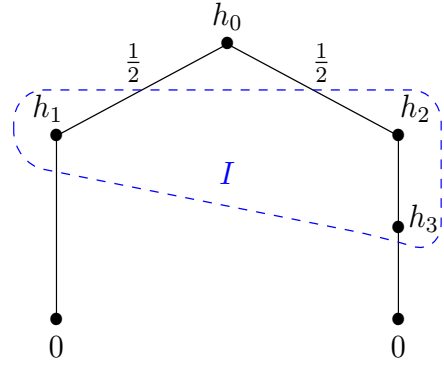

In Figure \ref{fig:adgame}, we describe the Absentminded Driver game. In Figure \ref{fig:sbgame}, we describe the Sleeping Beauty problem, which is less of a decision problem, but rather a probability puzzle.

\section{On Section \ref{sec:belief systems}: Comparing Generalized Thirding and Generalized Double Halving}
\label{app:comparing GT with GDH}

This section stays close to the exposition of \citet{Oesterheld22:Can}.

Let $\Gamma$ be a single-player extensive-form game with imperfect recall, $\mu$ be the strategy with which the player entered the game, and let the player find herself at an info set $I \in \infs_*$ with $\Prob(I \mid \mu) > 0 \iff \Fr(I \mid \mu) > 0$.

Recall that then, GT determines the probability of being at player node $h \in \nds_*$ specifically as 
\[
    \Prob_{\GT}(h \mid \mu, I) = \chi(h \in I) \cdot \frac{\Prob(h \mid \mu)}{\Fr(I \mid \mu)} \, .
\]
The number of times info set $I$ occurs in the history of a terminal node~$z$ can be described with $| I \cap \hist(z) |$. Let ``at $h$ in $\hist(z)$'' be the event that the player is currently at node $h$ and on the root-to-end path $\hist(z)$ for some given terminal node $z$. Then we get the following GT probabilities for being in the history $\hist(z)$ of a specified terminal node $z \in \term$:
\begin{align*}
    \Prob_{\GT}( \, \hist(z) \mid \mu, I \,) &= \sum_{h \in \nds_*} \Prob_{\GT}( \, \text{at} \, h \, \text{in } \hist(z)  \mid \mu, I \,)
    \\
    &:= \sum_{h \in \nds_* \cap \hist(z)} \Prob_{\GT}( h \mid \mu, I \,) \cdot \Prob(z \mid \mu, h)
    \\
    &= \sum_{h \in I \cap \hist(z)} \frac{\Prob(h \mid \mu)}{\Fr(I \mid \mu)} \cdot \Prob(z \mid \mu, h)
    \\
    &= | I \cap \hist(z) | \cdot \frac{\Prob(z \mid \mu)}{\Fr(I \mid \mu)}
\end{align*}
Compare this to the definition of GDH:
\[
    \Prob_{\GDH}( \, \hist(z) \mid \mu, I \,) = \chi(I \cap \hist(z) \neq \emptyset) \cdot \frac{\Prob(z \mid \mu)}{\Prob(I \mid \mu)}
\]
So GT assigns a history $\hist(z)$ of a terminal node $z$ a higher probability if the reach probability of $z$ under $\mu$ is higher (same as GDH) and if $\hist(z)$ visits info set $I$ very often (GDH only cares about $I$ being visited at least once).

As we can see, GT assigns being in the history of a terminal node $z$ a higher probability if the reach probability $\Prob(z \mid \mu)$ under $\mu$ is higher and/or if the history $\hist(z)$ enters info set $I$ very often.

When it comes to events of the form ``at $h$ in $\hist(z)$'', then GDH will just uniformly distribute the probability of being in $\hist(z)$ among all those nodes in $\hist(z)$ that are also in $I$. With this, GDH can also assign probabilities to the event of being at a specified node $h \in \nds_*$, namely, as
\begin{align*}
    \Prob_{\GDH}&( h \mid \mu, I \,)
    \\
    &= \sum_{z \in \term} \Prob_{\GDH}( \, \text{at} \, h \, \text{in } \hist(z)  \mid \mu, I \,)
    \\
    &= \sum_{\substack{ z \in \term \\ h \in I \cap \hist(z)}} \Prob_{\GDH}( \, \text{at} \, h \, \text{in } \hist(z)  \mid \mu, I \,)
    \\
    &:= \sum_{\substack{ z \in \term \\ h \in I \cap \hist(z)}} \frac{\Prob_{\GDH}( \, \hist(z) \mid \mu, I \,)}{| I \cap \hist(z) |}
    \\
    &= \chi(h \in I) \cdot \sum_{\substack{ z \in \term \\ h \in \hist(z)}} \frac{\Prob(z \mid \mu)}{\Prob(I \mid \mu)} \cdot \frac{1}{| I \cap \hist(z) |}
    \\
    &= \chi(h \in I) \cdot \sum_{\substack{ z \in \term \\ h \in \hist(z)}} \frac{\Prob(h \mid \mu) \cdot \Prob(z \mid \mu, h)}{\Prob(I \mid \mu) \cdot | I \cap \hist(z) |}
    \\
    &= \chi(h \in I) \cdot \frac{\Prob(h \mid \mu)}{\Prob(I \mid \mu)} \cdot \sum_{\substack{ z \in \term \\ h \in \hist(z)}} \frac{\Prob(z \mid \mu, h)}{| I \cap \hist(z) |}
\end{align*}

\section{On Section \ref{sec:eq def}: Proofs}
\label{app: cdt eqs vs edt eqs}

This section stays close to the ideas of \citet{Oesterheld22:Can}.

\paragraph{Mixed vs Pure Action Deviations in (CDT,GT).}

We have
\begin{align}
\label{cdt util are linear combi}
\begin{aligned}
    &\EU_{\CDT,\GT}(\alpha \mid \mu, I) 
    \\
    &= \sum_{h \in I} \Prob_{\GT}(h \mid \mu, I) \cdot \Big( \sum_{a \in A_I} \alpha(a) \cdot \U(\mu \mid h \circ a) \Big) 
    \\
    &=  \sum_{a \in A_I} \alpha(a) \cdot \Big( \sum_{h \in I} \Prob_{\GT}(h \mid \mu, I) \cdot \U(\mu \mid h \circ a) \Big) \\
    &= \sum_{a \in A_I} \alpha(a) \cdot \EU_{\CDT,\GT}(a \mid \mu, I) \, .
\end{aligned}
\end{align}
Therefore, $\alpha^*\in \argmax_{\alpha\in \Delta(A_I)} \EU_{\CDT,\GT}(\alpha \mid \mu, I)$ if and only if it mixes only over optimal pure actions, i.e., $\alpha^*(a^*)>0 \implies a^*\in \argmax_{a\in A_I} \EU_{\CDT,\GT}(a \mid \mu, I)$.

\paragraph{EDT Deviation does not affect Reach Probabilities.}
Observe that a node $h \in I^{\first}$ in $\Gamma$ satisfies $\Prob(h \mid \mu) = \Prob(h \mid \mu_{I \, \mapsto \alpha})$ because info set $I$ - in which $\mu$ and $\mu_{I \, \mapsto \alpha}$ differ - never appeared in the history of $h$. Therefore,
\begin{align}
\begin{aligned}
\label{eq reach prob after alpha change}
    \Prob(I \mid \mu) &= \sum_{h \in I^{\first}} \Prob(h \mid \mu) = \sum_{h \in I^{\first}} \Prob(h \mid \mu_{I \, \mapsto \alpha})
    \\
    &= \Prob(I \mid \mu_{I \, \mapsto \alpha}) \, .
\end{aligned}
\end{align}

\paragraph{Proof of Lemma~\ref{lem:no-absent}: Without absentmindedness, (CDT,GT) equals (EDT,GDH).}

First, we shall show two facts that hold in games with imperfect recall (with or without absentmindedness). For any strategy $\mu'$ of $\Gamma$, node $h \in \nds \setminus \term$, and terminal node $z \in \term$, we have
\begin{align*}
&\Prob(z \mid \mu', h) 
\\
&= \chi(h \in \hist(z)) \cdot \prod_{k=d(h)}^{d(z) - 1} \mu' \big( \act(z, k) \mid I_{\nu(z, k)} \big)
\\
&= \chi(h \in \hist(z)) \cdot \mu' \big( \act(z, d(h)) \mid I_{\nu(z, d(h))} \big) \cdot
\\
&\, \quad \quad \prod_{k=d(h) + 1}^{d(z) - 1} \mu' \big( \act(z, k) \mid I_{\nu(z, k)} \big)
\\
&= \sum_{a' \in A_{I_h}} \chi(h \in \hist(z)) \cdot \chi\Big(a' = \act(z, d(h))\Big) \cdot \mu' \big( a' \mid I_h \big) \cdot
\\
&\, \quad \quad \prod_{k=d(h) + 1}^{d(z) - 1} \mu' \big( \act(z, k) \mid I_{\nu(z, k)} \big)
\\
&= \sum_{a' \in A_{I_h}} \mu' \big( a' \mid I_h \big) \cdot \chi(h \circ a' \in \hist(z)) \cdot 
\\
&\, \quad \quad \prod_{k=d(h \circ a)}^{d(z) - 1} \mu' \big( \act(z, k) \mid I_{\nu(z, k)} \big)
\\
&= \sum_{a' \in A_{I_h}} \mu' \big( a' \mid I_h \big) \cdot \Prob(z \mid \mu', h \circ a') \, .
\end{align*}

Moreover, we can obtain
\begin{align*}
&\U(\mu' \mid h) = \sum_{z \in \term} \Prob(z \mid \mu', h) \cdot u(z)
\\
&= \sum_{z \in \term} \sum_{a' \in A_{I_h}} \mu' \big( a' \mid I_h \big) \cdot \Prob(z \mid \mu', h \circ a') \cdot u(z)
\\
&= \sum_{a' \in A_{I_h}} \mu' \big( a' \mid I_h \big) \cdot \sum_{z \in \term} \Prob(z \mid \mu', h \circ a') \cdot u(z)
\\
&= \sum_{a' \in A_{I_h}} \mu' \big( a' \mid I_h \big) \cdot \U(\mu' \mid h \circ a') \, .
\end{align*}

Now let $\Gamma$ be a single-player extensive-form game with imperfect recall and without absentmindedness. Let $\mu$ be the strategy with which the player entered the game, and let the player find herself at info set $I \in \infs_*$ with $\Prob(I \mid \mu) > 0 \iff \Fr(I \mid \mu) > 0$.

CDT and EDT address how a players choices at the current node affect the players choice at other nodes of the same path and of the same info set $I$. Without absentmindedness, this consideration becomes obsolete because for any root-to-end path in $\Gamma$, the player can arrive in $I$ at most once on that path. Thus, EDT deviations and CDT deviations have the same effect. 

Moreover, that lack of absentmindedness means $I^{\first} = I$. Hence, $\Prob(I \mid \mu) = \Fr(I \mid \mu)$ and $\chi(I \cap \hist(z) \neq \emptyset) = | I \cap \hist(z) |$ for any terminal node $z \in \term$. Therefore, by Appendix~\ref{app:comparing GT with GDH}, we have for any terminal node $z$:
\begin{align*}
    \Prob_{\GT}( \, \hist(z) \mid \mu, I \,) &= | I \cap \hist(z) | \cdot \frac{\Prob(z \mid \mu)}{\Fr(I \mid \mu)}
    \\
    &= \chi(I \cap \hist(z) \neq \emptyset) \cdot \frac{\Prob(z \mid \mu)}{\Prob(I \mid \mu)} \\
    &= \Prob_{\GDH}( \, \hist(z) \mid \mu, I \,)
\end{align*}

and for any node $h \in \nds_*$:
\begin{align*}
    \Prob_{\GDH}&( h \mid \mu, I \,) 
    \\
    &= \chi(h \in I) \cdot \frac{\Prob(h \mid \mu)}{\Prob(I \mid \mu)} \cdot \sum_{\substack{ z \in \term \\ h \in \hist(z)}} \frac{\Prob(z \mid \mu, h)}{| I \cap \hist(z) |}
    \\
    &= \chi(h \in I) \cdot \frac{\Prob(h \mid \mu)}{\Fr(I \mid \mu)} \cdot \sum_{\substack{ z \in \term \\ h \in \hist(z)}} \Prob(z \mid \mu, h) 
    \\
    &= \chi(h \in I) \cdot \frac{\Prob(h \mid \mu)}{\Fr(I \mid \mu)} \cdot 1
    \\
    &= \Prob_{\GT}( h \mid \mu, I \,)
\end{align*}

Finally, we show that (CDT,GT) equals (EDT,GDH) compute the same expected utilities from a deviation to a mixed action $\alpha \in \Delta(A_I)$. Since $I^{\first} = I$, we have $\Prob(h \mid \mu) = \Prob(h \mid \mu_{I \, \mapsto \alpha})$ for all $h \in I$. With the general facts showed in the beginning, we can derive for games without absentmindedness:
\begin{align*}
    &\EU_{\CDT,\GT}(\alpha \mid \mu, I) 
    \\
    &= \sum_{h \in I} \Prob_{\GT}(h \mid \mu, I) \cdot \Big( \sum_{a \in A_I} \alpha(a) \cdot \U(\mu \mid h \circ a) \Big) 
    \\
    &= \sum_{h \in I} \chi(h \in I) \cdot \frac{\Prob(h \mid \mu)}{\Fr(I \mid \mu)} \cdot  \sum_{a \in A_I} \alpha(a) \cdot \U(\mu \mid h \circ a)  
    \\
    &= \sum_{h \in I} 1 \cdot \frac{\Prob(h \mid \mu)}{\Prob(I \mid \mu)} \cdot  \sum_{a \in A_I} \mu_{I \, \mapsto \alpha}( a \mid I) \cdot \U(\mu_{I \, \mapsto \alpha} \mid h \circ a) 
    \\
    &= \sum_{h \in I} \frac{\Prob(h \mid \mu_{I \, \mapsto \alpha})}{\Prob(I \mid \mu_{I \, \mapsto \alpha})} \cdot  \U(\mu_{I \, \mapsto \alpha} \mid h )  
    \\
    &= \sum_{h \in I} \frac{\Prob(h \mid \mu_{I \, \mapsto \alpha})}{\Prob(I \mid \mu_{I \, \mapsto \alpha})} \cdot 
    \\
    &\, \quad \quad \sum_{z \in \term} \Prob(z \mid \mu_{I \, \mapsto \alpha}, h) \cdot \chi(h \in \hist(z)) \cdot u(z)
    \\
    &= \sum_{z \in \term}  u(z) \cdot \sum_{h \in I \cap \hist(z)} \frac{\Prob(h \mid \mu_{I \, \mapsto \alpha})}{\Prob(I \mid \mu_{I \, \mapsto \alpha})} \cdot  \Prob(z \mid \mu_{I \, \mapsto \alpha}, h)
    \\
    &= \sum_{z \in \term}  u(z) \cdot \sum_{h \in I \cap \hist(z)} \frac{\Prob(z \mid \mu_{I \, \mapsto \alpha})}{\Prob(I \mid \mu_{I \, \mapsto \alpha})}
    \\
    &= \sum_{z \in \term}  u(z) \cdot | I \cap \hist(z) | \cdot \frac{\Prob(z \mid \mu_{I \, \mapsto \alpha})}{\Prob(I \mid \mu_{I \, \mapsto \alpha})}
\end{align*}
\begin{align*}
    &= \sum_{z \in \term}  u(z) \cdot \chi(I \cap \hist(z) \neq \emptyset) \cdot \frac{\Prob(z \mid \mu_{I \, \mapsto \alpha})}{\Prob(I \mid \mu_{I \, \mapsto \alpha})}
    \\
    &= \sum_{z \in \term}  u(z) \cdot \Prob_{\GDH}( \, \hist(z) \mid \mu_{I \, \mapsto \alpha}, I \,)
    \\
    &= \EU_{\EDT,\GDH}(\alpha \mid \mu, I)
\end{align*}

As a consequence, we get Lemma \ref{lem:no-absent}:

\begin{lemma*}
In games without absentmindedness, a strategy $\mu$ is a (CDT,GT)-equilibrium if and only if it is an (EDT,GDH)-equilibrium.
\end{lemma*}

\section{On Section \ref{sec:eq ex ante charact}: Proofs}
\label{app: helping lemmas}

The proofs of this section stay conceptually close to \citet{PiccioneR73} and \citet{Oesterheld22:Can}.

\subsection{Proof of Lemma \ref{derivative equals CDT deviation}}

We identified the strategy space $\bigtimes_{i = 1}^\ninfs \Delta(A_{I_i})$ of a game $\Gamma$ with $\bigtimes_{i = 1}^\ninfs \Delta^{m_i - 1}$ and established that the strategy utility function $\U$ extends to all $\bigtimes_{i = 1}^\ninfs \R^{m_i}$. In order to discuss differentiabity, we view elements $\mu \in \bigtimes_{i = 1}^\ninfs \R^{m_i}$ as flattened vectors in $\R^{\sum_{i = 1}^\ninfs m_i}$. However, for notational convenience, we keep the vector description $\mu = (\mu_{ij})_{i,j = 1}^{\ninfs, m_i} \in \bigtimes_{i = 1}^\ninfs \R^{m_i}$ because it is indexed by a (info set, action)-pair for our game-theoretic perspective. Therefore, the basis vectors that span $\bigtimes_{i = 1}^\ninfs \R^{m_i}$ are the vectors $e_{i j}$, for $i \in [\ninfs]$ and $j \in [m_{i}]$, with $(i',j')$ entries
\begin{align*}
    (e_{i j})_{i'j'} :=
    \begin{cases}
        1 &\textnormal{if } i' = i \textnormal{ and } j' = j
        \\
        0 &\textnormal{otherwise.}
    \end{cases}
\end{align*}

Then, polynomial $\U$ has one partial derivative $\nabla_{i j} \, \U$ for each coordinate direction $e_{i j}$ which is defined as
\begin{align}
    \label{dir deriv on product space}
    \nabla_{i j} \, \U(\mu) := \lim_{\epsilon \to 0} \frac{1}{\epsilon} \cdot \Big( \U( \mu + \epsilon \cdot e_{i j} ) - \U(\mu) \Big) \, .
\end{align}

We are ready to prove Lemma \ref{derivative equals CDT deviation}.

\begin{lemma*}
    Let $I_i$ be an info set, $a_j \in A_{I_i}$ an action, and $\mu \in \bigtimes_{i = 1}^\ninfs \Delta(A_{I_i})$ a strategy of the game. Then:
    \begin{enumerate}[nolistsep] 
        \item $\nabla_{ij} \, \U(\mu) = 0 \quad $ if \, $\Fr(I_i \mid \mu) = 0$, and 
        \item $\nabla_{ij} \, \U(\mu) = \Fr(I_i \mid \mu) \cdot \EU_{\CDT,\GT}(a_j \mid \mu, I_i)$ otherwise.
    \end{enumerate}
\end{lemma*}

\begin{proof}
     Using (\ref{dir deriv on product space}), let us first get an expression for $\U(\mu + \epsilon \cdot e_{ij})$. By Section \ref{sec: Q is poly fct}, we have
    \begin{align}
    \label{Q in dir modif strat}
        &\U(\mu + \epsilon \cdot e_{ij}) \nonumber
        \\
        &= \sum_{z \in \term} u(z) \cdot 
        \\
        &\, \quad \quad \prod_{k=0}^{d(z) - 1} \Big( \mu \big( \act(z, k) \mid I_{\nu(z, k)} \big) + \epsilon \cdot e_{ij} \big( \act(z, k) \mid I_{\nu(z, k)} \big) \Big) \nonumber
    \end{align}
    where we use $e_{ij} \big( \act(z, k) \mid I_{\nu(z, k)} \big)$ to indicate the entry of coordinate direction $e_{ij}$ at the info set index of $I_{\nu(z, k)}$ and the action index of $\act(z, k)$. We continue the equation chain of (\ref{Q in dir modif strat}) by sorting the product of sums by their order in $\epsilon$:
    \begin{align*}
        &= \sum_{z \in \term} \Bigg( u(z) \prod_{k=0}^{d(z) - 1} \mu \big( \act(z, k) \mid I_{\nu(z, k)} \big) \Bigg) + 
        \\
        &\, \quad \quad \sum_{z \in \term} \Bigg( u(z) \sum_{h \in \hist(z)} \bigg[ \epsilon \cdot e_{ij} \Big( \act(z, d(h)) \mid I_h \Big) \cdot 
        \\
        &\, \quad \quad \prod_{\substack{k=0 \\ k \neq d(h)}}^{d(z) - 1} \mu \big( \act(z, k) \mid I_{\nu(z, k)} \big) \bigg] \Bigg) + \mathcal{O}(\epsilon^2)
        \\
        &= \U(\mu) + \epsilon \cdot \sum_{z \in \term} \bigg[ u(z) \cdot \sum_{h \in \hist(z)} e_{ij} \Big( \act(z, d(h)) \mid I_h \Big) \cdot
        \\
        &\, \quad \quad \Prob( h \mid \mu) \cdot \Prob \Big(z \mid \mu, h \circ \act(z, d(h)) \Big) \bigg] + \mathcal{O}(\epsilon^2)
        \\
        &\overset{(*)}{=} \U(\mu) + \mathcal{O}(\epsilon^2) +
        \\
        &\, \quad \quad \epsilon \cdot \sum_{z \in \term} \sum_{h \in \hist(z) \cap I_i} \chi \Big( a_j \neq \act(z, d(h)) \Big) \cdot u(z) \cdot 
        \\
        &\, \quad \quad \Prob(h \mid \mu) \cdot \Prob \Big(z \mid \mu, h \circ \act(z, d(h)) \Big)
        \\
        &\overset{(**)}{=} \U(\mu) + \mathcal{O}(\epsilon^2) +
        \\
        &\, \quad \quad \epsilon \cdot \underbrace{\sum_{z \in \term} \sum_{h \in I_i} u(z) \cdot \Prob(h \mid \mu) \cdot \Prob \Big(z \mid \mu, h \circ a_j \Big)}_{\textnormal{Denote this term as } (\dag)} 
    \end{align*}
    In equation line $(*)$, we use that $e_{ij}$ is zero in any info set $\neq I_i$ or any action $\neq a_j \in A_{I_i}$. In equation line $(*)$, we use that $\Prob \Big(z \mid \mu, h \circ a_j)$ is zero if $h \neq \hist(z)$ or $a_j \neq \act(z, d(h))$.

    Term $(\dag)$ is constant in $\epsilon$. Once we derived a better expression for $(\dag)$, we can obtain the statement of the lemma from
    \begin{align*}
        \nabla_{ij} \, \U(\mu) &= \lim_{\epsilon \to 0} \frac{1}{\epsilon}\big( \U(\mu + \epsilon \cdot e_{ij}) - \U(\mu) \big) 
        \\
        &= \lim_{\epsilon \to 0} \frac{1}{\epsilon} \Big( \U(\mu) + \epsilon \cdot (\dag) + \mathcal{O}(\epsilon^2) - \U(\mu) \Big) \\
        &= (\dag) \, .
    \end{align*}
    Consider the case where $0 = \Fr(I_i \mid \mu) = \sum_{h \in I_i} \Prob(h \mid \mu)$. Recall that $\mu \in \bigtimes_{i = 1}^\ninfs \Delta^{m_i - 1}$. Therefore, all reach probabilities are non-negative. In particular, we obtain $\Prob(h \mid \mu) = 0$ for all $h \in I_i$. Hence, $(\dag) = 0$.
    
    Consider the other case, namely, where $\Fr(I_i \mid \mu) > 0$. Then, we can simplify:
    \begin{align*}        
        (\dag) &= \sum_{h \in I_i} \sum_{z \in \term} u(z) \cdot \Prob(h \mid \mu) \cdot \Prob \Big(z \mid \mu, h \circ a_j) \Big)
        \\        
        &= \frac{\Fr(I_i \mid \mu)}{\Fr(I_i \mid \mu)} \cdot \sum_{h \in I_i} \Prob(h \mid \mu) \cdot \sum_{z \in \term} u(z) \cdot \Prob \Big(z \mid \mu, h \circ a_j) \Big)
        \\        
        &= \Fr(I_i \mid \mu) \cdot \sum_{h \in I_i} \chi(h \in I_i) \cdot \frac{\Prob(h \mid \mu)}{\Fr(I_i \mid \mu)} \cdot \U(\mu \mid h \circ a_j)
        \\        
        &= \Fr(I_i \mid \mu) \cdot \sum_{h \in I_i} \Prob_{\GT}(h \mid \mu, I_i) \cdot \U(\mu \mid h \circ a_j)
        \\        
        &= \Fr(I_i \mid \mu) \cdot \EU_{\CDT,\GT}(a_j \mid \mu, I_i)
    \end{align*}
\end{proof}

\subsection{Proof of Lemma \ref{EDT eq to NE of Poly}}

\begin{lemma*}
    Strategy $\mu \in \bigtimes_{i = 1}^\ninfs \Delta(A_{I_i})$ of a game $\Gamma$ is an (EDT,GDH)-equilibrium if and only if for all $i \in [\ninfs]$:
    \begin{align*}
        \mu_{i \cdot} \in \argmax_{y \in \Delta(A_{I_i})} \, \, \U(\mu_{1 \cdot}, \ldots, \mu_{i-1 \cdot}, y, \mu_{i + 1 \cdot}, \ldots, \mu_{\ninfs \cdot}) \, .
    \end{align*}
\end{lemma*}

\begin{proof}
    We start with the definition (EDT,GDH)-expected utilities (Definition \ref{GDH exp util}). Say, the player entered the game with strategy $\mu \in \bigtimes_{i = 1}^\ninfs \Delta(A_{I_i})$, and arrived at an info set $I_i$ with $\Prob(I_i \mid \mu) > 0$. Let $\alpha \in \Delta(A_{I_i})$. Then
    \begin{align}
    \label{GDH utilities reformulation}
        &\EU_{\EDT,\GDH}(\alpha \mid \mu, I_i) \nonumber
        \\
        &= \sum_{z \in \term} \Prob_{\GDH}( \, \hist(z) \mid \mu_{I_i \, \mapsto \alpha}, I_i \,) \cdot u(z) \nonumber
        \\
        &= \sum_{z \in \term} \chi(I \cap \hist(z) \neq \emptyset) \cdot \frac{\Prob(z \mid \mu_{I_i \, \mapsto \alpha}) }{\Prob(I_i \mid \mu_{I_i \, \mapsto \alpha})} \cdot u(z) \nonumber
        \\
        &= \frac{1}{\Prob(I_i \mid \mu_{I_i \, \mapsto \alpha})} \cdot \sum_{\substack{z \in \term \\ I_i \cap \hist(z) \neq \emptyset}} \Prob(z \mid \mu_{I_i \, \mapsto \alpha}) \cdot u(z) 
    \end{align}
    
    Continue with the definition of an (EDT,GDH)-equilibrium (Definition \ref{EDT eq}). When taking the argmax of (\ref{GDH utilities reformulation}) over $\alpha \in \Delta(A_{I_i})$, we can rescale (\ref{GDH utilities reformulation}) by strictly positive factors and add terms to it without changing the solution set to the argmax (as long as the factors and terms are independent of $\alpha$). First, multiply (\ref{GDH utilities reformulation}) by $\alpha$-independent factor $\Prob(I_i \mid \mu_{I_i \, \mapsto \alpha}) \overset{(\ref{eq reach prob after alpha change})}{=} \Prob(I_i \mid \mu) > 0$. Second, observe that for any $z \in \term$ for which info set $I$ does not occur in $\hist(z)$, we have $\Prob(z \mid \mu_{I_i \, \mapsto \alpha}) = \Prob(z \mid \mu)$. Hence, we can add $\alpha$-independent term 
    \[
        \sum_{\substack{z \in \term \\ (I \textnormal{ not in } \hist(z))}} \Prob(z \mid \mu_{I_i \, \mapsto \alpha}) \cdot u(z) 
    \]
    to (\ref{GDH utilities reformulation}) afterwards. These two steps yield
    \begin{align*}
        &\argmax_{\alpha \in \Delta(A_{I_i})} \EU_{\EDT, \GDH}(\alpha \mid \mu, I_i)
        \\
        &\overset{(\ref{GDH utilities reformulation})}{=} \argmax_{\alpha \in \Delta(A_{I_i})} \frac{1}{\Prob(I_i \mid \mu_{I_i \, \mapsto \alpha})} \cdot \sum_{\substack{z \in \term \\ I \cap \hist(z) \neq \emptyset}} \Prob(z \mid \mu_{I_i \, \mapsto \alpha}) \cdot u(z) 
        \\
        &= \argmax_{\alpha \in \Delta(A_{I_i})} \sum_{\substack{z \in \term \\ I \cap \hist(z) \neq \emptyset}} \Prob(z \mid \mu_{I_i \, \mapsto \alpha}) \cdot u(z) 
        \\
        &= \argmax_{\alpha \in \Delta(A_{I_i})} \sum_{z \in \term} \Prob(z \mid \mu_{I_i \, \mapsto \alpha}) \cdot u(z) = \argmax_{\alpha \in \Delta(A_{I_i})} \U(\mu_{I_i \, \mapsto \alpha}) 
        \\
        &= \argmax_{\alpha \in \Delta(A_{I_i})} \U (\mu_{1 \cdot}, \ldots, \mu_{i - 1 \cdot}, \alpha, \mu_{i + 1 \cdot}, \ldots, \mu_{\ninfs \cdot} ) \, .
    \end{align*}
    
    where we use that strategy $\mu_{I_i \, \mapsto \alpha}$ has vector description $(\mu_{1 \cdot}, \ldots, \mu_{i - 1 \cdot}, \alpha, \mu_{i + 1 \cdot}, \ldots, \mu_{\ninfs \cdot})$. 
    
    All in all, we obtain that $\mu$ is an (EDT,GDH)-equilibrium
    \begin{align*}
        &\iff \forall i \in [\ninfs] \textnormal{ with } \Prob(I_i \mid \mu) > 0 \, :
        \\
        &\, \quad \quad \quad \mu(\cdot \mid I_i) \in \argmax_{\alpha \in \Delta(A_{I_i})} \EU_{\EDT, \GDH}(\alpha \mid \mu, I_i)
        \\
        &\iff \forall i \in [\ninfs] \textnormal{ with } \Prob(I_i \mid \mu) > 0 \, :
        \\
        &\, \quad \quad \quad \mu_{i \cdot} \in \argmax_{\alpha \in \Delta(A_{I_i})} \U ( \mu_{1 \cdot}, \ldots, \mu_{i - 1 \cdot}, \alpha, \mu_{i + 1 \cdot}, \ldots, \mu_{\ninfs \cdot}) 
        \\
        &\overset{(*)}{\iff} \forall i \in [\ninfs]:
        \\
        &\, \quad \quad \quad \mu_{i \cdot} \in \argmax_{\alpha \in \Delta(A_{I_i})} \U (\mu_{1 \cdot}, \ldots, \mu_{i - 1 \cdot}, \alpha, \mu_{i + 1 \cdot}, \ldots, \mu_{\ninfs \cdot}) 
    \end{align*}
    
    The last equivalence is based on the following argument: If info set $I_i$ has reach probability $\Prob(I_i \mid \mu) = 0$, then function $\U (\mu_{1 \cdot}, \ldots, \mu_{i - 1 \cdot}, \alpha, \mu_{i + 1 \cdot}, \ldots, \mu_{\ninfs \cdot}) $ is constant in the action choice $\alpha$ at info set $I_i$. Therefore, any $\alpha' \in \Delta(A_{I_i})$ satisfies 
    \[
        \alpha' \in \argmax_{\alpha \in \Delta(A_{I_i})} \U (\mu_{1 \cdot}, \ldots, \mu_{i - 1 \cdot}, \alpha, \mu_{i + 1 \cdot}, \ldots, \mu_{\ninfs \cdot}) \, .
    \]
    This is why we can add or remove this generally true statement as a condition requirement.
\end{proof}

\section{Solutions to Simple Games can be Irrational}
\label{app:irrational solutions} 

We here show that the solutions (\textit{ex ante} optimal policy, and (EDT,GDH)- and (CDT,GT)-equilibria) to simple games can be irrational. In fact, we will show that they are sometimes are not even expressible in radicals.

As a starting point, consider the polynomial equation
\begin{equation*}
    x^5-x-1=0.
\end{equation*}
This equation has a unique real-valued solution $x^*\approx 1.1673$ that cannot be expressed in radicals, i.e., that cannot be expressed using sums, products, divisions and roots \cite{Lang94}[P.121]. In particular, it is irrational.

\begin{figure}[t]
    \centering
    \includegraphics[width=\linewidth]{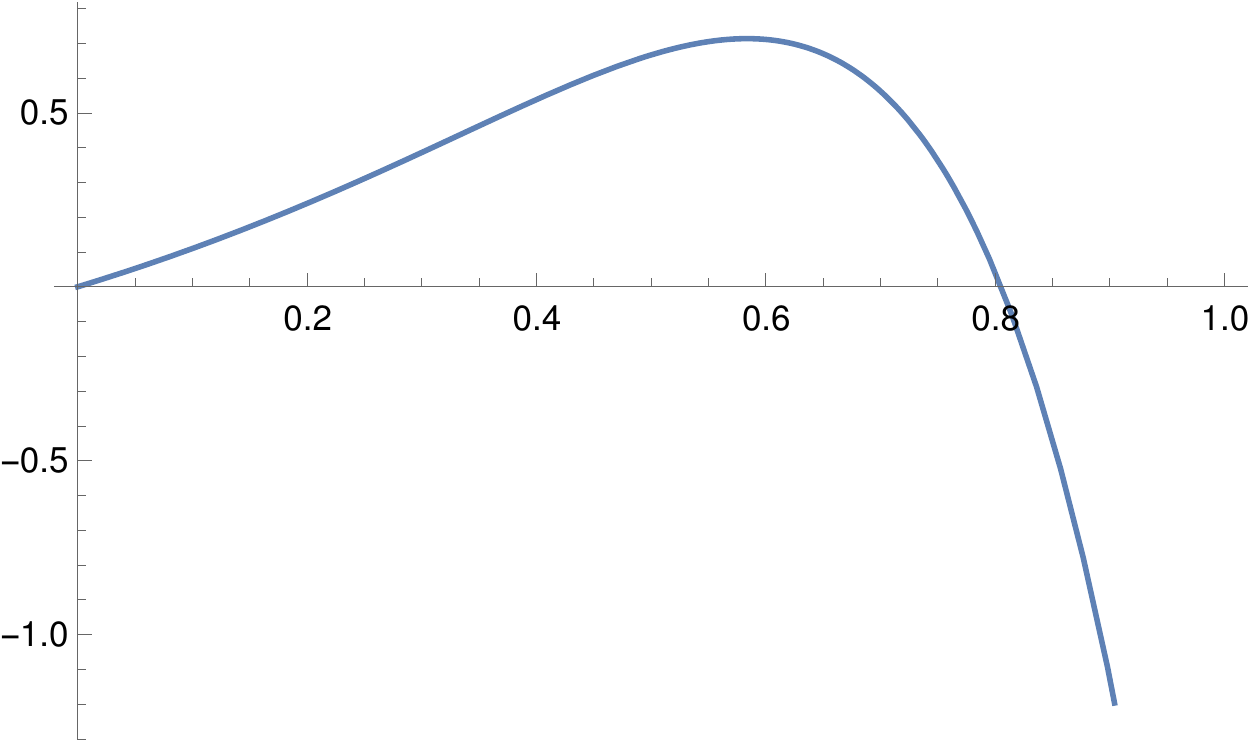}
    \caption{A plot of the function $p(x)=-\frac{16}{3}x^6 +x^2 +x$.}
    \label{fig:nonradicalmax}
\end{figure}

From the above polynomial equation we will construct a polynomial whose unique KKT point on $[0,1]$ is $x^*/2$. First, note that the equation $32x^5-2x-1=(2x)^5-2x-1=0$, obtained by substituting $2x$ for $x$ in the above, has only one solution, $\tilde x=x^*/2\approx 0.58365$. Now consider the polynomial
\begin{equation*}
    p(x)=-\frac{16}{3}x^6 +x^2 +x.
\end{equation*}
Notice that the derivative of this is exactly $-32x^5+2x+1$. Thus $p$'s only critical point is $\tilde x$. The function is plotted in Figure \ref{fig:nonradicalmax}. This point is a local maximum, and the function has no other local maximum in the compact interval $[0,1]$ (or elsewhere).

\begin{figure}[t]
    \centering
\centering
\begin{tikzpicture}[thin,
  level 1/.style={sibling distance=30mm},
  level 2/.style={sibling distance=15mm},
  level 3/.style={sibling distance=15mm},
  every circle node/.style={minimum size=1.5mm,inner sep=0mm}]
  
  \node[circle,fill,label=above:$h_0$] (h0) {}
    child { node (L1) [circle,fill,label=above left:$l_1$] {}
      child{
        node {$0$}
        edge from parent
          node[left] {$a_2$}}
      child { 
         node (L2) [circle,fill,label=right:$l_2$] {}
         child{
            node (L3) [circle,fill,label=left:$l_3$] {}
            child{
                node {$0$}
                edge from parent
                node[left] {$a_1$}
            }
            child{
                node (L4) [circle,fill,label=right:$l_{4}$] {}
                child{
                    node (L5) [circle,fill,label=left:$l_{5}$] {}
                    child{
                        node {$0$}
                        edge from parent
                        node[left] {$a_2$}
                    }
                    child{
                        node (L6) [circle,fill,label=left:$l_{6}$] {}
                        child{
                            node {$-16$}
                            edge from parent
                            node[left] {$a_1$}
                        }
                        child{
                            node {$0$}
                            edge from parent
                            node[right] {$a_2$}
                        }
                        edge from parent
                        node[right] {$a_1$}
                    }
                    edge from parent
                    node[left] {$a_1$}
                }
                child{
                    node {$0$}
                    edge from parent
                    node[right] {$a_2$}
                }
                edge from parent
                node[right] {$a_1$}
            }
            edge from parent
            node[left] {$a_1$}
         }
         child{
            node {$0$}
            edge from parent
            node[right] {$a_2$}
         }
         edge from parent 
            node[above right] {$a_1$}
    }
      edge from parent
        node[above] {$\frac{1}{3}$}}
    child {
      node (m1) [circle,fill, label=above left:$m_1$] {}
        child {
            node {$0$}
            edge from parent
              node[left] {$a_2$}}
          child {
             node (m2) [circle,fill, label=right:$m_2$] {}
             child {
                node {$0$}
                edge from parent
                  node[left] {$a_2$}}
              child {
                 node {$3$}
                 edge from parent
                   node[right] {$a_1$}
              }
             edge from parent
               node[right] {$a_1$}
          }
       edge from parent
         node[right] {$\frac{1}{3}$}}
  child { 
         node (r1) [circle,fill,label=above right:$r_1$] {}
         child{
            node {$0$}
            edge from parent
            node[left] {$a_2$}
         }
         child{
            node {$3$}
            edge from parent
            node[right] {$a_1$}
         }
         edge from parent
         node[above] {$\frac{1}{3}$}
    };
  \draw [dashed] (L1) to node[below] {$I_1$} (m1);
  \draw [dashed] (m1) to (r1);
  \draw [dashed] (L1) to[bend right=30] (L2);
  \draw [dashed] (m1) to[bend right=30] (m2);
  \draw [dashed] (L2) to[bend right=-30] (L3);
  \draw [dashed] (L3) to[bend right=30] (L4);
  \draw [dashed] (L4) to[bend right=-30] (L5);
  \draw [dashed] (L5) to[bend right=30] (L6);

\end{tikzpicture}
    \caption{A single-player game of imperfect recall whose only solution cannot be expressed in radicals.}
    \label{fig:nooptinradicalsgame}
\end{figure}
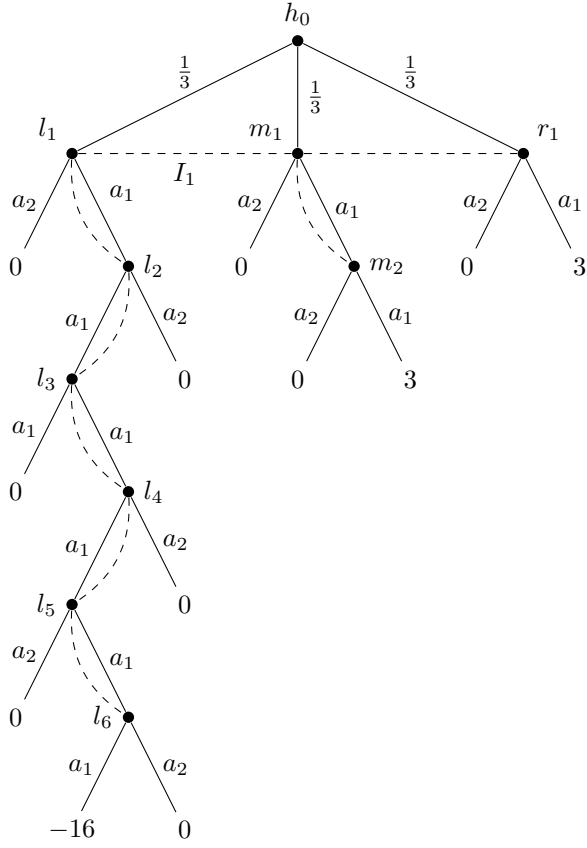

Now consider the game in Figure \ref{fig:nooptinradicalsgame}. Let $\U\colon \mathbb{R}^2 \rightarrow \mathbb{R}\colon (x,y)\mapsto-\frac{16}{3}x^6 +x^2 +x=p(x)$ be the \textit{ex ante} expected utility function extended to $\mathbb{R}^2$. Clearly, the unique optimal policy is $x=\tilde x,y=1-\tilde x$. Next we will show that the unique (CDT,GT)-equilibrium is also $x=\tilde x$. It is easy to see that neither $x=0$ nor $x=1$ induces a (CDT,GT)-equilibrium. So for any (CDT,GT)-equilibrium it has to be the case that $\mathrm{EU}_{\mathrm{CDT},\mathrm{GT}}(a_1\mid x)=\mathrm{EU}_{\mathrm{CDT},\mathrm{GT}}(a_2\mid x)$. By Lemma~\ref{derivative equals CDT deviation}, this means that we must have $\frac{d}{dx}\U(x,y)=\frac{d}{dy}\U(x,y)=0$. Therefore, the only (CDT,GT)-equilibrium is at $x=\tilde x$. By Lemma~\ref{equilibrium hierarchy}, this means that the only (EDT,GDH)-equilibrium is also at $x=\tilde x$. 

All in all, we have given an easy to represent game for which the only solution $(\tilde x, 1 - \tilde x)$ has entries which are irrational (or even more, cannot be expressed in radicals).

\section{Proofs of the Main Results}
\label{app: proofs of main results}

\subsection{Proof of Theorem \ref{CDT GT eq iff KKT point}}

Recall the problem of maximizing the ex-ante utility in a game $\Gamma$:
\begin{align}
\label{exantemaxprobl explicit constraints}
\begin{aligned}  
    &\max_{\mu \in \bigtimes_{i = 1}^\ninfs \R^{m_i}} &&\U(\mu) 
    \\
    &\textnormal{s.t.} && \mu_{ij} \geq 0 \quad \forall i \in [\ninfs], \forall j \in [m_i]
    \\
    &\, &&\sum_{j = 1}^{m_i} \mu_{ij} = 1 \quad \forall i \in [\ninfs]
\end{aligned}
\end{align}

Then, the KKT conditions for (\ref{exantemaxprobl explicit constraints}) and a point $\mu \in \bigtimes_{i = 1}^\ninfs \R^{m_i}$ are: There exist KKT multipliers $\{ \tau_{ij} \in \R \}_{i,j=1}^{\ninfs,m_i}$ and $\{ \kappa_i \in \R \}_{i = 1}^\ninfs$ such that
\begin{align}
\label{KKT conditions for CDT GT eq}
\begin{aligned}  
    &\mu_{ij} \geq 0 \quad \forall i \in [\ninfs], \forall j \in [m_i]
    \\    
    &\sum_{j = 1}^{m_i} \mu_{ij} = 1 \quad \forall i \in [\ninfs]
    \\
    &\tau_{ij} \geq 0 \quad \forall i \in [\ninfs], \forall j \in [m_i]
    \\
    &\tau_{ij} = 0 \quad \textnormal{or} \quad \mu_{ij} = 0 \quad \forall i \in [\ninfs], \forall j \in [m_i]
    \\
    &\nabla_{ij} \, \U(\mu) = - \tau_{ij} + \kappa_i  \quad \forall i \in [\ninfs], \forall j \in [m_i] \, .
\end{aligned}
\end{align}

We are ready to prove Theorem \ref{CDT GT eq iff KKT point}.

\begin{thm*}
    Strategy $\mu \in \bigtimes_{i = 1}^\ninfs \Delta(A_{I_i})$ of $\Gamma$ is a (CDT,GT)-equilibrium if and only if $\mu$ is a KKT point of (\ref{exantemaxprobl explicit constraints}).
\end{thm*}

\begin{proof}
    ``$\implies$'':
    \\
    Suppose $\mu \in \bigtimes_{i = 1}^\ninfs \Delta(A_{I_i})$ is a (CDT,GT)-equilibrium of $\Gamma$. Then the first two KKT conditions of (\ref{KKT conditions for CDT GT eq}) are satisfied by assumption. Let $i \in [\ninfs]$ and $\forall j \in [m_i]$. 
    
    If $\Fr(I_i \mid \mu) = 0$, then by Lemma \ref{derivative equals CDT deviation} we have $\nabla_{ij} \, \U(\mu) = 0$. Therefore, choose $\tau_{ij} = 0$ and $\kappa_i = 0$ to satisfy the last three KKT conditions for the respective $i$ and $j$.
    
    Suppose $\Fr(I_i \mid \mu) > 0$. Choose 
    \[
        \kappa_i = \max_{j' \in [m_i]} \nabla_{ij'} \, \U(\mu) \, .
    \]
    The choice of $\tau_{ij}$ depends on $\mu_{ij}$. If $\mu_{ij} > 0$, then by Lemma \ref{derivative equals CDT deviation} and by the equivalent characterization of a (CDT,GT)-equilibrium right below Definition \ref{CDT GT eq def}, we have 
    \begin{align*}
        \nabla_{ij} \, \U(\mu) &= \Fr(I_i \mid \mu) \cdot \EU_{\CDT,\GT}(a_j \mid \mu, I_i) 
        \\
        &= \max_{j' \in [m_i]} \big\{ \Fr(I_i \mid \mu) \cdot \EU_{\CDT,\GT}(a_{j'} \mid \mu, I_i) \big\} 
        \\
        &= \max_{j' \in [m_i]} \nabla_{ij'} \, \U(\mu) = \kappa_i \, .
    \end{align*}
    Therefore choose $\tau_{ij} = 0$. If $\mu_{ij} = 0$, on the other hand, then the above equation chain would yield inequality $\nabla_{ij} \, \U(\mu) \leq \kappa_i$ instead (when transitioning to the $\max$). Thus, choose $\tau_{ij} = \kappa_i - \nabla_{ij} \, \U(\mu)$. These choices of $\kappa_i$ and $\tau_{ij}$ satisfy the last three KKT conditions for the respective $i$ and $j$.
    \\
    
    ``$\impliedby$'':
    \\
    Suppose $\mu$ is a KKT point of problem (\ref{exantemaxprobl explicit constraints}), that is, it satisfies the KKT conditions (\ref{KKT conditions for CDT GT eq}) for some KKT multipliers $\{ \tau_{ij} \in \R \}_{i,j=1}^{\ninfs,m_i}$ and $\{ \kappa_i \in \R \}_{i = 1}^\ninfs$. Then, the first two KKT conditions ensure that $\mu$ makes a valid strategy for $\Gamma$ (recall that we always identify $\mu(a_j \mid I_i) := \mu_{ij}$ for info set $I_i \in \infs_{*}$ and action $a_j \in I_i$). So let us check equivalent characterization of a (CDT,GT)-equilibrium right below Definition \ref{CDT GT eq def}. Let $i \in [\ninfs]$ and $j \in [m_i]$ be such that $\Fr(I_i \mid \mu) > 0$ and $\mu_{ij} > 0$. Then, Lemma \ref{derivative equals CDT deviation} and the last three KKT conditions give us for all $j' \in [m_i]$:
    \begin{align*}
        &\EU_{\CDT,\GT}(a_j \mid \mu, I_i) = \frac{\Fr(I_i \mid \mu)}{\Fr(I_i \mid \mu)} \cdot \EU_{\CDT,\GT}(a_j \mid \mu, I_i) 
        \\
        &= \frac{1}{\Fr(I_i \mid \mu)} \cdot \nabla_{ij} \, \U(\mu) = \frac{1}{\Fr(I_i \mid \mu)} \cdot (- \tau_{ij} + \kappa_i) 
        \\
        &\overset{\tau_{ij} = 0}{=} \frac{1}{\Fr(I_i \mid \mu)} \cdot \kappa_i \geq \frac{1}{\Fr(I_i \mid \mu)} \cdot (- \tau_{ij'} + \kappa_i) 
        \\
        &= \frac{1}{\Fr(I_i \mid \mu)} \cdot \nabla_{ij'} \, \U(\mu) = \EU_{\CDT,\GT}(a_{j'} \mid \mu, I_i)
    \end{align*}
    This implies $\EU_{\CDT,\GT}(a_j \mid \mu, I_i) = \max_{a \in A_{I_i}} \EU_{\CDT,\GT}(a \mid \mu, I_i)$. Overall, we get that $\mu$ is a (CDT,GT)-equilibrium for $\Gamma$
\end{proof}

\subsection{Reproof of Lemma \ref{equilibrium hierarchy}}

The next lemma is due to \citet{Oesterheld22:Can} [cf.~\citeauthor{PiccioneR73}, \citeyear{PiccioneR73}].

\begin{lemma*}
    An ex-ante optimal strategy of a game $\Gamma$ is also an (EDT,GDH)-equilibrium. An (EDT,GDH)-equilibrium is also a (CDT,GT)-equilibrium.
    
    In particular, any single-player extensive-form game $\Gamma$ with imperfect recall admits an (EDT,GDH)-equilibrium and a (CDT,GT)-equilibrium.
\end{lemma*}
\begin{proof}
    An ex-ante optimal strategy is defined as a maximum to the maximization problem (\ref{exantemaxprobl}). Therefore, it will in particular satisfy the characterization of an (EDT,GDH)-equilibrium according to Lemma \ref{EDT eq to NE of Poly}. 
    
    Let us show that an (EDT,GDH)-equilibrium is a (CDT,GT)-equilibrium by using their respective ex-ante characterizations from Lemma \ref{EDT eq to NE of Poly} and Theorem \ref{CDT GT eq iff KKT point}. So let $\mu = \in \bigtimes_{i = 1}^\ninfs \Delta(A_{I_i})$ satisfy for all $i \in [\ninfs]$
    \begin{align*}
        \mu_{i \cdot}^* \in \argmax_{y \in \Delta(A_{I_i})} \, \, \U(\mu_{1 \cdot}^*, \ldots, \mu_{i-1 \cdot}^*, y, \mu_{i + 1 \cdot}^*, \ldots, \mu_{\ninfs \cdot}^*) \, .
    \end{align*}
    For $i \in [\ninfs]$ denote the function 
    \begin{align*}
        q_{i, \mu} : \Delta^{m_i - 1} &\to \R
        \\
        \quad y &\mapsto \U(\mu_{1 \cdot}, \ldots, \mu_{i-1 \cdot}, y, \mu_{i + 1 \cdot}, \ldots, \mu_{\ninfs \cdot}) \, .
    \end{align*}
    Then, for $j \in [m_i]$, we have 
    \[
        \nabla_{j} \, q_{i, \mu}(y) = \nabla_{ij} \, \U (\mu_{1 \cdot}, \ldots, \mu_{i - 1 \cdot}, y, \mu_{i + 1 \cdot}, \ldots, \mu_{\ninfs \cdot}) \, .
    \]
    By assumption on $\mu$, each $\mu_{i \cdot}$ is a solution to the maximum problem
    \begin{align*}
        \max_{y \in \Delta^{m_i - 1}} \, \, q_{i, \mu}(y) \, .
    \end{align*}
    Global optimum $\mu_{i \cdot}$ is therefore in particular a KKT point of that problem. So there exist KKT multipliers $\{ \tau_{ij} \in \R \}_{j=1}^{m_i}$ and $\kappa_i$ such that
    \begin{align*} 
        &\mu_{ij} \geq 0 \quad \forall j \in [m_i]
        \\    
        &\sum_{j = 1}^{m_i} \mu_{ij} = 1
        \\
        &\tau_{ij} \geq 0 \quad \forall j \in [m_i]
        \\
        &\tau_{ij} = 0 \quad \textnormal{or} \quad \mu_{ij} = 0 \quad \forall j \in [m_i]
        \\
        &\nabla_{ij} \, \U (\mu) = \nabla_{j} \, q_{i, \mu}(\mu_{i \cdot}) = - \tau_{ij} + \kappa_i  \quad \forall j \in [m_i] \, .
    \end{align*}
    Since $i \in [\ninfs]$ was arbitrary, we can use these multipliers to obtain that $\mu$ satisfies KKT conditions (\ref{KKT conditions for CDT GT eq}).

    Lastly, since there always exists an ex-ante optimal strategy (as a maximum (\ref{exantemaxprobl}) of a continuous polynomial function over a compact domain), there also always exists an (EDT,GDH)-equilibrium and (CDT,GT)-equilibrium.
    \end{proof}

\subsection{Approximate and Well-Supported (CDT,GT)-equilibria}

Before we get to the proof of Theorem \ref{CLS hardness}, we should discuss two definitions of being close to a (CDT,GT)-equilibrium that are useful to computatational considerations:

\begin{defn}
    Let $\Gamma$ be a single-player extensive-form game with imperfect recall, $\mu$ be a strategy for $\Gamma$, and $\epsilon > 0$ be a precision parameter. Then:
    \begin{itemize}
        \item $\mu$ is called an $\epsilon$-approximate (CDT,GT)-equilibrium if for all $I \in \infs_{*}$ with $\Fr(I \mid \mu) > 0$:
        \begin{align*}
        &\EU_{\CDT,\GT}\big( \mu( \cdot \mid I) \mid \mu, I \big) 
        \\
        &\, \quad \quad \quad \quad \geq \max_{a' \in A_I} \EU_{\CDT,\GT}(a' \mid \mu, I) - \epsilon \, .
        \end{align*}
        
        \item $\mu$ is called an $\epsilon$-well-supported (CDT,GT)-equilibrium if for all $I \in \infs_{*}$ with $\Fr(I \mid \mu) > 0$ and all $a \in A_I$ with $\mu(a \mid I) > 0$:
        \[ 
            \EU_{\CDT,\GT}(a \mid \mu, I) \geq \max_{a' \in A_I} \EU_{\CDT,\GT}(a' \mid \mu, I) - \epsilon \, .
        \]
    \end{itemize}
\end{defn}

Identity (\ref{cdt util are linear combi}) directly implies the following relationship:

\begin{lemma}
\label{lemma well supp to approx}
    Let $\Gamma$ be a single-player extensive-form game with imperfect recall and $\epsilon > 0$. Then any $\epsilon$-well-supported (CDT,GT)-equilibrium $\mu$ is also an $\epsilon$-approximate (CDT,GT)-equilibrium.
\end{lemma}

We can also prove a polynomial-time reduction in the reverse direction for all those games that have admit the following characteristic:
\begin{defn}
    \label{def lower bound on freqs}
    Ia single-player extensive-form game $\Gamma$ (with imperfect recall), a value $\lambda > 0$ is said to be a lower bound on positive visit frequencies in $\Gamma$ if it satisfies for all info sets $I$ and all strategies $\mu$ in $\Gamma$ 
    \[
        \Fr(I \mid \mu) = 0 \quad \textnormal{or} \quad \Fr(I \mid \mu) \geq \lambda \, .
    \] 
\end{defn}
Since $\Fr(I \mid \mu)$ as a function in $\mu$ is a polynomial function, and therefore continuous over the connected strategy set $\bigtimes_{i = 1}^\ninfs \Delta(A_{I_i})$, this definition becomes equivalent to
\begin{align}
    \label{easy char of lower bound on freqs}
    \forall \, I \in \infs_* \, : \, \, \Fr(I \mid \mu) = 0 \, \forall \mu \, \, \textnormal{ or } \, \, \Fr(I \mid \mu) \geq \lambda \, \forall \mu \, .
\end{align}

While the next lemma seems to be novel for imperfect-recall games, the proof borrows its main ideas from the equivalency of $\epsilon$-approximate Nash equilibria and $\epsilon$-well-supported Nash equilibria, shown by \citet{Chen06}. In the below formulation and proof, any roots $\sqrt{x}$ taken shall always refer to the positive valued root.

\begin{lemma}
\label{lemma approx to well supp}
    Let $\Gamma$ be a single-player extensive-form game with imperfect recall with a lower bound $\lambda \in \Q$ on its positive visit frequencies.
    Then we can compute a Lipschitz constant $L \geq 1$ w.r.t. the infinity norm $||\cdot||_{\infty}$ for all functions $\big( \EU_{\CDT,\GT}(a \mid \cdot, I) \big)_{I \in \infs, \Fr(I \mid \cdot) \geq \lambda, a \in A_I}$ in $\mu$ over the strategy space $\bigtimes_{\ihat = 1}^\ninfs \Delta(A_{I_{\ihat}})$. Let $\epsilon > 0$ such that $\epsilon < \frac{1}{m_i}$ for all info sets $I_i \in \infs_{*}$. 
    
    Then, given an $\epsilon$-approximate (CDT,GT)-equilibrium $\mu$ of $\Gamma$, we can compute a $(3 L |H| \sqrt{\epsilon})$-well-supported (CDT,GT)-equilibrium of $\Gamma$ in polynomial time.
\end{lemma}

\begin{proof}
    Let us observe how to compute such a Lipschitz constant~$L$ in polynomial time. We first describe the general process of computing a Lipschitz constant~$L_q$ of any polynomial function $q : \R^n \to \R^m$ w.r.t. the infinity norm and over the hypercube $[0,1]^n$. One possible Lipschitz constant is 
    \begin{align*}
        &\max_{x \in [0,1]^n} \{ || \nabla q (x) ||_{\infty} \} = \max_{x \in [0,1]^n} \max_{i \in [n]} | \nabla_i q (x) | 
        \\
        &= \max_{i \in [n]} \max_{x \in [0,1]^n} | \nabla_i q (x) | \, .
    \end{align*}
    Note that $\nabla_i q (x)$ is a polynomial function itself and that the factors $x_i^k$ in its monomials are $<1$. Thus, $\max_{x \in [0,1]^n} | \nabla_i q (x) |$ can be upper bounded by the absolute value sum $\sum_{\lambda_D \in \nabla_i q } |\lambda_D| $ of all coefficients present in $\nabla_i q$. Taking the maximum over index $i$ over those upper bound sums gives us a Lipschitz value $L_q$ that can be computed in polynomial time in the description size of $q$.
    
    Going back to the setup of this lemma, take an info set $I_i$ of $\Gamma$ with $\Fr(I_i \mid \cdot) \geq \lambda$ and an action $a_j \in A_i$. Then we can derive as possible Lipschitz constant for $\EU_{\CDT,\GT}(a_j \mid \cdot, I_i)$:
    \begin{align*}
        &\max_{\mu \in \bigtimes_{i = 1}^\ninfs \Delta(A_{I_i})} \{ || \nabla_{\mu} \, \EU_{\CDT,\GT}(a_j \mid \mu, I_i) ||_{\infty} \} 
        \\
        &\leq \max_{\mu \in [0,1]^{\sum_i m_i}} \{ || \nabla_{\mu} \, \EU_{\CDT,\GT}(a_j \mid \mu, I_i) || \} 
        \\
        &= \max_{\mu \in [0,1]^{\sum_i m_i}} \Big\{ \Big|\Big| \nabla_{\mu} \, \Big( \frac{\nabla_{ij} \, U(\mu)}{\Fr(I_i \mid \mu)} \Big) \Big|\Big| \Big\} 
        \\
        &= \max_{\mu \in [0,1]^{\sum_i m_i}} 
        \\
        &\Big\{ \Big|\Big| \frac{\Fr(I_i \mid \mu) \cdot \nabla_{\mu} \, \nabla_{ij} \, U(\mu) - \nabla_{ij} \, U(\mu) \cdot \nabla_{\mu} \, \Fr(I_i \mid \mu)}{\Fr(I_i \mid \mu)^2} \Big|\Big| \Big\} 
        \\
        &\leq \max_{\mu} \frac{1}{\Fr(I_i \mid \mu)^2} \cdot \Big( \max_{\mu} \Fr(I_i \mid \mu) \cdot \max_{\mu} || \nabla_{\mu} \, \nabla_{ij} \, U(\mu) || 
        \\
        &\, \quad + \max_{\mu} | \nabla_{ij} \, U(\mu) | \cdot \max_{\mu} || \nabla_{\mu} \, \Fr(I_i \mid \mu) || \Big)
        \\
        &\leq \frac{1}{\lambda^2} \cdot \Big( |\nds| \cdot \max_{\mu} || \nabla_{\mu} \, \nabla_{ij} \, U(\mu) || 
        \\
        &\, \quad + \max_{\mu} | \nabla_{ij} \, U(\mu) | \cdot \max_{\mu} || \nabla_{\mu} \, \Fr(I_i \mid \mu) || \Big) \, .
    \end{align*}    
    The terms $\max_{\mu} || \nabla_{\mu} \, \nabla_{ij} \, U(\mu) ||$ and $\max_{\mu} || \nabla_{\mu} \, \Fr(I_i \mid \mu) ||$ as well $\max_{\mu} | \nabla_{ij} \, U(\mu) |$ can be bounded above with the method described in the previous paragraph. The final value can be chosen as our Lipschitz constant $L_{ij}$ for $\EU_{\CDT,\GT}(a_j \mid \cdot, I_i)$. Finally, we can choose
    \[ 
        L = \max_{I_i \in \infs_*, \Fr(I_i \mid \cdot) \geq \lambda, a_j \in A_I} \{ 1, \max_{i,j} L_{ij} \}
    \]
    as the desired mutual Lipschitz constant. This construction takes polynomial time in the encoding of $\Gamma$ and $\lambda$.

    It remains to prove the first result. Denote with $\U_{ij}(\mu) := \EU_{\CDT,\GT}(a_j \mid \mu, I_i)$ the (CDT,GT)-expected utility function in $\mu$ from being in info set $I_i$ and using action $a_j$. 
    Let $\mu$ be an $\epsilon$-approximate (CDT,GT)-equilibrium of $\Gamma$, and take an info set $I_i$ with $\Fr(I_i \mid \cdot ) > 0$, thus $\Fr(I_i \mid \cdot ) \geq \lambda$. As the first step, let us show that any action that sufficiently suboptimal will also be played with only very low probability. Assume $a_{j'} \in A_{I_i}$ is a suboptimal action, that is, there exists some $a_j \in A_{I_i}$ such that $\U_{ij}(\mu) \geq \U_{ij'}(\mu) + \sqrt{\epsilon}$.
    Consider the mixed action $\alpha \in \Delta(A_{I_i})$ defined as $\alpha_{j'} = 0$, $\alpha_j = \mu_{ij} + \mu_{ij'}$, and $\alpha_{\jhat} = \mu_{i\jhat}$ for all $\jhat \neq j, j'$. Since $\mu$ is assumed to be an $\epsilon$-approximate (CDT,GT)-equilibrium, we get
    \begin{align*}
        \sum_{\jhat \in [m_i]} &\mu_{i\jhat} \cdot \U_{i \jhat}(\mu) \overset{(\ref{cdt util are linear combi})}{=} \EU_{\CDT,\GT}( \mu(\cdot \mid I_i) \mid \mu, I_i)
        \\
        &\geq \max_{k \in [m_i]} \EU_{\CDT,\GT}(a_{k} \mid \mu, I_i) - \epsilon 
        \\
        &= - \epsilon + \sum_{\jhat \in [m_i]} \alpha_{\jhat} \cdot \max_{k \in [m_i]} \U_{ik}(\mu)
        \\
        &\geq - \epsilon + \sum_{\jhat \in [m_i]} \alpha_{\jhat} \cdot \U_{i \jhat}(\mu)
        \\
        &= - \epsilon + (\mu_{ij} + \mu_{ij'}) \cdot \U_{ij}(\mu) + \sum_{\jhat \neq j, j'} \mu_{i\jhat} \cdot \U_{i\jhat}(\mu) \, .
    \end{align*}
    Rearranging yields
    \begin{align*}
        \mu_{i j'} \cdot \U_{i j'}(\mu) \geq \mu_{i j'} \cdot \U_{i j}(\mu) - \epsilon \, .
    \end{align*}    
    Using $\U_{ij}(\mu) \geq \U_{ij'}(\mu) + \sqrt{\epsilon}$, we therefore get
    \begin{align*}
        \mu_{i j'} \leq \frac{\epsilon}{\U_{i j}(\mu) - \U_{i j'}(\mu)} \leq \frac{\epsilon}{\sqrt{\epsilon}} = \sqrt{\epsilon} \, .
    \end{align*}    
    In other words, a suboptimal action $a_{j'} \in  A_{I_i}$ in an $\epsilon$-approximate (CDT,GT)-equilibrium will be played with low probability. Denote with $\textnormal{low}_i \subseteq [m_i]$ the set of all those action indices $j'$ such that $a_{j'}$ is played with $\leq \sqrt{\epsilon}$ probability in $\mu$. Since we assumed $\sqrt{\epsilon} < \frac{1}{m_i}$, there will be at least one action index $j \in [m_i] \setminus \textnormal{low}_i$. We can therefore create a new candidate strategy $\pi$ from $\mu$ by redistributing the probability of the $\textnormal{low}_i$ actions to the others and achieve well-supportedness for $\pi$. More precisely, define $\pi \in \bigtimes_{i = 1}^\ninfs \Delta^{m_i - 1}$ as follows:
    \begin{itemize}
        \item for all $I_i$ with $\Fr(I_i \mid \cdot) = 0$ and all $a_j \in  A_{I_i}$ 
        \\
        set $\pi_{ij} := \mu_{ij}$,
        \item for all $I_i$ with $\Fr(I_i \mid \cdot) \geq \lambda$ and all $a_j \in  A_{I_i}$ with $\mu_{ij} \leq \sqrt{\epsilon}$ set $\pi_{ij} = 0$, and
        \item for all $I_i$ with $\Fr(I_i \mid \cdot) \geq \lambda$ and all $a_j \in  A_{I_i}$ with $\mu_{ij} > \sqrt{\epsilon}$ set 
        \[
        \pi_{ij} = \mu_{ij} + \frac{\sum_{\jhat \in \textnormal{low}_i} \mu_{i \jhat}}{m_i - |\textnormal{low}_i|} \, .
        \] 
    \end{itemize}
    This is easily computable. Let us prove that $\pi$ is a $(3 L |H| \sqrt{\epsilon})$-well-supported (CDT,GT)-equilibrium: First note $||\pi - \mu ||_{\infty} \leq \sqrt{\epsilon} \cdot |H|$. Let $I_i \in \infs_{*}$ again be an info set with $\Fr(I_i \mid \cdot) > 0$, hence $\Fr(I_i \mid \cdot) \geq \lambda$, and let an action $a_j \in A_{I_i}$ have $\pi(a_j \mid I_i) > 0$. Recall that $L$ serves as a Lipschitz constant on $\U_{ij}$. By construction of $\pi$, we get $\mu_{ij} > \sqrt{\epsilon}$ which implies that $a_j$ could not have been suboptimal for $\mu$, that is, 
    \begin{align}
    \label{action not suboptimal}
        \U_{ij}(\mu) \geq \max_{\jhat \in [m_i]} \U_{i\jhat}(\mu) - \sqrt{\epsilon} \, .
    \end{align}
    Therefore,
    \begin{align*}
        \EU&_{\CDT,\GT}(a_j \mid \pi, I_i) 
        \\
        &= \U_{ij}(\pi) = \U_{ij}(\pi) - \U_{ij}(\mu) + \U_{ij}(\mu) 
        \\
        &\geq - L \cdot ||\pi - \mu || + \U_{ij}(\mu)
        \\
        &\overset{(\ref{action not suboptimal})}{\geq} - L \cdot \sqrt{\epsilon} \cdot |H| + \max_{\jhat \in [m_i]} \U_{i\jhat}(\mu) - \sqrt{\epsilon}
        \\
        &\geq - 2 L |H| \sqrt{\epsilon} + \max_{\jhat \in [m_i]} \{ \U_{i\jhat}(\mu) - \U_{i \jhat}(\pi) + \U_{i \jhat}(\pi) \}
        \\
        &\geq - 2 L |H| \sqrt{\epsilon} + \max_{\jhat \in [m_i]} \{ - L \cdot ||\pi - \mu || + \U_{i \jhat}(\pi) \}
        \\
        &\geq - 3 L |H| \sqrt{\epsilon} + \max_{\jhat \in [m_i]}  \U_{i \jhat}(\pi)
        \\
        &= \max_{a_{\jhat} \in A_{I_i}}  \EU_{\CDT,\GT}(a_{\jhat} \mid \pi, I_i) - 3 L |H| \sqrt{\epsilon}
    \end{align*}

\end{proof}

\subsection{Proof of Theorem \ref{CLS hardness}}

\begin{thm*}
    {\sc (CDT,GT)-equilibrium} is CLS-hard. CLS-hardness holds even for games restricted to:
    \begin{enumerate}[nolistsep]
        \item[(1.)] a tree depth of $6$ and the player has $2$ actions per info set,
        \item[(2.)] no absentmindedness and a tree depth of $6$, and
        \item[(3.)] no chance nodes, a tree depth of $5$, and only one info set.
    \end{enumerate}
    \smallskip
    The problem is in CLS for the subclass of problem instances of {\sc (CDT,GT)-equilibrium} where a lower bound on positive visit frequencies in $\Gamma$ is easily obtainable.
\end{thm*}

We prove this result in four parts.

\subsubsection{CLS Membership of restricted {\sc (CDT,GT)-equilibrium}}

By Theorem \ref{CDT GT eq iff KKT point}, (CDT,GT)-equilibria of $\Gamma$ coincide with the KKT points of the strategy utility function $\U$ of $\Gamma$. Finding an approximate KKT point of a continuously differentiable function over a compact domain was shown to be CLS-complete by \citet{FGHS23}:
\begin{defn}
\label{general KKT comp problem}
    An instance of the problem {\sc KKT} consists of (1) a precision parameter $\epsilon > 0$ encoded in binary, (2) a matrix $A \in \R^{m \times n}$ and a vector $b \in \R^m$ defining a bounded non-empty domain $D = \{x \in \R^n : Ax \leq b\}$, (3) two well-behaved arithmetic circuits $f : \R^n \to \R$ and $\nabla f : \R^n \to \R^n$, and (4) a Lipschitz constant $L > 0$. 
    
    A solution consists of a point $x \in D$ such that there exist $\mu_1, \ldots, \mu_m \geq 0$ such that $|| \nabla f(x) + A^T \mu || \leq \epsilon$ and $\mu^T(Ax - b) = 0$. Alternatively, we also accept points $x,y \in D$ as a solution that show that one of the following is true: (i) $f$ or $\nabla f$ is not $L$-Lipschitz, or (ii) $\nabla f$ is not the gradient of $f$.
\end{defn}

\begin{lemma}[\citet{FearnleyGHS21}]
    {\sc KKT} is CLS-complete.
\end{lemma}

\citet{FearnleyGHS21} note that any norm can be used in Definition \ref{general KKT comp problem}. We will henceforth use the infinity norm $|| \cdot ||_{\infty}$.

Let us outline the reduction from the restricted {\sc (CDT,GT)-equilibrium} to {\sc KKT}. Let $(\Gamma, \epsilon)$ be an instance of {\sc (CDT,GT)-equilibrium} for which a lower bound $\lambda$ on positive visit frequencies in $\Gamma$ can be computed in polynomial time. Construct the strategy utility function $\U$ of $\Gamma$ and its gradient $\nabla \U$. Then $-\U$ and $-\nabla \U$ as polynomial functions will make well-behaved arithmetic circuits $f$ and $\nabla f$ (the negative sign comes in because {\sc KKT} is about minimizing $f$ whereas {(CDT,GT)-equilibrium} is about utility maximization). The domain $D$ will be the Cartesian product of simplices $\bigtimes_{i = 1}^l \Delta(A_{I_i})$ described with inequalities as in (\ref{exantemaxprobl explicit constraints}). Next, we can construct a Lipschitz constant $L$ for the polynomial functions $-\U$ and $-\nabla \U$ over $\bigtimes_{i = 1}^l \Delta(A_{I_i})$ as described in the proof of Lemma \ref{lemma approx to well supp}. Finally, set the precision parameter $\delta$ of the {\sc KKT} instance to 
$\delta = \frac{1}{2} \lambda \epsilon$,
which is polynomial time computable in the encoding size of $(\Gamma, \epsilon)$.

Get a solution to the {\sc KKT} instance. We have by construction that $L$ is a valid Lipschitz constant for $f$ and $\nabla f$, and that $\nabla f$ is indeed the gradient of $f$. Thus, the solution has to be a $\delta$-KKT point $\mu$, i.e. have KKT multipliers $\{ \tau_{ij} \in \R \}_{i,j=1}^{\ninfs,m_i}$ and $\{ \kappa_i \in \R \}_{i = 1}^\ninfs$ such that
\begin{align*}
    &\mu_{ij} \geq 0 \quad \forall i \in [\ninfs], \forall j \in [m_i]
    \\    
    &\sum_{j = 1}^{m_i} \mu_{ij} = 1 \quad \forall i \in [\ninfs]
    \\
    &\tau_{ij} \geq 0 \quad \forall i \in [\ninfs], \forall j \in [m_i]
    \\
    &\tau_{ij} = 0 \quad \textnormal{or} \quad \mu_{ij} = 0 \quad \forall i \in [\ninfs], \forall j \in [m_i]
    \\
    &| \nabla_{ij} \, \U(\mu) - ( - \tau_{ij} + \kappa_i ) | \leq \delta \quad \forall i \in [\ninfs], \forall j \in [m_i] \, .
\end{align*}
The point $\mu$ forms a valid strategy of $\Gamma$. Let us show that it is also an $\epsilon$-well-supported (CDT,GT)-equilibrium of $\Gamma$. Let $I_i$ be an info set with $\Fr(I_i \mid \mu) >0$, hence $\Fr(I_i \mid \mu) \geq \lambda$, and let $a_j \in A_{I_i}$ be an action with $\mu_{ij} = \mu( a_j \mid I_i) > 0$. Then, for any other action $a_{j'} \in A_{I_i}$, we have by Lemma \ref{derivative equals CDT deviation} and the KKT conditions:
\begin{align*}
    \EU&_{\CDT,\GT}(a_j \mid \mu, I_i) = \frac{\nabla_{ij} \, \U(\mu)}{\Fr(I_i \mid \mu)} \geq \frac{ - \tau _{ij} + \kappa_i - \delta}{\Fr(I_i \mid \mu)}
    \\
    &= \frac{\kappa_i - \delta}{\Fr(I_i \mid \mu)} \geq \frac{- \tau _{ij'} + \kappa_i - \delta}{\Fr(I_i \mid \mu)} \geq \frac{\nabla_{ij'} \, \U(\mu) - 2 \delta}{\Fr(I_i \mid \mu)}
    \\
    &= \EU_{\CDT,\GT}(a_{j'} \mid \mu, I_i) - \frac{2}{\Fr(I_i \mid \mu)} \cdot \delta 
    \\
    &\geq \EU_{\CDT,\GT}(a_{j'} \mid \mu, I_i) - \frac{2}{\lambda} \cdot \delta
    \\
    &= \EU_{\CDT,\GT}(a_{j'} \mid \mu, I_i) - \epsilon
\end{align*}
Thus $\mu$ forms an $\epsilon$-well-supported (CDT,GT)-equilibrium for~$\Gamma$, and therefore, also an $\epsilon$-(CDT,GT)-equilibrium for $\Gamma$.

\subsubsection{First CLS Hardness Result of Theorem \ref{CLS hardness}}

We will derive our first CLS hardness result from a KKT problem studied by \citet{BabichenkoR21}. Consider the maximization of a polynomial function $p : \R^\ninfs \to \R$ over the hypercube:
\begin{align}
\label{polynomial over hypercube max}
\begin{aligned}  
    &\max_{x \in \R^\ninfs} \, &&p(x)
    \\
    &\textnormal{s.t.} && x \in [0,1]^\ninfs \, .
\end{aligned}
\end{align}
Here, $p$ is again assumed to be represented in the Turing (bit) model $p(x) = \sum_{D \in \MB( d, \ninfs ) } \lambda_D \cdot \prod_{i \in [\ninfs]} x_i^{D_i}$.

\begin{defn}
    An instance of the problem {\sc KKTPolyOverCube} consists of a polynomial function $p : \R^\ninfs \to \R$ together with a precision value $\epsilon > 0$. A solution consists of a point $x \in [0,1]^\ninfs$ that is an $\epsilon$-KKT point of the maximization problem (\ref{polynomial over hypercube max}):
    \begin{align}
    \label{no multiplier KKT point hypercube}
    \begin{aligned}
        &\forall i \in [\ninfs]: \quad x_i > 0 \implies \nabla_i \, p(x) \geq -\epsilon
        \\
        &\forall i \in [\ninfs]: \quad x_i < 1 \implies \nabla_i \, p(x) \leq \epsilon
    \end{aligned}
    \end{align}
\end{defn}

\citet{BabichenkoR21} call this problem {\sc GD-FixedPoint}. 

\begin{lemma}[\citet{BabichenkoR21}]
    {\sc KKTPolyOverCube} is CLS-complete. Hardness holds even\footnote{Hardness even holds if each summand $\lambda_D \cdot \prod_{i \in [\ninfs]} x_i^{D_i}$ for $\sum_{D \in \MB( 5, \ninfs ) }$ has degree $5$ and is component-wise concave (and therefore the polynomial is also component-wise concave).} for polynomials in which every monomial has degree $5$.
\end{lemma}

We can now derive CLS-hardness of {\sc (CDT,GT)-equilibrium} by reducing from {\sc KKTPolyOverCube}.

Take an instance $(p : \R^{\ninfs} \to \R, \, \epsilon)$ of {\sc KKTPolyOverCube} where $p$ is known to only have degree $5$ monomials. There are at most $\binom{\ninfs  + 5 - 1}{5} = \mathcal{O}(\ninfs^5)$ many such monomials. Consider the modified polynomial
\begin{align}
\label{modified polynomial to 1-simplices}
\begin{aligned}  
    &\hat{p} : \bigtimes_{i = 1}^\ninfs \R^2 &&\to \R
    \\
    &\quad \Big( (x_{i1}, x_{i2}) \Big)_{i = 1}^\ninfs &&\mapsto p(x_{11}, x_{21}, \ldots, x_{\ninfs 1}) \, .
\end{aligned}
\end{align}
Create a game $\Gamma$ out of $\hat{p}$ as described in Appendix \ref{app: poly fcts to games} with the second variant, which is constructed in polynomial time in the encoding of $p$ since its degree is known to be $5$. Let $|\nds|$ be the number of nodes in $\Gamma$. Recall that for any info set $I_i$ of $\Gamma$ its visit frequency $\Fr( I_i \mid \mu)$ is constant in the used strategy $\mu$ and easily computable by (\ref{poly fct to game v2 freqs}). Thus, we are able to get a lower bound on the positive visit frequencies, which allows us to use Lemma \ref{lemma approx to well supp} later on. Note moreover that the visit frequencies are always bounded above by $|H|$.  Set
\[
\delta := \min \{ \frac{1}{3}, \frac{\epsilon^2}{(3 L |\nds|^2)^2} \}
\]
which is computable in polynomial time in the encoding of $p$ and $\epsilon$.

Recall from Appendix \ref{app: poly fcts to games} that the strategy utility function $\U$ of $\Gamma$ has the property $\U(\mu) = \hat{p}(\mu) = p(\mu_{11}, \ldots, \mu_{\ninfs 1})$ for any point $\mu \in \bigtimes_{i = 1}^\ninfs \R^2$. Therefore, $\nabla_{i1} \, \U(\mu) = \nabla_{i} \, p (\mu_{11}, \ldots, \mu_{\ninfs 1})$ and $\nabla_{i2} \, \U(\mu) = 0$ for all $\mu \in \bigtimes_{i = 1}^\ninfs \R^2$ and $i \in [\ninfs]$. 

Let $\pi^* \in \bigtimes_{i = 1}^\ninfs \Delta^1$ be a a solution to the {\sc (CDT,GT)-equilibrium}-instance $(\Gamma, \delta)$. By Lemma \ref{lemma approx to well supp}, we can compute from it a $(3 L |\nds| \sqrt{\delta})$-well-supported (CDT,GT)-equilibrium $\mu^* = (\mu_{ij}^*)_{i,j=1}^{\ninfs,2} \in \bigtimes_{i = 1}^\ninfs \Delta^1$ of $\Gamma$ in polynomial time. For us $\mu^*$ is, in particular, a $\frac{\epsilon}{|\nds|}$-well-supported (CDT,GT)-equilibrium because we set $\delta$ such that $3 L |\nds| \sqrt{\delta} \leq \frac{\epsilon}{|\nds|}$. We can now show that the point $x^* := (\mu_{i1}^*)_{i = 1}^\ninfs \in [0,1]^\ninfs$ is a solution to the {\sc KKTPolyOverCube}-instance $(p : \R^\ninfs \to \R, \, \epsilon)$.

Let us first consider any index $i \in [\ninfs]$ with $\Fr(I_i \mid \mu^*) = 0$. Then, by Lemma \ref{derivative equals CDT deviation}, we get $0 = \nabla_{i1} \, \U(\mu^*) = \nabla_{i} \, p(x^*)$. Thus, such indices $i$ always satisfy the conditions (\ref{no multiplier KKT point hypercube}) independent of the value $x_i^*$.

Now consider an index $i \in [\ninfs]$ with $\Fr(I_i \mid \mu^*) > 0$ and $0 < x_i^* = \mu_{i1}^*$. Then, due to $\mu^*$ being a $\frac{\epsilon}{|\nds|}$-well-supported (CDT,GT)-equilibrium, we get
\begin{align*}
    &\EU_{\CDT,\GT}(a_1 \mid \mu^*, I_i) \geq \max_{j = 1,2} \EU_{\CDT,\GT}(a_j \mid \mu^*, I_i) - \frac{\epsilon}{|\nds|}
    \\
    &\iff \EU_{\CDT,\GT}(a_1 \mid \mu^*, I_i) \geq \EU_{\CDT,\GT}(a_2 \mid \mu^*, I_i) - \frac{\epsilon}{|\nds|}
\end{align*}

With Lemma \ref{derivative equals CDT deviation}, we can therefore derive
\begin{align*}
    &\nabla_{i} \, p(x^*) = \nabla_{i1} \, \U(\mu^*) 
    \\
    &= \Fr(I_i \mid \mu^*) \cdot \EU_{\CDT,\GT}(a_1 \mid \mu^*, I_i) 
    \\
    &\geq \Fr(I_i \mid \mu^*) \cdot \Big( \EU_{\CDT,\GT}(a_2 \mid \mu^*, I_i) - \frac{\epsilon}{|\nds|} \Big) 
    \\
    &= \Fr(I_i \mid \mu^*) \cdot \EU_{\CDT,\GT}(a_2 \mid \mu^*, I_i) - \Fr(I_i \mid \mu^*) \cdot \frac{\epsilon}{|\nds|}
    \\
    &= \nabla_{i2} \, \U(\mu) - \Fr(I_i \mid \mu^*) \cdot \frac{\epsilon}{|\nds|} = - \Fr(I_i \mid \mu^*) \cdot \frac{\epsilon}{|\nds|}
    \\
    &\geq - |\nds| \cdot \frac{\epsilon}{|\nds|} = - \epsilon
\end{align*}

Now consider an index $i \in [\ninfs]$ with $\Fr(I_i \mid \mu^*) > 0$ and $1 > x_i^* = \mu_{i1}^* = 1 - \mu_{i2}^*$, i.e., $\mu_{i2}^* > 0$. Then with analogous arguments, we get
\begin{align*}
    &\EU_{\CDT,\GT}(a_2 \mid \mu^*, I_i) \geq \max_{j = 1,2} \EU_{\CDT,\GT}(a_j \mid \mu^*, I_i) - \frac{\epsilon}{|\nds|}
    \\
    &\iff \EU_{\CDT,\GT}(a_1 \mid \mu^*, I_i) \leq \EU_{\CDT,\GT}(a_2 \mid \mu^*, I_i) + \frac{\epsilon}{|\nds|}
\end{align*}

Similarly to the other case, we can therefore derive
\begin{align*}
    &\nabla_{i} \, p(x^*) = \nabla_{i1} \, \U(\mu^*) 
    \\
    &= \Fr(I_i \mid \mu^*) \cdot \EU_{\CDT,\GT}(a_1 \mid \mu^*, I_i) 
    \\
    &\leq \Fr(I_i \mid \mu^*) \cdot \Big( \EU_{\CDT,\GT}(a_2 \mid \mu^*, I_i) + \frac{\epsilon}{|\nds|} \Big)
    \\
    &= \nabla_{i2} \, \U(\mu) + \Fr(I_i \mid \mu^*) \cdot \frac{\epsilon}{|\nds|} = \Fr(I_i \mid \mu^*) \cdot \frac{\epsilon}{|\nds|}
    \\
    &\leq |\nds| \cdot \frac{\epsilon}{|\nds|} = \epsilon
\end{align*}

Therefore, all in all, point $x^*$ makes a solution to the {\sc KKTPolyOverCube}-instance $(p : \R^\ninfs \to \R, \, \epsilon)$. This finishes the polynomial time reduction from the search problem of {\sc Degree-5-KKTPolyOverCube} to the search problem {\sc (CDT,GT)-equilibrium}.

Note that in the above reduction, the constructed game $\Gamma$ has a tree depth of $6$. The root is a chance node with a number of outgoing edges that equals the number of monomials in $p$. Any other node of $\Gamma$ will have two outgoing actions. Therefore, CLS-hardness of {\sc (CDT,GT)-equilibrium} remains even if the game instance has a tree depth of $6$ and the player only has $2$ actions per info set.

\subsubsection{Second CLS Hardness Result of Theorem \ref{CLS hardness}}

The second and third CLS hardness results will both rely on the following game-theoretic problem studied by \citet{BabichenkoR21} (again).

Let $G$ be a $n$-player simultaneous game with action sets $\{A_i\}_{i \in [n]}$ of size $m_i := |A_i|$ and with utility functions $V_i: \bigtimes_{\ihat = 1}^n \Delta(A_{\ihat}) \to \R$ for each player $i$. A mixed action profile $x = (x_{ij})_{i,j = 1}^{n, m_i} \in \bigtimes_{i = 1}^n \Delta(A_{I_i})$ is called an $\epsilon$-Nash equilibrium of $G$ if for every player $i \in [n]$ and every action $a_j \in A_i$ of player $i$, we have $V_i(x) \geq V_i(x_{1 \cdot}, \ldots, x_{i-1 \cdot}, a_j, x_{i + 1 \cdot}, \ldots, x_{n \cdot}) - \epsilon$. In standard game theory notation, this would be phrased as $V_i(x) \geq V_i(a_j, x_{-i \cdot}) - \epsilon$. For us, it will be helpful to use the notation of a EDT deviation, in which this condition becomes $V_i(x) \geq V_i(x_{I_i \mapsto a_j}) - \epsilon$.

A $c$-polytensor game is a multiplayer simultaneous game presented in terms of payoff tables, one for each subset of $c$ players. Given a pure-strategy action profile of all players, a player's payoff consists of the sum of payoffs they get from the payoff tables of the $c$-subsets they belong to. For any constant $c$, such games with $n \geq c$ players and up to $m$ pure actions per player have a polynomial-sized representation in $n$ and $m$ because there are $ \leq n \cdot \binom{n}{c} \cdot m^c = \mathcal{O}(n^{c+1} \cdot m^5)$ many payoff entries overall. A $c$-polytensor identical interest game is a $c$-polytensor game in which every subgame associated with a c-subset yields the participating player the same utility. 

\begin{defn}
\label{def:cpii}
    An instance of the problem $c$-{\sc Polytensor-IdenticalInterest} consists of a $c$-polytensor identical interest game together with a precision value $\epsilon > 0$. A solution consist of an $\epsilon$-Nash equilibrium of the game. 
\end{defn}

\begin{lemma}[\citet{BabichenkoR21}]
    $5$-{\sc Polytensor-IdenticalInterest} is CLS-complete.
\end{lemma}

Without changing the complexity, we may assume that in instances of $5$-{\sc Polytensor-IdenticalInterest} every player has exactly $m$ actions (by possibly copying actions) and that the payoffs lie in $[0,1]$ (by shifting and rescaling utilities them).

We can now prove the second CLS hardness of {\sc (CDT,GT)-equilibrium} by reducing from $5$-{\sc Polytensor-IdenticalInterest}.

Take an instance $(G, \epsilon)$ of $5$-{\sc Polytensor-IdenticalInterest}. Let $n$ be the number of players of $G$, and $m$ be the number of pure strategies of player $i$.
Then, each subset of 5 players of $G$ has a table of  
$m^5$ payoffs in $[0,1]$ one for each of their possible pure profiles. Moreover, there will be $\binom{n}{5}$ many subsets of $5$ players among $n$ players. Collect all those subsets to $\Lambda := \{ i_{[5]} \subset [n] : |i_{[5]}| = 5 \}$. If $i_{[5]} \in \Lambda$ is a 5-subset and the players of $i_{[5]}$ play strategy profile $\mu_{i_{[5]}} \in \bigtimes_{i \in i_{[5]}} \Delta(A_{I_i})$, then denote the payoff that these players get in that subgame by $u^{i_{[5]}}(\mu_{i_{[5]}})$. Then, the overall utility function $V_i$ of a player $i \in [n]$ takes as input a strategy profile $\mu$ for the game $\Gamma$ and returns the payoff: 
\[
    V_i(\mu) = \sum_{i_{[5]} \in \Lambda : \, i \in i_{[5]}} u^{i_{[5]}} ( \mu_{i_{[5]}} )
\]

Construct an instance $(\Gamma, \delta)$ of {\sc (CDT,GT)-equilibrium} as follows. There will be $n$ info sets $I_1, \ldots, I_n$. The root of the tree of $\Gamma$ is a chance node having $\binom{n}{5}$ subtrees, one for each set $i_{[5]}$ of 5 players of $G$. The actions towards these subtrees are chosen uniform randomly, that is, with the same probability $1/\binom{n}{5}$. Consider a subtree $T_{i_{[5]}}$ associated with a set $i_{[5]}=\{i_1,\ldots,i_5\} \in \Lambda$. Then $T_{i_{[5]}}$ shall have depth 5 with depth layer $k-1$ for $k \in [5]$ corresponding to player $i_k$. Every node of $T_{i_{[5]}}$ of depth $k_1$ shall be assigned to info set $I_{i_k}$ and have $m$ outgoing edges labelled by the pure strategies of player $i_k$. Nodes of depth $5$ shall be terminal nodes. If terminal node $z$ of $\Gamma$ has action history $(i_{[5]}, j_1, \ldots, j_5)$, then it shall yield a payoff equal to $u^{i_{[5]}}(j_1, \ldots, j_5)$, that is, the payoff of the subgame associated with $i_{[5]}$ from the pure action profile $(j_1, \ldots, j_5)$. Define the precision parameter as $\delta := \epsilon/\binom{n-1}{4}$. The instance $(\Gamma, \delta)$ can be constructed in polynomial time in the encoding of $(G, \epsilon)$. Note that the game $\Gamma$ has tree depth $6$ and no absentmindedness.

Take an info set $I_i$ of $\Gamma$. The player reaches info set $I_i$ with probability one in every subtree $T_{i_{[5]}}$ with $i \in i_{[5]}$, independent of her strategy choice $\mu$. Thus, the overall reach probability of each info set $I_i$ of $\Gamma$ is exactly the number subsets $i_{[5]} \in \Lambda$ that contain $i$, divided by the number of $5$-subsets overall, which is $\binom{n-1}{4} / \binom{n}{5} = \frac{5}{n}$.\footnote{Note that this value would also serve as a lower bound on positive frequencies.} In particular, this reach probability is non-zero, so an approximate (CDT,GT)-equilibrium must be approximately optimal in every info set $I_i$ of $\Gamma$.

Since there is no absentmindedness in $\Gamma$, we get by Appendix \ref{app: cdt eqs vs edt eqs} for all info sets $I_i$, all strategies $\mu$, and all mixed actions $\alpha \in \Delta(A_{I_i})$:
\begin{align*}
    &\EU_{\CDT,\GT}(\alpha \mid \mu, I_i) 
    \\
    &= \EU_{\EDT,\GDH}(\alpha \mid \mu, I_i) 
    \\
    &\overset{(\ref{GDH utilities reformulation})}{=} \frac{1}{\Prob(I_i \mid \mu_{I_i \, \mapsto \alpha})} \cdot \sum_{\substack{z \in \term \\ I_i \cap \hist(z) \neq \emptyset}} \Prob(z \mid \mu_{I_i \, \mapsto \alpha}) \cdot u(z)
    \\
    &= \frac{n}{5} \cdot \sum_{i_{[5]} \in \Lambda : \, i \in i_{[5]}} \, \sum_{z \in T_{i_{[5]}}} \Prob(z \mid \mu_{I_i \, \mapsto \alpha}) \cdot u(z)
    \\
    &= \frac{n}{5} \cdot \sum_{i_{[5]} \in \Lambda : \, i \in i_{[5]}} \, \, \sum_{j_{[5]} \in \bigtimes_{\ihat \in i_{[5]}} A_{I_{\ihat}}} 
    \\
    &\, \quad \quad \quad \quad \quad \quad \quad \Prob(z_{i_{[5]}, \, j_{[5]}} \mid \mu_{I_i \, \mapsto \alpha}) \cdot u^{i_{[5]}}(j_{[5]})
    \\
    &\overset{(*)}{=} \frac{n}{5} \cdot \sum_{i_{[5]} \in \Lambda : \, i \in i_{[5]}} \, \frac{1}{\binom{n}{5}} \cdot u^{i_{[5]}} \Big( (\mu_{I_i \, \mapsto \alpha})_{i_{[5]}} \Big)
    \\
    &= \frac{n}{5} \cdot \frac{1}{\binom{n}{5}} \cdot V_i(\mu_{I_i \, \mapsto \alpha})
    \\
    &= \frac{1}{\binom{n-1}{4}} \cdot V_i(\mu_{I_i \, \mapsto \alpha})
\end{align*}

where in $(*)$ we used that for any $i_{[5]} = \{ i_1, \ldots, i_5\} \in \Lambda$ and strategy $\mu$ that 
\begin{align*}
    &u^{i_{[5]}} ( \mu_{i_{[5]}} ) 
    \\
    &=  \sum_{(j_k)_{k = 1}^5 \in \bigtimes_{k = 1}^5 A_{I_{i_k}}} u^{i_{[5]}}(j_1, \ldots, j_5) \cdot \prod_{k=1}^5 \mu_{i_{[5]}}( j_k \mid I_{i_k}) 
    \\
    &= \sum_{j_{[5]} \in \bigtimes_{i' \in i_{[5]}} A_{I_{i'}}} \binom{n}{5} \cdot \Prob(z_{i_{[5]}, \, j_{[5]}} \mid \mu_{I_i \, \mapsto \alpha}) \cdot u^{i_{[5]}}(j_{[5]}) \, .
\end{align*}

Let $\mu^* \in \bigtimes_{i = 1}^\ninfs \Delta(A_{I_i})$ be a solution to the {\sc (CDT,GT)-equilibrium}-instance $(\Gamma, \delta)$. Then we can show that the mixed action profile $(\mu_{i \cdot}^*)_{i=1}^{n}$ is a solution to $5$-{\sc Polytensor-IdenticalInterest} instance $(G, \epsilon)$, that is, an $\epsilon$-Nash equilibrium of $G$. Note that $\mu_{I_i \, \mapsto \mu_{i \cdot}^*}^* = \mu^*$. We obtain for all player $i \in [n]$ and all pure actions $j \in [m_i]$ of player $i$:
\begin{align*}
    &V_i(\mu^*) = 
    \\
    &= \frac{\binom{n-1}{4}}{\binom{n-1}{4}} \cdot V_{i} ( \mu_{I_i \, \mapsto \mu_{i \cdot}^*}^* )
    \\
    &= \binom{n-1}{4} \cdot \EU_{\CDT,\GT}(\mu_{i \cdot}^* \mid \mu^*, I_i)
    \\
    &\geq \binom{n-1}{4} \cdot \Big[ \EU_{\CDT,\GT}(a_j \mid \mu^*, I_i) - \delta \Big]
    \\
    &= \frac{\binom{n-1}{4}}{\binom{n-1}{4}} \cdot V_{i} ( \mu_{I_i \, \mapsto a_j}^* ) - \binom{n-1}{4} \cdot \delta
    \\
    &= V_{i} ( \mu_{I_i \, \mapsto a_j}^* ) + \epsilon \, .
\end{align*}

Therefore, a $\delta$-(CDT,GT)-equilibrium in $\Gamma$ makes an $\epsilon$-Nash equilibrium in $G$.

\subsubsection{Third CLS Hardness Result of Theorem \ref{CLS hardness}}

We prove the third CLS hardness of Theorem \ref{CLS hardness} by again reducing from $5$-{\sc Polytensor-IdenticalInterest}.

Take an instance $(G, \epsilon)$ of $5$-{\sc Polytensor-IdenticalInterest}. Use the same notation as in the last reduction proof from $5$-{\sc Polytensor-IdenticalInterest}. In particular, $G$ has $n$ players with $m$ actions, and payoffs lie in $[0,1]$. Moreover, we can assume $\epsilon < 1$.

\paragraph{Corresponding Game} Construct an instance $(\Gamma, \delta)$ of {\sc (CDT,GT)-equilibrium} as follows. $\Gamma$ is a game tree of depth $5$ and all nodes of $\Gamma$ will be player nodes that belong to the same info set $I$. The action set $A_I$ is defined as $\{ (i,j) : i\in[n], j\in[m] \}$. Nodes of depth $\leq 4$ all have out-degree $n \cdot m$. At depth $5$, we have $(n \cdot m)^5$ many nodes that shall be terminal nodes. If the path to a terminal node $z$ is $(i_{[5]}, j_{[5]}) = ((i_1,j_1),\ldots,(i_5,j_5))$, and if $\eta(i_{[5]}) \in \{1, \ldots, 5\}$ is the number of distinct values present\footnote{Note our shift in notation from the previous reduction proof. Now, tuple $i_{[5]}$ may contain duplicates from the set of players $[n]$.} in $i_{[5]}$ of $z$, then $z$ shall yield the player a payoff of \begin{enumerate}
    \item $M_2 \cdot \eta(i_{[5]})$ if $\eta(i_{[5]}) \leq 4$, and
    \item $M_2 \cdot \eta(i_{[5]}) + u^{i_{[5]}}(j_{[5]})$ if $\eta(i_{[5]}) = 5$.
\end{enumerate}
Note that we have constant visit frequency $\Fr(I \mid \mu) = \Fr(I) = 5$ independent of the strategy $\mu$ used in $\Gamma$. This makes Lemma \ref{lemma approx to well supp} applicable to $\Gamma$. Let $L$ be the Lipschitz constant $L$ from Lemma \ref{lemma approx to well supp}. Then we can define the values
\begin{align*}
    M_1 &:= 2 \cdot 100 \cdot n^9 m^4 \cdot \frac{1}{\epsilon}
    \\
    \delta_1 &:= \frac{1}{5} \Big( \frac{1}{n} - \frac{1}{M_1} \Big)^4 \cdot \frac{\epsilon}{2}
    \\
    \delta_2 &:= \Big(\frac{\delta_1}{3 L \sum_{k = 0}^{5} (nm)^k} \Big)^2
    \\
    M_2 &:= (\delta_1 + n^4) \cdot M_1^4
\end{align*}
where the sum $\sum_{k = 0}^{5} (nm)^k$ represents the number of nodes $|H|$ in $\Gamma$. To give some intuition, $\delta_2$ is chosen such that Lemma \ref{lemma approx to well supp} can be applied to get a $\delta_1$-well-supported (CDT,GT)-equilibrium. $M_2$ is chosen large enough (in comparison to the payoffs in $[0,1]$) such that for equilibrium play, the player of $\Gamma$ mostly cares about mixing up the players $i$ from which she chooses an action in an approximate uniform fashion. $M_1$ is chosen large enough to counterbalance the errors in this approximate uniform mixing, and $\delta_1$ is chosen small enough to recover an $\epsilon$-Nash equilibrium at the end.

Let us show that a $\delta_2$-(CDT,GT)-equilibrium of $\Gamma$ gives rise to an $\epsilon$-Nash equilibrium of $G$. For that, we first have to collect some further observations about $\Gamma$. 

\paragraph{Ex-ante Utility in $\Gamma$} Let us describe the expected utility $\U(\mu)$ of a strategy $\mu = (\mu_{ij})_{i,j = 1}^{n,m} \in \Delta(A_I)$ of $\Gamma$. Split it into $\U(\mu) = M_2 \phi(\mu) + \psi(\mu)$ where the first part shall come from the $M_2 \cdot \eta(i_{[5]})$ and the second part from the $u^{i_{[5]}}(j_{[5]})$. The latter is simply
\begin{align}
\label{cls hardness psi fct}
    \psi(\mu) = \sum_{(i_{[5]}, j_{[5]}) : \eta(i_{[5]}) = 5} u^{i_{[5]}}(j_{[5]}) \prod_{k=1}^5 \mu_{i_k j_k} \, .
\end{align}
For the former, denote with $p_i = \sum_j \mu_{ij}$ the total weight on actions that came from player $i$ of the original game. Consider the stochastic experiment of drawing $5$ items out of the set of players $[n]$ according to the probability distribution $p = (p_i)_i$. Let $\chi_i \in \{0,1\}$ be the random variable that denotes whether player $i$ was drawn at least once in the stochastic experiment. Then we can calculate the expected number of distinct players that are drawn in the experiment as
\begin{align*}
    \phi(\mu) &= \mathbb{E}[\sum_{i \in [n]} \chi_i] = \sum_{i \in [n]} \mathbb{E}[\chi_i] 
    \\
    &= \sum_{i \in [n]} \Prob(\chi_i = 1) = \sum_{i \in [n]} \Big( 1 - \Prob(\chi_i = 0) \Big)
    \\
    &= n - \sum_{i \in [n]} (1 - p_i)^5 \, .
\end{align*}

\paragraph{De Se Utility in $\Gamma$} Next, we describe $\EU_{\CDT,\GT}(a_{i j} \mid \mu, I)$ for an action $a_{i j}$ at $I$ with the help of Lemma \ref{derivative equals CDT deviation}.

\begin{align*}
    \EU&_{\CDT,\GT}(a_{i j} \mid \mu, I) = \frac{1}{\Fr(I)} \nabla_{i j} \, \U(\mu)  = \frac{1}{5} \nabla_{i j} \, \U(\mu) 
    \\
    &= \frac{M_2}{5} \cdot \nabla_{i j} \, \phi(\mu) + \frac{1}{5} \nabla_{i j} \, \psi(\mu)
    \\
    &= \frac{M_2}{5} \cdot 5 (1 - p_{i})^4 
    \\
    &\, \quad + \frac{1}{5} \sum_{\substack{(i_{[4]}, j_{[4]}) : \\ \eta(i_{[4]}) = 4 \, , i \notin i_{[4]}}} u^{(i , i_{[4]})}(j , j_{[4]}) \prod_{k=2}^5 \mu_{i_k j_k}
    \\
    &\overset{(*)}{\in} \Big[ M_2 (1 - p_{i})^4, M_2 (1 - p_{i})^4 + \frac{1}{5} \binom{n-1}{4} \Big]
    \\
    &\subseteq \Big[ M_2 (1 - p_{i})^4, M_2 (1 - p_{i})^4 + n^4 \Big] \, .
\end{align*}
In $(*)$, we used that all factors in the indexed sum lie in $[0,1]$.

\paragraph{Solution Recovery} Let $\mu'$ be a $\delta_2$-(CDT,GT)-equilibrium of $\Gamma$. By Lemma \ref{lemma approx to well supp}, we can then compute a $\delta_1$-well-supported (CDT,GT)-equilibrium $\mu$ of $\Gamma$ from $\mu'$. In the remaining part of the proof, we will only work with $\mu$.

\paragraph{In equilibrium, every player gets to play similarly often} Denote $p_i = \sum_j \mu_{ij}$ again. Take the two players $i^* \in \argmax_i p_i$ and $i_* \in \argmax_i p_i$ that are played the most and least often under $\mu$. Choose any actions $j^*,j_* \in [m]$ of players $i^*$ and $i_*$ respectively. Then we have

\begin{align*}
    \delta_1 &\geq \EU_{\CDT,\GT}(a_{i_* j_*} \mid \mu, I) - \EU_{\CDT,\GT}(a_{i^* j^*} \mid \mu, I) 
    \\
    &\geq  M_2 (1 - p_{i_*})^4 - M_2 (1 - p_{i^*})^4 - n^4 
    \\
    &\geq  M_2 \cdot \Big[ (1 - p_{i^*} + p_{i^*} - p_{i_*} )^4 - (1 - p_{i^*})^4 \Big] - n^4
    \\
    &\overset{(\dagger)}{\geq} M_2 \cdot (p_{i^*} - p_{i_*} )^4 - n^4 \, .
\end{align*}
In $(\dagger)$ we first used the binomial theorem for the components $1 - p_{i^*} \geq 0$ and $p_{i^*} - p_{i_*} \geq 0$ to split up the first exponent term. Next, the term $(1 - p_{i^*})^4$ canceled out and we dropped three other positive terms. From this, we conclude
\begin{align*}
    0 \leq p_{i^*} - p_{i_*} \leq \bigg| \sqrt[4]{\frac{\delta_1 + n^4}{M_2}} \, \bigg| = \frac{1}{M_1} \, .
\end{align*}
All in all we get that in a $\delta_1$-well-supported (CDT,GT)-equilibrium -- such as $\mu$ -- the actions $a_{ij}$ of each player $i \in [n]$ get played with summed probability
\[
    p_i \in \big[ p_{i_*} \, , \, p_{i^*} \big] \subseteq \Big[\frac{1}{n} - \frac{1}{M_1} \, , \, \frac{1}{n} + \frac{1}{M_1} \Big] \, .
\]

\paragraph{Recovering an $\epsilon$-Nash equilibrium} We now have everything needed to complete the proof. Define the corresponding strategy profile $\pi = (\pi_{i \cdot})_{i = 1}^n \in \bigtimes_{i = 1}^n \Delta^{m - 1}$ in $G$ to $\mu$ as $\pi_{ij} := \frac{\mu_{ij}}{p_i}$ for each player $i \in [n]$ and action $j \in [m]$. This is well-defined because $p_i \geq \frac{1}{n} - \frac{1}{M_1} > 0$. We show that $\pi$ is an $\epsilon$-Nash equilibrium of $G$ by contradiction. Assume $\pi$ is not an $\epsilon$-Nash equilibrium. Then, by \citet{Chen06}, it is also not an $\epsilon$-well-supported Nash equilibrium , that is, there exists a player $i \in [n]$ with actions $j,j' \in [m]$ such that action $j'$ is played with positive probability $\pi_{i j'} > 0$ and such that playing $j$ yields her more than $\epsilon$ more utility than playing $j'$, given that the other players play $\pi_{-i \cdot}$. Formally, this means
\begin{align}
\label{CLS hardness, not ws NE}
\begin{aligned}
    &\sum_{\substack{(i_{[4]}, j_{[4]}) : \\ \eta(i_{[4]}) = 4 \, , i \notin i_{[4]}}} u^{(i , i_{[4]})}(j , j_{[4]}) \prod_{k=2}^5 \pi_{i_k j_k}
    \\
    &\, \quad \quad > \epsilon + \sum_{\substack{(i_{[4]}, j_{[4]}) : \\ \eta(i_{[4]}) = 4 \, , i \notin i_{[4]}}} u^{(i , i_{[4]})}(j' , j_{[4]}) \prod_{k=2}^5 \pi_{i_k j_k} \, .
\end{aligned}
\end{align}
Since $\frac{1}{n} - \frac{1}{M_1} \leq p_{\ihat} \leq \frac{1}{n} + \frac{1}{M_1}$ for all $\ihat \in [n]$, we can derive
\begin{align*}
    \sum_{\substack{(i_{[4]}, j_{[4]}) : \\ \eta(i_{[4]}) = 4 \, , i \notin i_{[4]}}} &u^{(i , i_{[4]})}(j , j_{[4]}) \prod_{k=2}^5 \pi_{i_k j_k} 
    \\
    &=     \sum u^{(i , i_{[4]})}(j , j_{[4]}) \prod_{k=2}^5 \frac{\mu_{i_k j_k}}{p_{i_k}}
    \\
    &\leq \sum u^{(i , i_{[4]})}(j , j_{[4]}) \prod_{k=2}^5 \frac{\mu_{i_k j_k}}{\frac{1}{n} - \frac{1}{M_1}}
    \\
    &= \Big( \frac{1}{\frac{1}{n} - \frac{1}{M_1}} \Big)^4 \cdot \sum u^{(i , i_{[4]})}(j , j_{[4]}) \prod_{k=2}^5 \mu_{i_k j_k}  \, ,
\end{align*}
and analogously
\begin{align*}
    \sum_{\substack{(i_{[4]}, j_{[4]}) : \\ \eta(i_{[4]}) = 4 \, , i \notin i_{[4]}}} &u^{(i , i_{[4]})}(j' , j_{[4]}) \prod_{k=2}^5 \pi_{i_k j_k}
    \\
    &\geq \sum u^{(i , i_{[4]})}(j' , j_{[4]}) \prod_{k=2}^5 \frac{\mu_{i_k j_k}}{\frac{1}{n} + \frac{1}{M_1}}
    \\
    &= \Big( \frac{1}{\frac{1}{n} + \frac{1}{M_1}} \Big)^4 \cdot \sum u^{(i , i_{[4]})}(j' , j_{[4]}) \prod_{k=2}^5 \mu_{i_k j_k}  \, .
\end{align*}
Observe the connection to the previously defined function $\psi$ here:
\[
    \sum_{\substack{(i_{[4]}, j_{[4]}) : \\ \eta(i_{[4]}) = 4 \, , i \notin i_{[4]}}} u^{(i , i_{[4]})}(j , j_{[4]}) \prod_{k=2}^5 \mu_{i_k j_k} = \nabla_{i j} \, \psi(\mu) \, ,
\]
and
\[
    \sum_{\substack{(i_{[4]}, j_{[4]}) : \\ \eta(i_{[4]}) = 4 \, , i \notin i_{[4]}}} u^{(i , i_{[4]})}(j' , j_{[4]}) \prod_{k=2}^5 \mu_{i_k j_k} = \nabla_{i j'} \, \psi(\mu) \, .
\]
Inserting the bounds into (\ref{CLS hardness, not ws NE}) yields
\begin{align*}
    \Big( &\frac{1}{\frac{1}{n} - \frac{1}{M_1}} \Big)^4 \nabla_{i j} \, \psi(\mu) > \epsilon + \Big( \frac{1}{\frac{1}{n} + \frac{1}{M_1}} \Big)^4 \nabla_{i j'} \, \psi(\mu)
    \\
    &= \epsilon + \Big( \frac{1}{\frac{1}{n} - \frac{1}{M_1}} \Big)^4 \nabla_{i j'} \, \psi(\mu) 
    \\
    &\, \quad - \bigg[  \Big( \frac{1}{\frac{1}{n} - \frac{1}{M_1}} \Big)^4 - \Big( \frac{1}{\frac{1}{n} + \frac{1}{M_1}} \Big)^4 \bigg] \cdot \nabla_{i j'} \, \psi(\mu)
    \\
    &= \epsilon + \Big( \frac{1}{\frac{1}{n} - \frac{1}{M_1}} \Big)^4 \nabla_{i j'} \, \psi(\mu) 
    \\
    &\, \quad - n^4 M_1^4 \underbrace{\Big(  \frac{1}{(M_1 - n)^4} - \frac{1}{(M_1 + n)^4} \Big)}_{\textnormal{Denote this term as } (\dagger)} \cdot \nabla_{i j'} \, \psi(\mu)
    \\
    &= \ldots \textnormal{ to be continued under } (\circ) \, .
\end{align*}

We have $(\dagger) \geq 0$ as well as
\begin{align*}
    (\dagger) &=  \frac{(M_1 + n)^4 - (M_1 - n)^4}{(M_1 - n)^4 (M_1 + n)^4} 
    \\
    &= \frac{ 8 M_1^3 n + 8 M_1 n^3}{(M_1 - n)^4 (M_1 + n)^4} 
    \\
    &\leq \frac{ 16 M_1^3 n }{(M_1 - n)^4 (M_1 + n)^4} 
    \\
    &\leq \frac{ 16 M_1^3 n }{(M_1 - \frac{M_1}{2})^4 (M_1 + \frac{M_1}{2})^4}  
    \\
    &= \frac{16}{\frac{3^4}{2^8}} \frac{ M_1^3 n }{M_1^8} \leq 100 \frac{ n }{M_1^5}  \, .
\end{align*} 
Moreover, we have 
\begin{align*}
    0 \leq \nabla_{i j'} \, \psi(\mu) \leq \sum_{\substack{(i_{[4]}, j_{[4]}) : \\ \eta(i_{[4]}) = 4 \, , i \notin i_{[4]}}} 1 \cdot 1 \leq (nm)^4 \, .
\end{align*}
Therefore, we can continue the inequality chain at $(\circ)$ with
\begin{align*}
    (\circ) &\geq \epsilon + \Big( \frac{1}{\frac{1}{n} - \frac{1}{M_1}} \Big)^4 \nabla_{i j'} \, \psi(\mu) - n^4 M_1^4  \cdot 100 \frac{ n }{M_1^5}  \cdot (nm)^4
    \\
    &= \epsilon + \Big( \frac{1}{\frac{1}{n} - \frac{1}{M_1}} \Big)^4 \nabla_{i j'} \, \psi(\mu) - 100 \frac{n^9 m^4}{M_1}
    \\
    &= \epsilon + \Big( \frac{1}{\frac{1}{n} - \frac{1}{M_1}} \Big)^4 \nabla_{i j'} \, \psi(\mu) - \frac{\epsilon}{2} \, ,
\end{align*}
where we inserted for the value of $M_1$. All in all, we derived
\begin{align*}
    \Big( \frac{1}{\frac{1}{n} - \frac{1}{M_1}} \Big)^4 \nabla_{i j} \, \psi(\mu) > \Big( \frac{1}{\frac{1}{n} - \frac{1}{M_1}} \Big)^4 \nabla_{i j'} \, \psi(\mu) + \frac{\epsilon}{2} \, .
\end{align*}
Thus, we can conclude with our previous observations about the de se utility in $\Gamma$ that
\begin{align*}
    \EU&_{\CDT,\GT}(a_{i j} \mid \mu, I) = \frac{M_2}{5} \cdot \nabla_{i j} \, \phi(\mu) + \frac{1}{5} \nabla_{i j} \, \psi(\mu)
    \\
    &= \frac{M_2}{5} \cdot 5 (1 - p_{i})^4  + \frac{1}{5} \nabla_{i j} \, \psi(\mu)
    \\
    &= \frac{M_2}{5} \cdot \nabla_{i j'} \, \phi(\mu) + \frac{1}{5} \nabla_{i j} \, \psi(\mu)
    \\
    &> \frac{M_2}{5} \cdot \nabla_{i j'} \, \phi(\mu) + \frac{1}{5} \cdot \Big[ \nabla_{i j'} \, \psi(\mu) + \Big( \frac{1}{n} - \frac{1}{M_1} \Big)^4 \cdot \frac{\epsilon}{2} \Big]
    \\
    &= \frac{M_2}{5} \cdot \nabla_{i j'} \, \phi(\mu) + \frac{1}{5} \nabla_{i j'} \, \psi(\mu) + \delta_1
    \\
    &= \EU_{\CDT,\GT}(a_{i j'} \mid \mu, I) + \delta_1 \, .
\end{align*}
Hence, we derived that an action $(i,j')$ of $\Gamma$ has positive play probability
\[ 
    \mu_{i j'} = \pi_{i j'} \cdot p_i \geq \pi_{i j'} \cdot (\frac{1}{n} - \frac{1}{M_2}) > 0 \, ,
\]
under $\mu$  while simultaneously being more than $\delta_1$ dominated by another action $(i,j)$. This contradicts $\mu$ being a $\delta_1$-well-supported (CDT,GT)-equilibrium. The contradiction completes our proof that $\pi$ is an $\epsilon$-Nash equilibrium.

\subsection{Proof of Theorem \ref{dec probs of eqs are np hard}}

\begin{thm*}
    The following problems are all NP-hard. Unless NP = ZPP, there is also no FPTAS for these problems.
    \begin{enumerate}[nolistsep]
    \item[(1a.)] Given $\Gamma$ and $t \in \Q$, is there a (CDT,GT)-equilibrium of $\Gamma$ with ex-ante utility $\geq t$?
    \item[(1b.)] Given $\Gamma$, an info set $I$ of $\Gamma$ and $t \in \Q$, is there a (CDT,GT)-equilibrium $\mu$ such that $\Fr( I \mid \mu) > 0$, and such that the player has a (CDT,GT)-expected utility $\geq t$ upon reaching $I$?
    \item[(1c.)] Given $\Gamma$, an info set $I$ of $\Gamma$ and $t \in \Q$, is there a strategy $\mu$ of $\Gamma$ such that $\Fr( I \mid \mu) > 0$, and such that the player has a (CDT,GT)-expected utility $\geq t$ upon reaching $I$?
    \item[(2a.)] Given $\Gamma$ and $t \in \Q$, is there an (EDT,GDH)-equilibrium of $\Gamma$ with ex-ante utility $\geq t$?
    \item[(2b.)] Given $\Gamma$, an info set $I$ of $\Gamma$ and $t \in \Q$, is there an (EDT,GDH)-equilibrium $\mu$ such that $\Prob( I \mid \mu) > 0$, and such that the player has an (EDT,GDH)-expected utility $\geq t$ upon reaching $I$?
    \item[(2c.)] Given $\Gamma$, an info set $I$ of $\Gamma$ and $t \in \Q$, is there a strategy $\mu$ of $\Gamma$ such that $\Prob( I \mid \mu) > 0$, and such that the player has an (EDT,GDH)-expected utility $\geq t$ upon reaching $I$?
    \item[(3a.)] Given $\Gamma$ and $t \in \Q$, do all (EDT,GDH)-equilibria of $\Gamma$ have ex-ante utility $\geq t$?
    \item[(3b.)] Given $\Gamma$, an info set $I$ of $\Gamma$ and $t \in \Q$, do all (EDT,GDH)-equilibria $\mu$ with $\Prob( I \mid \mu) > 0$ yield the player an (EDT,GDH)-expected utility $\geq t$ upon reaching~$I$?
    \end{enumerate}
\end{thm*}

We reduce all those decision problems from the problem in Proposition \ref{hardness and inapprox of ex ante opt}, which we will henceforth call {\sc ExAnteOpt-D}.

First note that any single-player extensive-form game with imperfect recall $\Gamma$ has an ex-ante optimal strategy since the maximization problem (\ref{exantemaxprobl}) is about a continuous polynomial function over a compact domain. Moreover, observe that $(\Gamma, t)$ is a yes instance for {\sc ExAnteOpt-D} (by definition) if and only if there is a strategy $\mu$ in $\Gamma$ with $\U(\mu) \geq t$ if and only if there is an ex-ante optimal strategy $\mu^*$ in $\Gamma$ with $\U(\mu^*) \geq t$. 

(1a.)\ and (2a.): 
\\
Let $(\Gamma, t)$ be an instance for {\sc ExAnteOpt-D}. Without transforming, we can choose $(\Gamma, t)$ as the instance to the problems (1a.)\ and (2a.). Then, it will be a yes instance of (-a.)\ if and only if there is a (CDT,GT)-equilbrium or, resp., (EDT,GDH)-equilibrium in $\Gamma$ with ex-ante utility $\geq t$ if and only if (by Lemma \ref{equilibrium hierarchy}) there is an ex-ante optimal strategy in $\Gamma$ with with ex-ante utility $\geq t$ if and only if (by the comment above) $(\Gamma, t)$ is a yes instance for {\sc ExAnteOpt-D}.

(1b.), (1c.), (2b.), and (2c.): 
\\
Let $(\Gamma, t)$ be an instance for {\sc ExAnteOpt-D}.  Let us refer to the root of $\Gamma$ with $h_0$. We construct a new game $\Gamma'$ by adding a new artificial game start: Copy the game tree of $\Gamma$. Add a new node $h_{-1}$ to it which shall represent root of $\Gamma'$. Connect $h_{-1}$ to $h_0$ by one edge, called action $a_{-1}$, and assign $h_{-1}$ to a new info set $I_{-1}$ that is added to $\infs_*$. The corresponding instance shall be $(\Gamma', I_{-1}, t)$.

Observe that any strategy $\mu$ for $\Gamma$ corresponds to a strategy $\mu'$ in $\Gamma'$ that behaves like $\mu$ at info sets $I$ that were inherited from $\Gamma$ and that takes the only viable action $a_{-1}$ at info set $I_{-1}$ with a $100 \%$ certainty. Moreover, info set $I_{-1}$ has a visit frequency and reach probability of $1$ for any strategy $\mu'$ of $\Gamma'$ because they occur in the history of any terminal node exactly once in the beginning. Most crucially, we also get that the (CDT,GT)-expected utility (resp. (EDT,GDH)-expected utility) of $\mu'$ upon reaching info set $I_{-1}$ is equal to the ex-ante expected utility of $\mu'$ in $\Gamma'$ and of corresponding strategy $\mu$ in $\Gamma$. 

We obtain: \smallskip
\\
$\cdot$ \, $(\Gamma', I_{-1}, t)$ is a yes instance for problem (1b.)\ (resp. (2b.))
\begin{enumerate}[align=left]
\item[$\implies$] $(\Gamma', I_{-1}, t)$ it is a yes instance for problem (1c.)\ (resp. (2c.))
\item[$\implies$] there is a strategy $\mu'$ in $\Gamma'$ with (CDT,GT)-expected utility (resp. (EDT,GDH)-expected utility) $\geq t$ upon reaching $I_{-1}$
\item[$\implies$] corresponding strategy $\mu$ in $\Gamma$ has ex-ante utility $\geq t$
\item[$\implies$] $(\Gamma, t)$ is a yes instance for {\sc ExAnteOpt-D}.
\end{enumerate}
\noindent
and for the other direction: \medskip
\\
$\cdot$ \, $(\Gamma, t)$ is a yes instance for {\sc ExAnteOpt-D} 
\begin{enumerate}
\item[$\implies$] there exists an ex-ante optimal strategy $\mu^*$ of $(\Gamma, t)$ with ex-ante utility $\geq t$
\item[$\implies$] corresponding strategy $(\mu^*)'$ in $\Gamma'$ is ex-ante optimal for $\Gamma'$, and it has (CDT,GT)-expected utility (resp. (EDT,GDH)-expected utility) $\geq t$ upon reaching $I_{-1}$
\item[$\implies$] there exists a (CDT,GT)-equilibrium in $\Gamma'$ (resp. (EDT,GDH)-equilibrium) with (CDT,GT)-expected utility (resp. (EDT,GDH)-expected utility) $\geq t$ upon reaching $I_{-1}$
\item[$\implies$] $(\Gamma', I_{-1}, t)$ is a yes instance for problem (1b.)\ (resp. (2b.))
\item[$\implies$] $(\Gamma', I_{-1}, t)$ is a yes instance for problem (1c.)\ (resp. (2c.)).
\end{enumerate}

which completes the reduction. Note that $G'$ has one additional info set and tree depth in comparison to $G$.

(3a.)\ and (3b.):
\\
For (3a.), consider the reductions for (2a.)\ again and assume the starting instance $(\Gamma, t)$ of {\sc ExAnteOpt-D} has one info set only. By Proposition \ref{hardness and inapprox of ex ante opt} we know that even with such instances {\sc ExAnteOpt-D} is NP-hard and conditionally inapproximable. But an (EDT,GDH)-equilibrium of such an instance must be an ex-ante optimal strategy (seen by Lemma \ref{EDT eq to NE of Poly}). Therefore, each (EDT,GDH)-equilibrium of $\Gamma$ must have the same (ex-ante optimal) value in ex-ante utility. Thus, deciding whether all (EDT,GDH)-equilibrium exceeds a target ex-ante utility (problem (3a.)) coincides with deciding whether one (EDT,GDH)-equilibrium does that (problem (2a.)). 

An analogous argument holds for (3b.)\ by considering the reductions for (2b.)\ for starting instances $(\Gamma, t)$ with only one info set. There, we observe that (EDT,GDH)-equilibria of $\Gamma'$ correspond exactly to the ex-ante optimal strategies in $\Gamma$, and must therefore all promise the same (EDT,GDH)-expected utility upon reaching $I_{-1}$.

\end{document}